\documentclass[prd,showpacs,nofootinbib,preprintnumbers,preprint]{revtex4}

\usepackage{amsmath}
\usepackage{amsfonts}
\usepackage{amssymb}
\usepackage{mathrsfs}
\usepackage{graphicx}
\usepackage{hyperref}
\usepackage[caption=false]{subfig}
\usepackage{blindtext, subfig}

\graphicspath{{SD/}}

\begin{document}

\title{ 
Photon trapping in static axially symmetric spacetime}
\author{D. V. Gal'tsov} \email{galtsov@phys.msu.ru}
\affiliation{Faculty of Physics,
Moscow State University, 119899, Moscow, Russia,\\Kazan Federal University, 420008 Kazan, Russia}
\author{K.V.~Kobialko} \email{kobyalkokv@yandex.ru}
\affiliation{Faculty of Physics, Moscow State University, 119899, Moscow, Russia}

\begin{abstract} 
Recently, several new characteristics have been introduced to describe null 
geodesic structure of strong gravitational field, such as photon regions, transversely trapping surfaces and some generalizations. They give an alternative and concise way to describe lensing and shadow features of compact objects with strong gravitational field without recurring to complete integration of the geodesic equations. Here we test this construction in the case of the Weyl metrics when geodesic equations are non-separable, and thus can not be integrated analytically, while the above characteristic surfaces and regions can be described in a closed form. We develop further our formalism for a class of static axially symmetric spacetimes introducing more detailed specification of transversely trapping surfaces in terms of their principal curvatures. Surprisingly, we find in the static case without spherical symmetry certain features, such as photon regions, previously known in the Kerr space. These photon regions can be regarded as photon spheres, ``thickened''  due to oblateness of the metric.  
 
\end{abstract}

\maketitle


\setcounter{page}{2}

\setcounter{equation}{0}

\section{Introduction}
Currently, there are two main ways of describing images of BH and other dark compact objects \cite{Cardoso:2019rvt} resulting from the scattering of electromagnetic radiation on them. The first is the direct integration of geodesic equations for null rays, adopted in the theory of gravitational lensing and the formation of shadows \cite{Perlick:2004tq,Abdujabbarov:2012bn,A.deVries,Wei:2013kza,Atamurotov:2013sca,Abdikamalov:2019ztb,Takahashi:2005hy,Bambi:2008jg,Amarilla:2010zq,Amarilla:2013sj,Shaikh:2018lcc,Cunha:2015yba,Bohn:2014xxa,Cunha:2017wao,Cunha:2018acu}. In the standard black hole theory based on the no-hair theorem and the Kerr-Newman family of solutions \cite{Johannsen:2015mdd,Johannsen:2016uoh}, such integration is  possible in view of the separability of the Hamilton-Jacobi equation due to the existence of the third  Carter's  integral of motion  associated with the Killing-Yano tensor \cite{Kubiznak:2007kh} in addition to the energy and the azimuthal angular momentum associated with stationarity and axial symmetry. In the case of more general stationary and axially symmetric metrics that have been used recently \cite{Johannsen:2011dh,Hartle:1967he,Hartle:1968si} to extract more general predictions outside the Kerr paradigm, the third integral  no longer exist, and analytic solution of geodetic equations is impossible. Moreover, as one would expect for nonintegrable systems, the motion demonstrates areas of chaos \cite{Lukes,Dolan:2019gsr}, which complicates the task of visualization.

An alternative and/or complimentary way to characterize  optical properties of  nonstandard static and stationary metrics is to use characteristic surfaces related to trapping of certain classes of null geodesics. Recall that the event horizon in the Schwarzschild space $ r = 2M $ is an absolute trapping surface, which captures all null and timelike geodesics,  with any initial conditions. Other trapping surfaces, which we propose  to call ``relative trapping surfaces'' capture some of null geodesics. The most famous is the {\em photon sphere } (PS) \cite{Virbhadra:1999nm}. In the Schwarzschild case it is the sphere of the radius $ r = 3M $, which is the union of unstable circular photon orbits. It is a trapping surface for photons moving tangentially to it or inward. More precisely, tangential photons remain in the photon sphere forever, while those moving with a non-zero inward component of the momentum will be absorbed by the black hole.   

Such surfaces, however, are rare in more general spacetimes. Several uniqueness theorems were established for vacuum and electrovacuum \cite{Cederbaum,Yazadjiev:2015hda,Yazadjiev:2015mta,Yazadjiev:2015jza,Rogatko,Cederbaumo,Yoshino:2016kgi} stating that in the static case the only asymptotically flat metrics admitting PS are the known black holes, an existence of PS replacing the assumption of a regular event horizon.  In some non-spherical  static spacetimes, properties of the photon spheres are shared by the {\em photon surfaces} of non-spherical shape  \cite{Claudel:2000yi},   examples of spacetime admitting non-spherical  photon  surfaces are presented by asymptotically non-flat metrics
(vacuum C-metric, Melvin's solution of Einstein-Maxwell theory and its generalizations including the dilaton field \cite{Gibbons}). 

Meanwhile, in the stationary spacetimes photon surfaces generically do not exist, being, in some sense, disintegrated by rotation. The Taub-NUT metric, though belongs to this class, is an exception, whose existence is explained by local $SO(3)$ symmetry which is  preserved.  In the Kerr metric the circular photon orbits exist  in the equatorial plane \cite{Paganini:2016pct}. Non-equatorial orbits with constant Boyer-Lindquist radii no longer belong to any plane, but lie on the surface of a sphere instead (``spherical photon orbits'' \cite{Teo}). But such surfaces are not photon spheres, which by definition should be densely filled. In the Kerr case every spherical orbit corresponds to certain value of the impact parameter defined as ratio of the angular momentum to the energy. Altogether spherical orbits now will a three-dimensional photon region (PR) \cite{Grenzebach,Grenzebach:2015oea,Charbulak:2018wzb}, which can be regarded as ``thickened'' photon sphere. 
 Note that the closed photon orbits may exist in more general spacetimes in which case the name of fundamental or/and spheroidal photon orbits was suggested \cite{Pappas:2018opz,Glampedakis:2018blj,Cunha:2017eoe}.

  More general trapping surfaces, proposed in
\cite{Yoshino1} and further studied in \cite{Galtsov:2019bty}, were called transversely trapping surfaces (TTS). In Schwarzschild case TTS are the spheres  with the radii $ r \leq 3M $.
These are defined as surfaces  such that initially tangent photons either remain in them or move inward; these do exist in Kerr and more general stationary axially symmetric metrics.    The totality of TTSs form a three-dimensional region (TTR), which apparently will be invisible if one looks at the Schwarzschild black hole illuminated from behind. Further generalizations   suggested in \cite{Galtsov:2019bty} include {\em partial} TTS (PTTS) which are non-closed surfaces (contrary to the definition in \cite{Yoshino1}) of the shape of a spherical cap. Finally, it is reasonable to introduce {\em anti}-TTS and PTTS (abbreviated as ATTS and APPTS, the corresponding regions --- TTR and PTTR respoctively) replacing {\em inward} to {\em outward} in the above definitions. ATTSs and APTTSs {\em can} be seen by an asymptotic observer.

The above listed characteristic surfaces fill all the three-space of the Kerr-Newman metric thus providing complete description of photon trapping surfaces around the standard black hole.  As we have shown in \cite{Galtsov:2019bty}, the Kerr-Newman metrics with different values of three parameters $M, J, Q$ can be cast into four optical types depending on existence of various combinations of PR, TTS and PTTS characteristic surfaces.

From astrophysical perspective, the current interest extends to non-standard metrics for compact objects as well; such a research aims to extract wider predictions for their possible optical images which recently became subject of the observational study. Here we will investigate a popular family of oblate deformations of the Schwarzschild metric known as Zipoy-Voorhees (ZV) solutions or gamma-metrics \cite{Zipoy,Voorhees:1971wh,Kodama:2003ch,Malafarina:2004yw,Herrera:1998rj,Boshkayev:2015jaa,Griffiths}, which are the static axially symmetric but not spherically symmetric solutions belonging to the Weyl class.
This spacetime is non-separable for geodesic motion, so its description in terms of characteristic trapping surfaces and regions is especially relevant.

On general grounds, ZV family of solutions has no photon surfaces, and closed circular photon orbits exist only in the equatorial plane.  
As about TTS, it was proved  \cite{Yoshino1} that all closed TTS spatial sections in a static asymptotically flat space should have the topology of the sphere $ S^2 $, but not necessarily spherically symmetric. Thus, it is natural to assume that for static asymptotically flat spaces that go beyond uniqueness theorems, such as the Zipoy-Voorhees solution, and therefore not allowing the existence of photon spheres, it is natural to consider some deformed   transversely trapping surfaces of spherical topology.

 Mathematically, an important constructive property of the photon  surfaces is expressed by the theorem asserting that they are  conformally invariant and totally umbilical hypersurfaces in spacetime \cite{Okumura,Senovilla:2011np}, such that that their second fundamental form is pure trace, i.e. is proportional to the induced metric or other words hypersurface have equal principal curvatures. Therefore, contraction of the second fundamental form with null geodesic generator will be zero. In the case of TTS  such contractions are non-zero, and their sign serve an indicator for TTS and ATTS. This observation is essential for subsequent analysis.

In this paper, the problem of photon capture and the description of the optical structure of static axially symmetric vacuum spaces will be studied.  We will demonstrated that in spaces of this type new geometrical structures arise, such as non-spherical TTS surfaces, and a non-spherical photon region, having a direct analog with the same spherically symmetric structures
in Kerr-like solutions and admitting an elegant formulation in terms of principal curvatures of hypersurfaces. 

The plan of the paper is as follows. In Section~2, we define characteristic hypersurfaces in space-time based on the properties of the second fundamental form and principal curvatures. In Section~3, we describe the features and properties of the Zipoy-Voorhees and Curzon-Chazy solutions. Section~4 cover the main aspects of an explicit description of null geodesic and real observation.
In Section~5, we study the possibility of constructing spherically symmetric TTS hypersurfaces in the ZV metric and their connection with the optical properties of the solution. In Section~6 a new type of non-spherical TTS in the ZV is being studied. We study its connection with the explicit behavior of geodesics and develop construction methods both numerical and analytical. In Section~7 introduce the concept of a generalized photon region, also based on the properties of the principal curvatures of the hypersurfaces and analyze its connection with the behavior of isotropic geodesics too.  In Section~8 the optical properties of the ZV solution with with deformation parameter $\delta<1$ are studied in more detail.  In Section~9 we compare the predictions obtained from the analysis of characteristic photon surfaces in the ZV space with an explicit shadow structure for different values of the deformation parameter. In conclusion, the main results are summarized and some perspectives are discussed.

\setcounter{equation}{0}

\section{Transversely trapping surfaces in static axially symmetric spacetime}

Consider a static axially symmetric Weyl metric with two commuting Killing vectors $\partial_t$ and $\partial_\phi$ and signature ($-,+,+,+$) in adapted Erez-Rosen coordinates $x^\mu =t,\,r,\,\theta,\,\phi$ \cite{Herrera:2013hm}: 
\begin{equation}
d\hat{s}^2=\hat{g}_{\mu\nu}dx^\mu dx^\nu=\hat{g}_{tt}dt^2+\hat{g}_{rr}dr^2+\hat{g}_{\theta\theta}d\theta^2+\hat{g}_{\phi\phi}d\phi^2,
\label{a0} 
\end{equation} 
where $\hat{g}_{tt}$, $\hat{g}_{rr}$, $\hat{g}_{\theta\theta}$ and $\hat{g}_{\phi\phi}$ are functions of $r$ and $\theta$ only. Consider a timelike three-dimensional hypersurface in the parametric form, choosing coordinates  $\sigma^A=t,\,\theta,\,\phi$ on the hypesurface coinciding with the coordinates in the bulk:
\begin{equation} 
x^\mu=f^\mu(\sigma^A)=(t,f(\theta),\theta,\phi). 
\label{a01} 
\end{equation} 

 Let $g_{AB}=\hat{g}_{\mu\nu}f^{\mu}_{A}f^{\nu}_{B}$ and $H_{AB}=\hat{g}_{\mu\nu}(\hat{\nabla}_{f_A}f^{\mu}_{B})n^{\nu}$  be the {\em first } and the {\em second } fundamental forms of the hypersurface, where $f^\mu_A=\partial f^\mu/\partial \sigma^A$ are three linearly independent tangent vectors to the hypersurface, $\hat{\nabla}$ - is the covariant derivative associated with $\hat{g}$, and $n^{\nu}$ is the {\em outer} normal. In the explicit representation (\ref{a01}), the tangents vectors $f^\mu_a=\delta^\mu_a$, $a=(t,\phi)$, coincide with two commuting Killing vectors $\partial_a$ and thus $f^a_\theta=0$. Consequently, the unit normal vector to the hypersurface will have two non-zero components $n^{\nu}=(0,n^{r},n^{\theta},0)$ and the induced metric $g_{AB}$ and the second quadratic form $H_{AB}$ will be diagonal:
\begin{align} 
&g_{ab}=\hat{g}_{\mu\nu}f^{\mu}_{a}f^{\nu}_{b}=\hat{g}_{ab}=0, \quad  g_{a\theta}=\hat{g}_{\mu\nu}f^{\mu}_{a}f^{\nu}_{\theta}=\hat{g}_{aa}f^{a}_{\theta}=0,\\
&H_{ab}=\hat{g}_{\mu\nu}(f^\rho_a\hat{\nabla}_{\rho}f^{\mu}_{b})n^{\nu}=\hat{g}_{\mu\nu}(\hat{\Gamma}^\mu_{ab})n^{\nu}=0,\\
&H_{a\theta}=\hat{g}_{\mu\nu}(f^\rho_a\hat{\nabla}_{\rho}f^{\mu}_{\theta})n^{\nu}=\hat{g}_{\mu\nu}(\hat{\Gamma}^\mu_{a\rho}f^{\rho}_{\theta})n^{\nu}=\hat{g}_{\theta\theta}(\hat{\Gamma}^\theta_{a\theta}f^{\theta}_{\theta})n^{\theta}+\hat{g}_{rr}(\hat{\Gamma}^r_{a r}f^{r}_{\theta})n^{r}=0,
\end{align} 
since the Christoffel symbols for metric (\ref{a0}) have the properties $\hat{\Gamma}^{(r,\theta)}_{ab}=\hat{\Gamma}^{(r,\theta)}_{a(r,\theta)}=0$.

As was demonstrated in  \cite{Claudel:2000yi,Yoshino1,Galtsov:2019bty}, to analyze the optical properties of these surfaces, it suffices to study the value of the second fundamental form $\mathrm H$ on the null vectors $\dot{\gamma}$ in the hypersurface, tangent to affinely parameterized null geodesics: 
\begin{equation} 
\mathrm H(\dot{\gamma},\dot{\gamma})=H_{tt}\dot{t}^2+H_{\theta\theta}\dot{\theta}^2+H_{\phi\phi}\dot{\phi}^2.
\label{a02} 
\end{equation} 
Recall that the null geodesic initially touching the spatial section of the hypersurface remains on it if  $\mathrm H(\dot{\gamma},\dot{\gamma})=0$ for any tangent  null vectors, which corresponds to the definition of the photon surface in  \cite{Claudel:2000yi}. If we require the fulfillment of this condition only for {\em some} isotropic vectors (with a fixed ratio of the orbital angular momentum and the energy), we meet the definition of the photon region \cite{Galtsov:2019bty}. Otherwise, the null geodesic leaves the hypersurface spatial section in the opposite direction to the normal  $n^{\nu}$ if $\mathrm H(\dot{\gamma},\dot{\gamma})>0$ (inwards), corresponding to TTS \cite{Yoshino1}, and in the direction of the normal if $\mathrm H(\dot{\gamma},\dot{\gamma})<0$ (outwards), which meets the definition of ATTS  \cite{Galtsov:2019bty}. If such surfaces are not closed, we called  them {\em partial}. 

We now use the isotropy condition for the tangent vector $\dot{\gamma}$ 
\begin{equation} 
ds^2=g_{AB}d\sigma^A d\sigma^B=g_{tt}\dot{t}^2+g_{\theta\theta}\dot{\theta}^2+g_{\phi\phi}\dot{\phi}^2=0,
\label{a03} 
\end{equation} 
where $ g_{AB} $ is an induced metric satisfying $ g_{tt} <0 $, $ g_{\theta \theta}> 0 $ and $g_{\phi \phi}>0$ in the outer region. Then, eliminating $\dot{t}$ from (\ref{a02}), we get: 
\begin{equation} 
\mathrm H(\dot{\gamma},\dot{\gamma})=\tilde{H}_{\theta\theta}\dot{\theta}^2+\tilde{H}_{\phi\phi}\dot{\phi}^2, \quad \tilde{H}_{\theta\theta}=H_{\theta\theta}-\frac{g_{\theta\theta}}{g_{tt}} H_{tt}, \quad \tilde{H}_{\phi\phi}=H_{\phi\phi}-\frac{g_{\phi\phi}}{g_{tt}} H_{tt},
\label{a04}
\end{equation}  
with no restrictions on  $\dot{\theta}$ and $\dot{\phi}$. Consequently the sign of $\mathrm H(\dot{\gamma},\dot{\gamma})$ is determined by the signs of $\tilde{H}_{\theta\theta}$ and $\tilde{H}_{\phi\phi}$ for all relevant $\theta$, which we would like to keep separately as given in the Table \ref{Tip}. 

It will be useful to introduce more special characteristic surfaces (A)TTS${}_\theta$ for which the component $\tilde{H}_{\theta\theta}=0$ of the second fundamental form identically equals to zero for all relevant $\theta$ and, consequently, the $\tilde{H}_{\phi\phi}$ component is (negative) positive definite. Such surfaces contain closed {\em meridional} isotropic geodesics with   $\dot{\phi}=0$ (a special case of spheroidal or fundamental photon orbits \cite{Glampedakis:2018blj,Cunha:2017eoe}). These surfaces $\tilde{H}_{\theta\theta}=0$ trap tangential geodesics with  $\dot{\phi}\neq0$. 

By analogy, we can also introduce the surfaces (A)TTS${}_\phi$ with $\tilde{H}_{\phi\phi}=0$ which contain closed geodesic with $\dot{\theta}=0$, such as circular photon orbits. Though generically such non-degenerate fully three-dimensional hypersurfaces may not exist, such properties can be met on some two-dimensional timelike section or boundary of more general three-dimensional hypersurfaces. In this case, the unit normal field to such two-dimensional sections will be induced from the three-dimensional hypersurfaces by continuity. Such reduced two-dimensional surfaces (A)TTS${}_\phi$ do  not allow tangent  geodesics with $\dot{\theta}=0$ to propagate (against) along the normal, though they can leave them in a tangent direction. 

One can give an explicit geometric interpretation to these definitions as follows. In the case of diagonal fundamental forms with which we restrict here, the principal curvatures (eigenvalues of a pair of quadratic forms) of a hypersurface are defined as $\lambda_A=H_{AA}/g_{AA}$. Respectively,
\begin{equation} 
\tilde{H}_{\theta\theta}=g_{\theta\theta}(\lambda_\theta-\lambda_t), \quad \tilde{H}_{\phi\phi}=g_{\phi\phi}(\lambda_\phi-\lambda_t).
\label{a05}
\end{equation} 
Clearly, the photon surface is a hypersurface  of equal principal curvatures $\lambda_t=\lambda_\theta=\lambda_\phi$ for all relevant $\theta$, or a totally {\em umbilic} surface \cite{Claudel:2000yi,Okumura,Senovilla:2011np}. The (A)TTS${}_\theta$ - surfaces correspond to equality of only two principal curvatures $\lambda_t=\lambda_\theta$ while $\lambda_\phi\geq\lambda_t$($\lambda_\phi\leq\lambda_t$). Respective properties of (A)TTS${}_\phi$ are shown in Table \ref{Tip}. Below in the section VII we will show that there exists an one-parameter family of hypersurfaces containing both of them. The corresponding three-dimensional region can be interpreted as a {\em generalized photon region}.  
\begin{table}[h]
\caption{Types of TTS} 
\begin{center} Инн
\begin{tabular}{|c|c|c|c|c|c|c|c|}
\hline
  & (P)TTS & (P)ATTS &  TTS${}_\theta$ & ATTS${}_\theta$ & TTS${}_\phi$ & ATTS${}_\phi$ & PS\\
\hline
$\tilde{H}_{\theta\theta}$ & $\geq0$ & $\leq0$ & $=0$& $=0$& $\geq0$& $\leq0$& $=0$\\
\hline
$\tilde{H}_{\phi\phi}$ & $\geq0$ & $\leq0$ & $\geq0$ & $\leq0$& $=0$ & $=0$& $=0$\\
\hline
$\lambda_\theta$ & $\lambda_\theta\geq\lambda_t$& $\lambda_\theta\leq\lambda_t$ & $\lambda_\theta=\lambda_t$ & $\lambda_\theta=\lambda_t$ &$\lambda_\theta\geq\lambda_t$ & $\lambda_\theta\leq\lambda_t$ & $\lambda_\theta=\lambda_t$ \\
 $\lambda_\phi$ & $\lambda_\phi\geq\lambda_t$&  $\lambda_\phi\leq\lambda_t$ & $\lambda_\phi\geq\lambda_t$ & $\lambda_\phi\leq\lambda_t$ &  $\lambda_\phi=\lambda_t$ & $\lambda_\phi=\lambda_t$ & $\lambda_\phi=\lambda_t$ \\
\hline
\end{tabular}
\end{center}
\label{Tip}
\end{table}
  
To simplify the expressions  further, we rewrite the metric components (\ref{a0}) in an exponential form
\begin{equation}
d\hat{s}^2=\hat{g}_{\mu\nu}dx^\mu dx^\nu=-e^\alpha dt^2+e^\lambda dr^2 +e^\beta d\theta^2+e^\gamma d\phi^2,
\label{a1}
\end{equation} 
where $\alpha$, $\lambda$, $\beta$ and $\gamma$ are functions of $r$ and $\theta$. It is assumed that the solution is asymptotically flat at  $r\rightarrow\infty$, and we are in the domain of positivity of all  the metric functions. 

Explicit calculations for the metric parametrization (\ref{a1}) give: 
\begin{align}
&\xi\tilde{H}_{\theta\theta}=2f''-f'(\partial_\theta\alpha_\lambda+\partial_\theta\beta_\lambda+e^{\lambda_\beta}f'^2\partial_\theta \alpha_\lambda)+(f'^2(\partial_r\alpha_\beta+\partial_r\lambda_\beta)+e^{\beta_\lambda}\partial_r\alpha_\beta), \label{a3a}\\
&\zeta\tilde{H}_{\phi\phi}=\partial_r\alpha_\gamma-e^{\lambda_\beta}f'\partial_\theta\alpha_\gamma, 
\label{a3b}
\end{align} 
where $\alpha_\beta\equiv\alpha-\beta$ etc., $\xi$ and $\zeta$ are  strictly positive functions, the primes denoting derivatives $\partial_\theta$. 

The equations (\ref{a3a}, \ref{a3b}) are non-linear and the construction of an explicit solution of $f$ for specific metrics usually is problematic, though in the particular case of  $f(\theta)\equiv r_T={\rm const}$ we get the simple expressions:
 \begin{equation} 
\xi'\tilde{H}_{\theta\theta}=\partial_r\alpha_\beta, \quad \zeta\tilde{H}_{\phi\phi}=\partial_r\alpha_\gamma.
\label{a4}
\end{equation} 

 Finally, note that for the closed TTS, there exists an inequality of Penrose type \cite{Yoshino1}, which in our case takes the form:
 \begin{equation} 
2\pi\int^{\pi}_{0}\sqrt{e^\gamma (e^\beta+e^\lambda f'^2)}d\theta=S\leq 4\pi(3M)^2.
\label{a5}
\end{equation}  
An important characteristic will be   the ratio of the TTS area to its maximal allowable value
 \begin{equation} 
\kappa=S/(4\pi(3M)^2)\leq1.
\label{a6}
\end{equation}

\setcounter{equation}{0}

\section{Zipoy-Voorhees}

As a non-trivial example of a static axially symmetric asymptotically flat spacetime  not admitting  the standard photon surfaces, we will consider the Zipoy-Voorhees (ZV) vacuum solution \cite{Zipoy,Voorhees:1971wh,Griffiths,Kodama:2003ch} which in the Erez-Rosen coordinates (\ref{a1}) reads \cite{Malafarina:2004yw,Herrera:1998rj}:
\begin{align}
&\alpha=\delta \ln \left(1-\frac{2m}{r}\right), \quad \gamma=\ln \left((r^2-2mr)\sin^2\theta\right)-\alpha, \\
&\lambda=(\delta^2-1)\ln\left(r^2-2mr\right)+(1-\delta^2)\ln\left(r^2-2mr+m^2\sin^2\theta\right)-\alpha, \\
&\beta=\delta^2\ln\left(r^2-2mr\right)+(1-\delta^2)\ln\left(r^2-2mr+m^2\sin^2\theta\right)-\alpha,
\label{b1}
\end{align}
This solution can be interpreted as an axially symmetric deformation of the Schwarzschild metric with the deformation parameter $\delta\geq0$, to which it reduces for $\delta=1 $. For $\delta=2$ it can be interpreted as a two-center solution, a particular non-rotation version of the Tomimatsu-Sato metric \cite{Kodama:2003ch}. This metric does not allow for complete separation of  variables in the geodesic equations and exhibits features of geodesic chaos \cite{Lukes}. Here we wish to study its photon trapping properties using the above described tools.

In general, the solution has the following features \cite{Kodama:2003ch,Malafarina:2004yw,Herrera:1998rj}. For all $\delta>0$ there is a curvature singularity at $r= 0$ originating from the Schwarzschild singularity. For $\delta<2$, except for $\delta\neq0,1$, there is also a naked  singularity at $r=2m$ for any $\theta$.  In the case $\delta>2$ the Kretchmann scalar vanishes at   $r=2m$ for $\theta=0,\pi$ so there is no curvature singularity there. Moreover, at least for $\delta=2,3$ and $\delta\geq4$, these varieties are the Killing horizons (the case $\delta=2$  demanding more accurate analysis, see  \cite{Kodama:2003ch}). The Arnowitt-Deser-Misner mass is equal to  $M=m\delta$. The outer domain in which we are interested in extends as $r>2m$.

In the following, we will compare solutions of the same physical mass $ M $ by varying the value of the deformation parameter $ \delta $. These manifolds will be called  "ZV$\delta$" for brevity. We also include the limiting configuration    $\delta\rightarrow\infty$ with fixed $M$, which corresponds to Chazy-Curzon solution in spherical coordinates \cite{Griffiths,Chazy,Curzon,Montero-Camacho:2014bza}:
\begin{align}
&\alpha=-\frac{2M}{r}, \quad \gamma=\ln \left(r^2\sin^2\theta\right)-\alpha, \\
&\lambda=-\frac{M^2\sin^2\theta}{r^2}-\alpha, \quad  \beta=-\frac{M^2\sin^2\theta}{r^2}+\ln r^2-\alpha.
\label{b2}
\end{align} 
We will refer to this solutions as "ZVI". The evaluation of the Kretchmann scalar in this case gives:
\begin{align}
R^{\mu\nu\lambda\rho}R_{\mu\nu\lambda\rho}|_{r\rightarrow 0}=-\frac{8e^{2M(M \sin^2\theta-2r)/r^2}P(r,\theta)}{r^{12}}|_{r\rightarrow 0}=
\begin{cases}
0, \quad \theta=0,\pi \\
\infty, \quad \theta\neq 0,\pi.
\end{cases}
\end{align} 
where $P(r,\theta)$ is some polynomial. Thus curvature singularity has a complicated directionally dependent structure and can be clarified by making a more appropriate choice of coordinates. It was confirmed  that the curvature singularity has the structure of a ring actually. In addition, this space-time has an invariantly defined hypersurface $r_J= M$, on which the cubic invariant of the Weyl tensor $J$ vanishes \cite{Griffiths}. At any time, this has the topology of a 2-sphere which surrounds the singularity. Note that, as well as in the ZV spacetime, the chaos effects on the  Poincare  sections indicates the non-integrability of the corresponding geodesic system \cite{Dolan:2019gsr}. 

\setcounter{equation}{0}

\section{Geodesics and observers}
In what follows, we would like to establish the correspondence between the structure of the TTSs, an explicit behavior of geodesics and the optical appearance of the metric. By virtue of staticity and axial symmetry, there are two Killing vectors $\partial_t$, $\partial_\phi$ and, accordingly, two integrals of motion:
\begin{align} 
E=\dot{t}e^{\alpha}, \quad L=\dot{\phi}e^{\gamma}, \quad \rho=L/E,
\label{c1}
\end{align} 
namely the energy and the azimuthal component of the angular momentum, whose ratio $\rho$ is the azimuthal impact parameter. We now turn to the description of isotropic geodesics in an explicit form. Computing the Christoffel symbols for the metric (\ref{a1}), we obtain the following two-dimensional system of partial differential equations for $r,\,\theta$:
\begin{align}
&\ddot{r}+\frac{1}{2}e^{-\lambda}\left(e^{\lambda}\partial_r\lambda\dot{r}^2+2 e^{\lambda}\partial_\theta\lambda\dot{r}\dot{\theta}-e^{\beta}\partial_r\beta\dot{\theta}^2+e^{-\alpha}E^2\partial_r\alpha-e^{-\gamma}L^2\partial_r\gamma\right)=0,\\
&\ddot{\theta}+\frac{1}{2}e^{-\beta}\left(-e^{\lambda}\partial_\theta\lambda\dot{r}^2+2 e^{\beta}\partial_r\beta\dot{r}\dot{\theta}+e^{\beta}\partial_\theta\beta\dot{\theta}^2+e^{-\alpha}E^2\partial_\theta\alpha-e^{-\gamma}L^2\partial_\theta\gamma\right)=0,
\label{c2}
\end{align}
with suitable initial conditions. An explicit from of these equations for the ZV metric, as well as some details on the porperties of geodesics motion can be found in \cite{Herrera:1998rj,Boshkayev:2015jaa}. The initial conditions can be related to some observer located at the point $(r_{O},\theta_{O})$, supposedly in the asymptotic region. The light rays are then traced back to the compact object whose image we are interested in. Following
 \cite{Grenzebach},  introduce an orthonormal frame:
\begin{align}
e_0=e^{\alpha/2}dt, \quad e_1=e^{\beta/2}d\theta, \quad e_2=e^{\gamma/2}d\phi, \quad e_3=-e^{\lambda/2}dr,
\label{c3}
\end{align}
using which any  null   vector $\dot{\xi}$ can be presented as:
\begin{align}
\dot{\xi}=Q(-e^0+\sin\vartheta \cos\psi e^1+\sin\vartheta \sin\psi e^2+\cos\vartheta e^3),
\label{c4}
\end{align}
where $0\leq\vartheta\leq\pi$, $0\leq\psi\leq2\pi$  are coordinates on the observer’s celestial sphere with a north pole directed to the origin $ r = 0 $. Eliminating $Q$, it is easy to get the following explicit expressions:  
\begin{align}
&\dot{r}=-E e^{-(\alpha+\lambda)/2}\cos\vartheta, \quad \dot{\theta}=E  e^{-(\alpha+\beta)/2}\sin\vartheta\cos\psi,\\
&\rho=e^{(\gamma-\alpha)/2}\sin\vartheta\sin\psi.
\label{c4}
\end{align}  
It is also convenient to enter the coordinates $ (X, Y) $ of the stereographic projection of the celestial sphere onto the plane \cite{Grenzebach}:
\begin{align}
X=-2 \tan \left(\vartheta/2\right)\sin(\psi), \quad Y=-2 \tan \left(\vartheta/2\right)\cos(\psi).
\label{c5}
\end{align}  
The point $X=0, Y=0$ corresponds to the north pole on the celestial sphere and, accordingly, the direction to the origin $r=0$. It is easy to see that the line $X=0$ corresponds to the set of initial conditions for geodesics with a zero value of the impact parameter $\rho=0$ and, accordingly, with a zero orbital momentum $L=0$. It is clear from the equations (\ref{c1}) that such geodesics always lie in the plane $ \phi = {\rm const} $.  
Since the  solutions are axially symmetric,  for any observation point the image symmetry holds: $ X \rightarrow-X $. 

In the case of the Zipoy-Voorhees solution, due to $\mathbb Z_2 $ symmetry for the observer in the equatorial plane, there will obviously be another symmetry: $ Y \rightarrow-Y $. Hence, in particular, one can estimate the maximum of transverse size of the shadow $ \Delta X $ for such observers. It follows that the maximum or at least the extremum of the shadow size must lie at the axis of symmetry $Y = 0 $. Then from the last equation (\ref{c4}) and the relations (\ref{c5}) we immediately get the following estimates of the shadow transverse sizes:
\begin{align}
\vartheta_{\rm max}=\arcsin \rho_{\rm max} e^{(\alpha-\gamma)/2}|_{\theta_{O}=\pi/2}, \quad \Delta X=4\tan \left(\arcsin \rho_{\rm max} e^{(\alpha-\gamma)/2}/2\right)|_{\theta_{O}=\pi/2}.
\label{c6}
\end{align}  
The way how $ \rho_{\rm max} $ can be found for the shadow of the Zipoy-Voorhees solution will be clear shortly.

\setcounter{equation}{0}

\section {Spherical TTS}

In the case of Kerr-Newman family of metrics, as we have shown in  \cite{Galtsov:2019bty} the set of spherical (P)(A)TTS together with the photon region creates a complete filling of their three-dimensional spatial section and characterizing the basic properties of the solution optical structure. The question arises whether such a construction is possible in the case of spherically asymmetric static metrics. 

In view of the  restriction on the geometry of closed TTS to have a spatial section of spherical topology \cite{Yoshino1}, let us consider as a first approximation the hypersurfaces of the form $f(\theta)\equiv r_T = {\rm const} $ in Erez-Rosen coordinates\cite{Herrera:2013hm}. As we will see below, this choice leads to the coarsest, but at the same time the  simplest possible description of the optical properties of ZV spacetme. From (\ref{a4}) for (P)TTS it is easy to get the following conditions:
\begin{align} 
&r_T(r_T-2m) (r_T-m (1+2\delta)) + 
 m^2\delta (r_T \delta - m (\delta+2)) \sin\theta^2\leq0, \label{d1a}\\
&r_T-m(1+2\delta)\leq0. 
\label{d1b}
\end{align} 
In particular, for $ \theta = 0, \pi /2 $ we get:
\begin{align}
(r_T-m \delta) (r_T-m (\delta + 2)) \leq0, \quad r_T-m (1 + 2 \delta) \leq0,
\label{d2}
\end{align}
whence it is easy to see that closed TTSs will exist in the following range of values Table \ref{ZV1}.

\begin{table}[ht]
\caption{Spherical TTSs}
\begin{center}
\begin{tabular}{|c|c|c|c|c|c|c|c|}
\hline
$\delta$ & $\delta\leq1/2$ & $\frac{1}{2}<\delta\leq1$ &  $1<\delta<2$ & $\delta\geq2$ & $\delta=\infty$ \\
\hline
$r_T$ & $-$ & $r_T\leq m+2m\delta$ & $r_T\leq2m+m\delta$ & $m\delta\leq r_T\leq 2m+m\delta$ & $r_T=M$\\
\hline
Тип & III & I & I & I & I\\
\hline
\end{tabular}
\end{center}
\label{ZV1}
\end{table}
 
\begin{figure}[tb]
\centering
\subfloat[][ZV1]{
  	\includegraphics[scale=0.41]{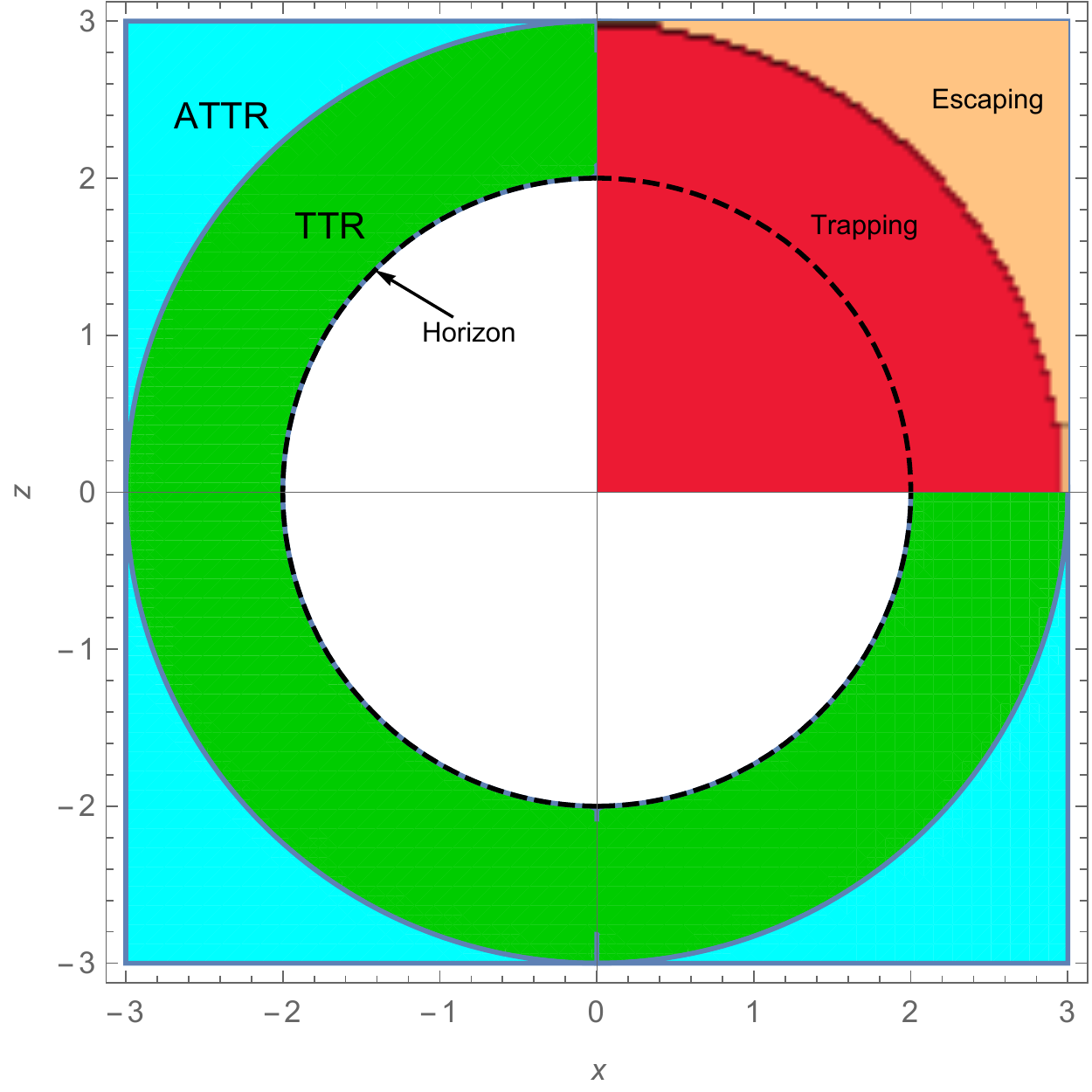}
		\label{TTSZV1S}
 }
\subfloat[][ZV2]{
  	\includegraphics[scale=0.41]{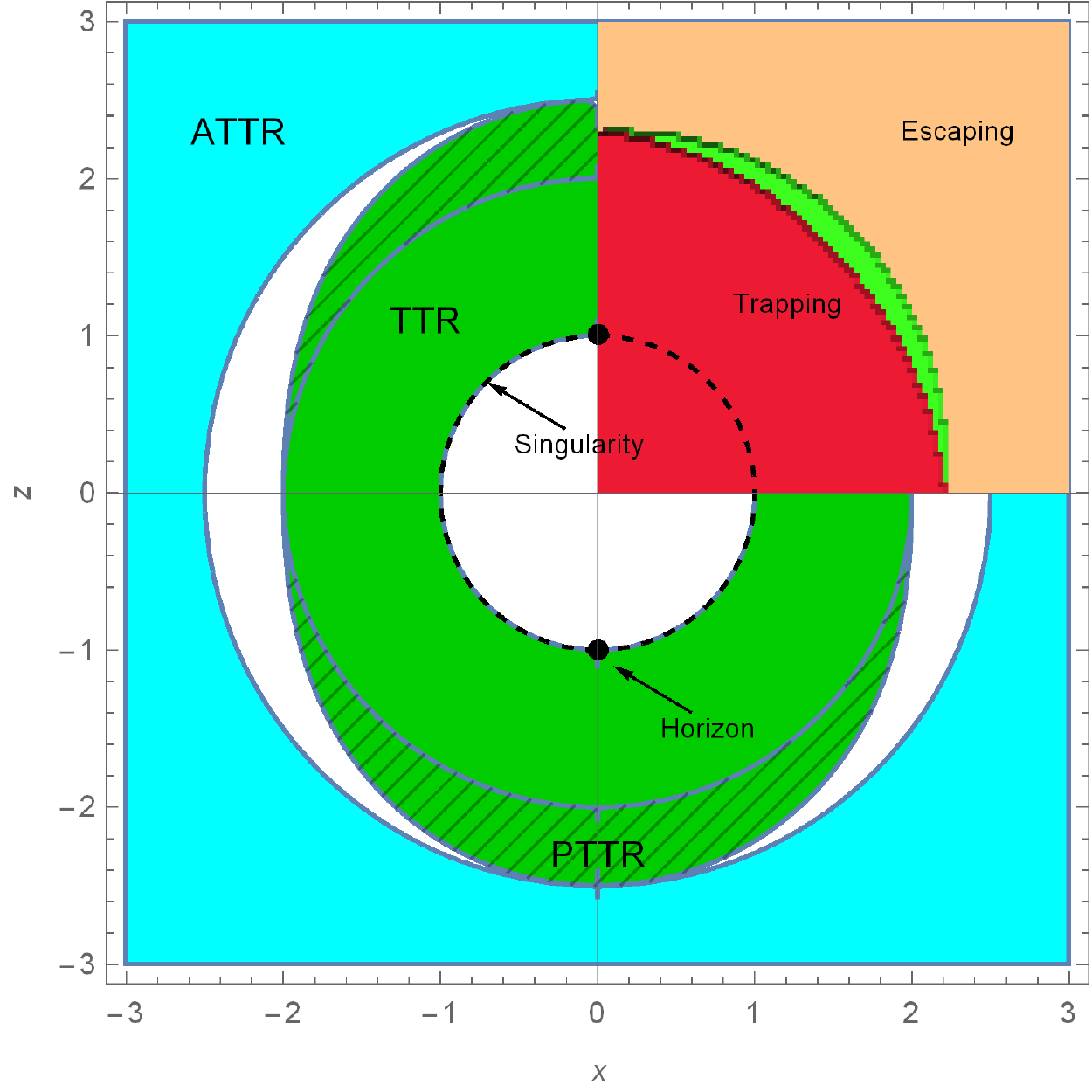}
		\label{TTSZV2S}
 }
 \subfloat[][ZVI]{
  		\includegraphics[scale=0.41]{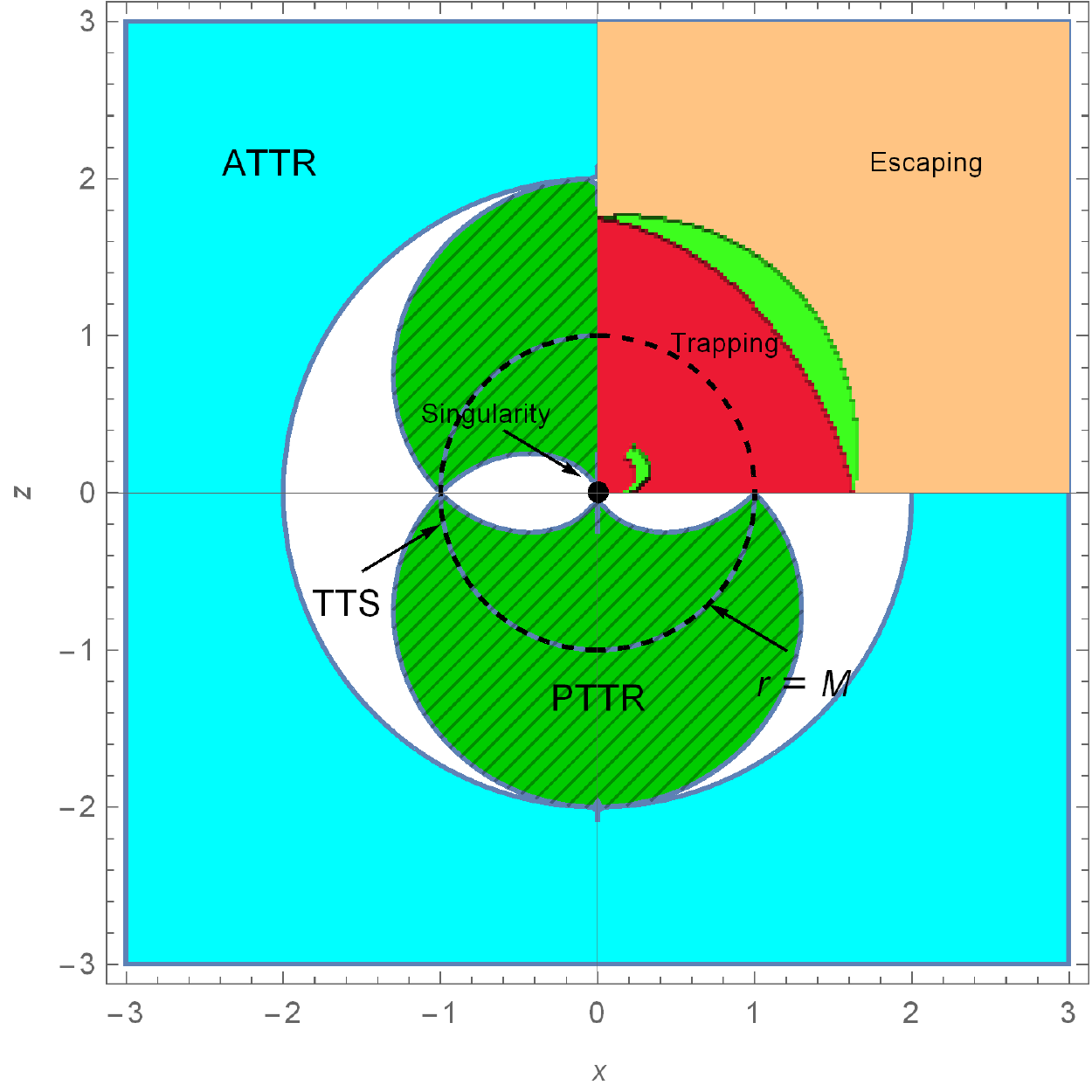}
		\label{TTSZVIS}
 }
\caption{Filling the spatial section of ZV with $ M = 1 $ with spherical TTSs and ATTSs. Green - (P)TTS, blue - ATTS. Area filled by PTTS is covered by blue mesh. In the upper right corner there is a graphic representation of the escape and capture regions for photons with zero orbital momentum. Brown: photons initially tangent to spheres go to infinity. Red: geodesic are trapped. Light green: they can either escape to infinity or be captured. The white region outside the singularity and/or the horizon indirectly determines the photon region in ZV space.}
\label{Kerr1}
\end{figure}

Note that in the Curzon-Chazy case, closed TTS coincide with invariantly hypersurface, on which the Weyl invarian $J$ (determinant of the Weyl five complex scalar functions) vanishes \cite{Montero-Camacho:2014bza}. Together with PTTSs (which are defined by the full equation (\ref{d1a})), the three-dimensional spatial section of ZV can be filled with surfaces of various types, as shown in the Figs. \ref{Kerr1}. The results of the numerical calculation of the photon capture and escape regions are presented in the upper right corner of each of the figures. The importance of studying the photon escape and the relation of these properties to photon surfaces in the Schwarzschild metric was demonstrated by Synge \cite{Synge}, and later in \cite{Semerak} for rotating solutions, where the photon escape cones and their relationship with the shadows were studied.

We note the main features and patterns. We see that all solutions with $\delta> 1/2$ belong to the first type I of the classification  introduced in \cite{Galtsov:2019bty}, as containing a set of closed TTSs. In addition, the presence of white areas indirectly indicates the possible existence of a photon region. Thus, it can be expected that, in general, the optical structure of such solutions does not differ significantly from similar Kerr solutions of type I. At the same time, solutions with $ \delta<1/2$ seem to belong to the third type III of super-extreme Kerr-Newman solutions (see also \cite{Abdikamalov:2019ztb} for comparison). When $ \delta> 2$, the TTS region exfoliates from the singularity, thus geodesics tangent to spheres from a small neighborhood of the singularity can go to infinity Fig. \ref{TTSZVIS}, which is clearly different from the case of the Schwarzschild metric \cite{Synge}.
   
Nevertheless, a number of drawbacks of this picture are obvious. It is easy to see that the photon capture region (red) Fig. \ref{Kerr1} is much larger than the region of existence of closed TTS (the difference increases with $\delta$, but has a finite limit for ZVI). This situation arose in the case of the Kerr metric due to the existence of the photon region and PTTS, however, unlike the case of Kerr-like solutions, now even touching the PTTS does not guarantee the capture of the null geodesic, since it can leave the PTTR region through its boundary and subsequently get to infinity, which was not the case with Kerr \cite{Galtsov:2019bty}. In particular, there is also a new intermediate type of trajectories going from infinity to singularity and tangents to PTTS at some point (green color).

In order to estimate how accurately the spherical closed TTS characterize the photon capture region, one can calculate the characteristic of the area ratio $ \kappa $. In the case of  Zipoy-Voorhees space with $ \delta \geq1 $, it is easy to obtain the following explicit expression:
\begin{equation} 
\kappa=\frac{1}{18\delta^2}\int^{\pi}_{0}\left(1+\frac{2}{\delta}\right)^{\delta}((\delta+2)\delta)^{(\delta^2+1)/2}((\delta+2)\delta+\sin^2\theta)^{(1-\delta^2)/2}\sin\theta d\theta,
\label{d3}
\end{equation}
where it is taken into account that $M=m\delta$. 
In the case of $ \delta \rightarrow \infty $ we get:
\begin{equation} 
\kappa=\frac{1}{9}e^{3/2}\sqrt{\frac{\pi}{2}}\rm Erfi \left(\frac{1}{\sqrt{2}}\right)=0.595048...
\label{d4}
\end{equation}
Thus, spherical closed TTSs cover about $ 0.6 $ of the maximum allowable capture region for large $ \delta $. For $ \delta \approx1 $ it is not difficult to show that $ 0.59 <\kappa \leq1 $. It is easy to verify that $ \kappa = 1 $ for $ \delta = 1 $. And so the spherical TTS are best suited only for this occasion. The question arises, is there another and better option for TTS filling? The answer is yes. 

\setcounter{equation}{0}

\section{Non-spherical TTS}

Our goal now is to find such a surfaces that would most accurately characterize the optical properties of ZV solutions. To do this, we need to consider the new notion of TTS${}_\theta$, namely, TTS with the equal two principal curvatures $\lambda_t=\lambda_\theta$. Spherically asymmetrical TTS${}_\theta$ surfaces by construction will contain closed non-equatorial geodesics with zero azimuthal impact parameter. Such geodesics are a special case of spheroidal \cite{Glampedakis:2018blj} and fundamental \cite{Cunha:2017eoe} photon orbits. At the same time, the very existence of spheroidal photon orbits, and therefore non-spherical TTS${}_\theta$, already can indicates the impossibility of separating variables in the corresponding geodesic system due to absorption of spherical orbits \cite{Pappas:2018opz,Glampedakis:2018blj}.  The exact expression for TTS${}_\theta$ can be found numerically. To do this, we consider the differential equation $\tilde{H}_{\theta\theta}=0$ (\ref{a3a}) separately:
\begin{align} 
&2f''-f'(\partial_\theta\alpha_\lambda+\partial_\theta\beta_\lambda+e^{\lambda_\beta}f'^2\partial_\theta \alpha_\lambda)+(f'^2(\partial_r\alpha_\beta+\partial_r\lambda_\beta)+e^{\beta_\lambda}\partial_r\alpha_\beta)=0. 
\label{f1}
\end{align}  
We must also add boundary conditions. It is clear from symmetry that the normal vector $n^\mu$ at the points $\theta=0,\pi/2$ must coincide with the radial direction. In addition, the function $ f (\theta) $ must have a period $ \pi $:
\begin{equation} 
f(0)=f(\pi),  \quad f'(0)=0.
\label{f2}
\end{equation} 
If we substitute an explicit expression for the components of the ZV metric into this equation with the given boundary-initial conditions, then the numerical solution can be easily obtained by the shooting method. To do this, we will choose some initial value $f(0)=r_I$ (best of all is to choose one of the approximate analytical values obtained below) and build the corresponding numerical solution of the equation (\ref{f1}). If the condition $ f (\pi) \approx r_I $ is not fulfilled with a predetermined accuracy, then a small value $ \epsilon = 1/2^m $, $ m> 1 $, is added or subtracted to $ r_I $, depending on what sign has the expression $ s = r_I-f (\pi) $. Repeating this procedure many times, with increased $ m $ every time when $ s $ changes sign, one can achieve any desired accuracy in determining TTS${}_\theta$.

\begin{figure}[tb]
\centering
 \subfloat[][ZV$\delta$]{
  		\includegraphics[scale=0.41]{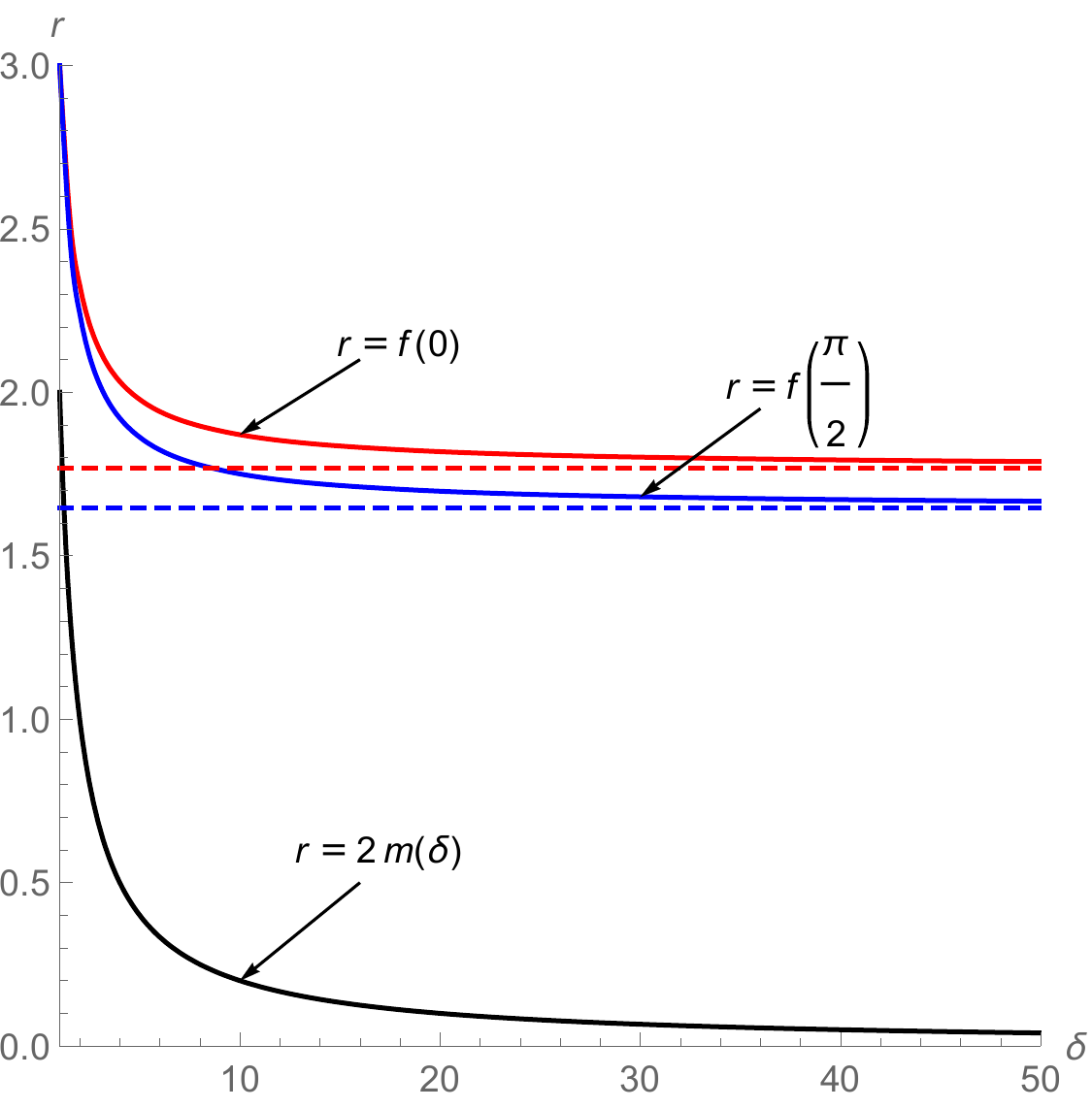}
		\label{ZVGRA1}
 }
\subfloat[][$\kappa$]{
  	\includegraphics[scale=0.41]{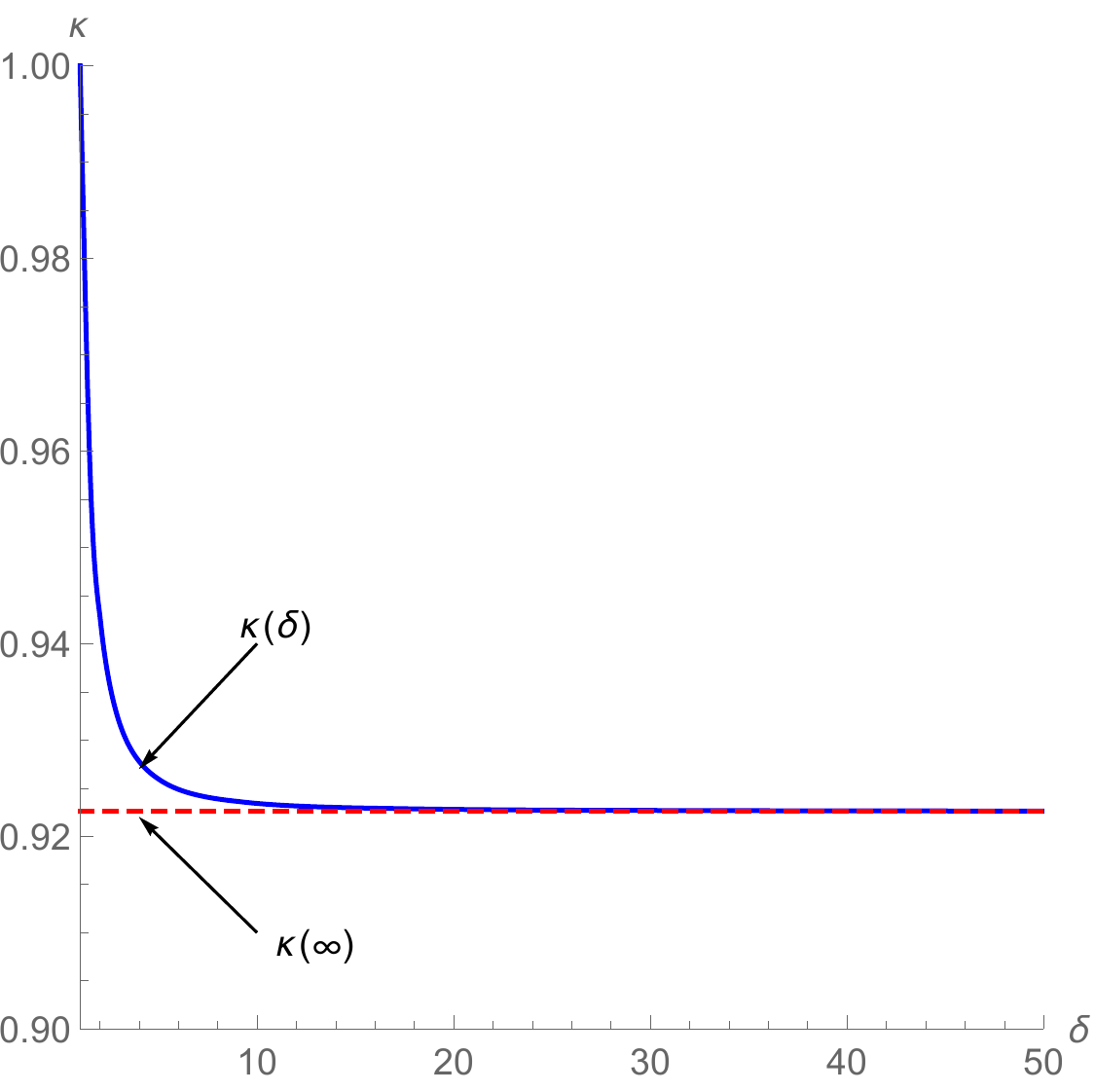}
		\label{ZVGRA2}
 }
\subfloat[][ZV$\delta$]{
  	\includegraphics[scale=0.41]{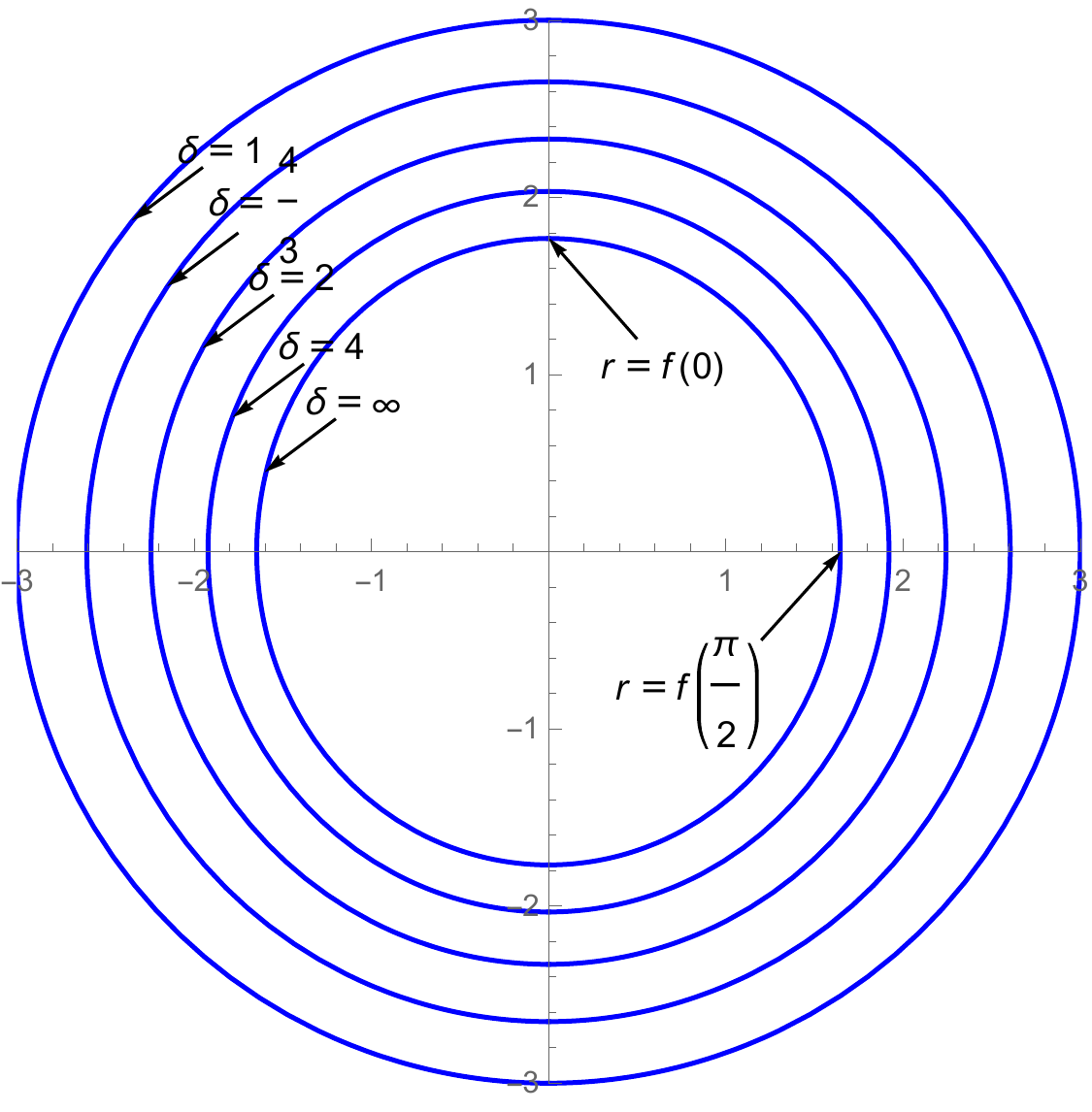}
		\label{ZVGRA}
 }
\caption{TTS${}_\theta$ in ZV space with $M=1$. The figure  (Fig. \ref{ZVGRA1}) depicts the dependence of the vertical (red) and horizontal (blue) size TTS${}_\theta$ on the value of the deformation parameter $ \delta $. The dashed lines represent the corresponding dimensions for the ZVI. The figure (Fig.  \ref{ZVGRA2}) depicts the dependence of $ \kappa $ (\ref{a6}) on the values of $ \delta $. The figure (Fig. \ref{ZVGRA}) depicts a set of sections of different TTS${}_\theta$ by the plane $ y = 0 $ in Cartesian coordinates.}
\label{K}
\end{figure}

The set of TTS${}_\theta$ (or, more precisely, section by the $ y = 0 $ plane in Cartesian coordinates of their spatial section) obtained by this method for different values of $  \delta \geq1 $, as well as the dependence of the pole and equatorial size, and the area ratio $ \kappa $ is shown in the Figs. \ref{K}. It is easy to see that these surfaces quickly tend to the same limit corresponding to the limiting solution ZVI (dashed lines). Moreover, the characteristic properties of the violation of spherical symmetry are strongly manifested already when $\delta\approx8$ when surfaces TTS${}_\theta$ become clearly spherically asymmetrical.   

\begin{figure}[tb!]
\centering
 \subfloat[][ZV2, $\pi/2$]{
  		\includegraphics[scale=0.41]{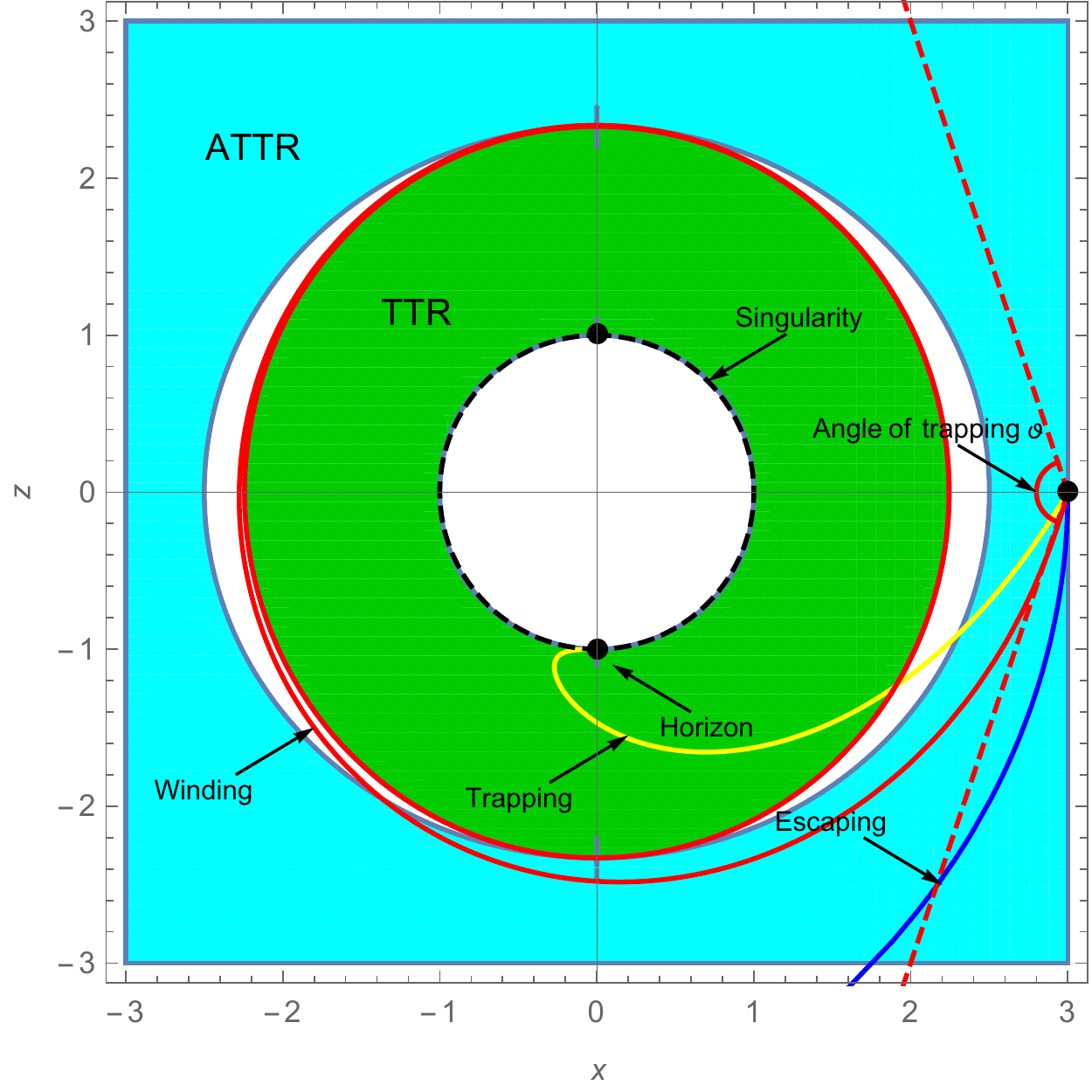}
		\label{ZVNS2} 
 }
 \subfloat[][ZVI, $\pi/2$]{
  		\includegraphics[scale=0.41]{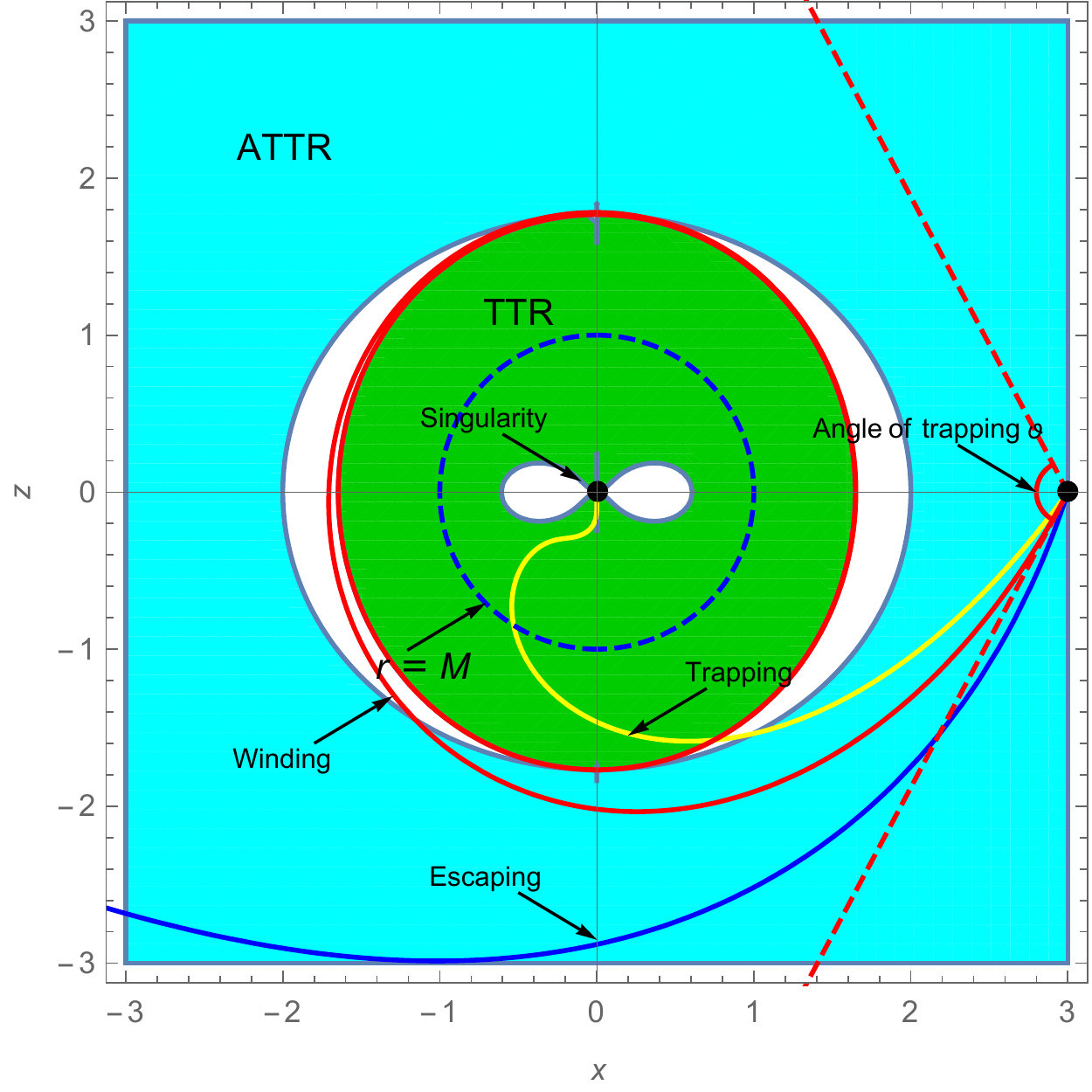}
		\label{ZVNSI}
 }
\\
\subfloat[][ZV2, $\pi/4$]{
  	\includegraphics[scale=0.45]{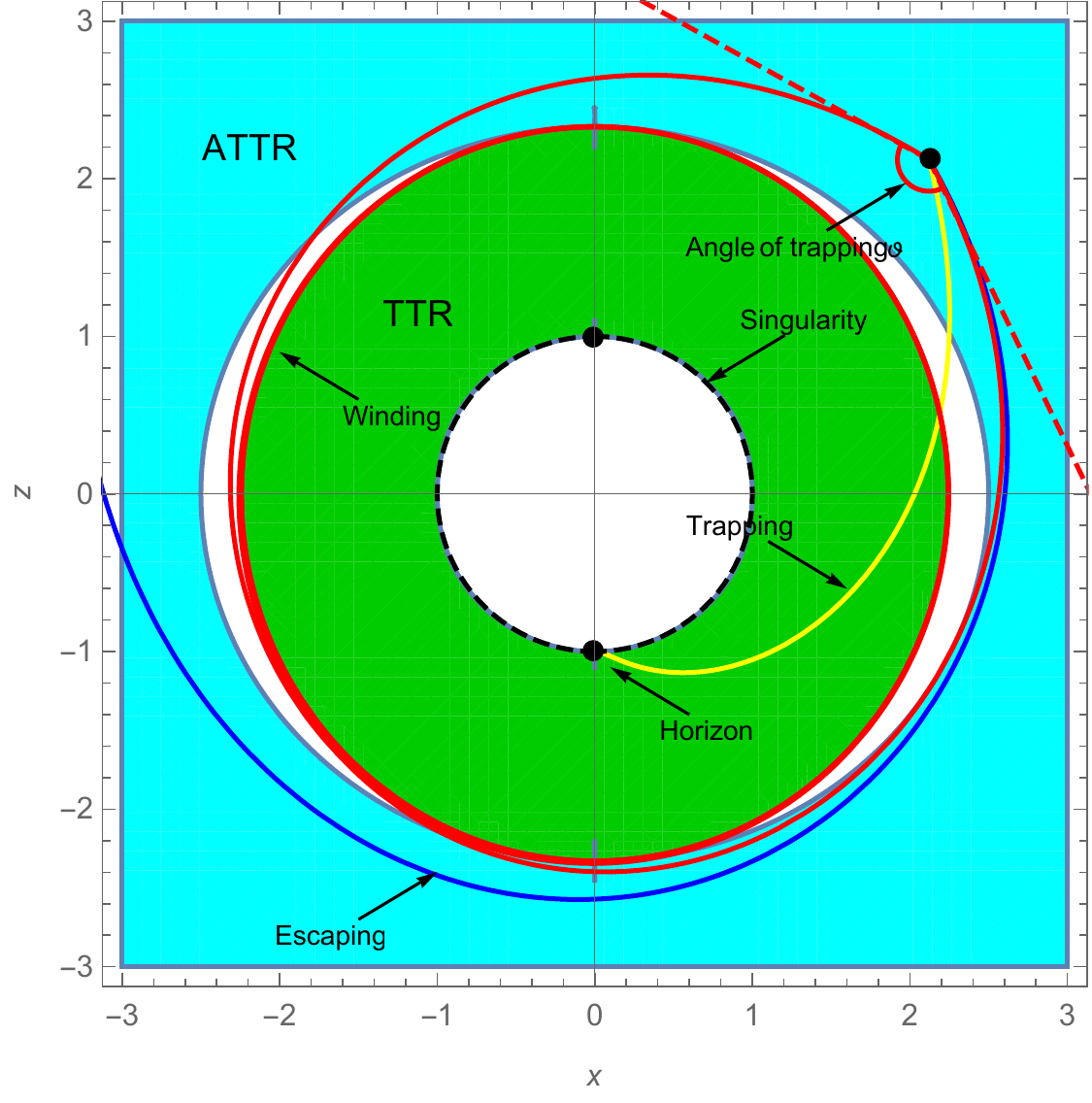}
		\label{ZVNS2NE}
 }
 \subfloat[][ZVI, $\pi/4$]{
  		\includegraphics[scale=0.41]{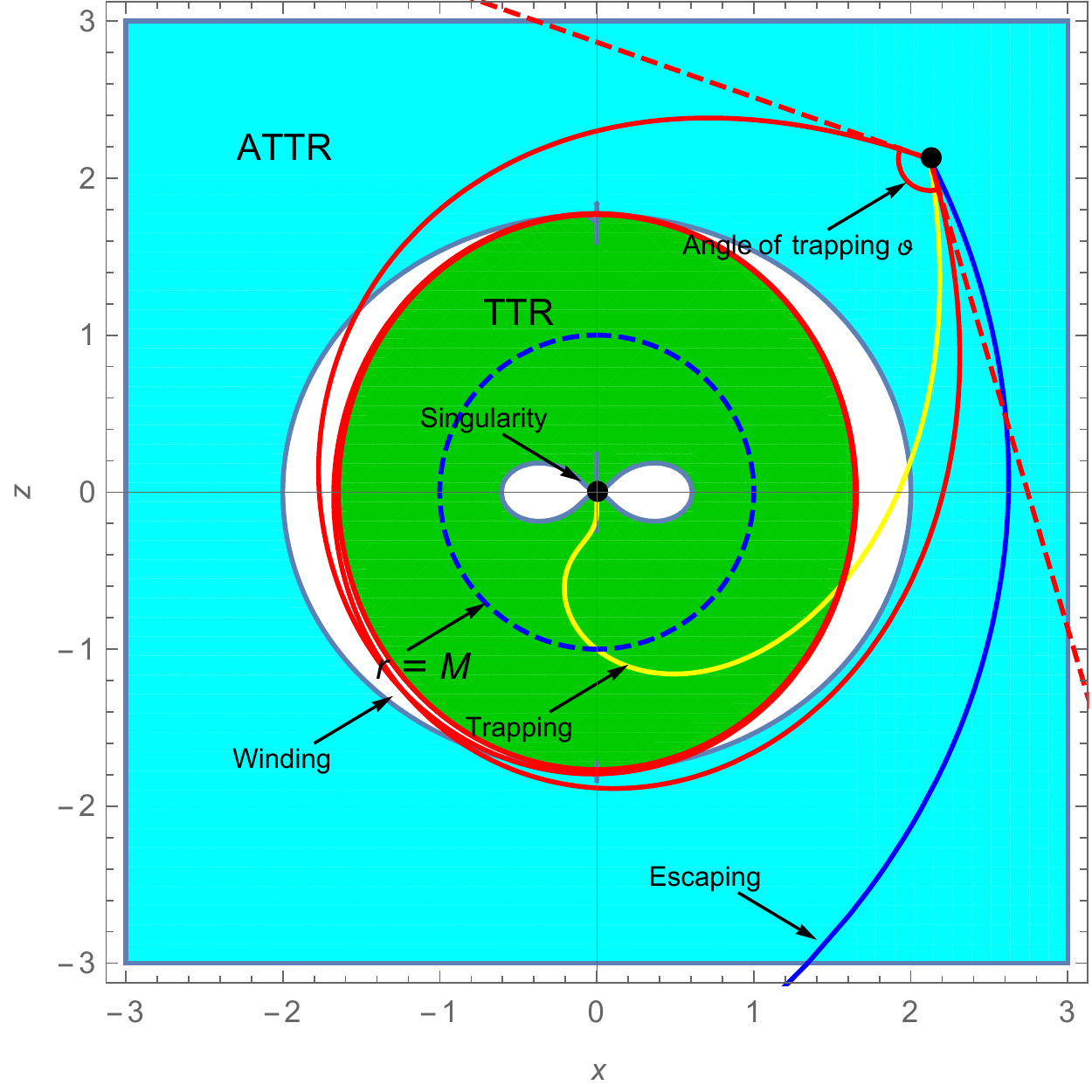}
		\label{ZVNSINE} 
 }
\caption{Filling the spatial section of the ZV solution with $ M = 1 $ for $ \delta> 1$ with the TTS${}_\theta$ family and geodesic scattering. Green - TTSs  surfaces, blue - ATTSs. The figures also show the set of isotopic geodesics with a zero impact parameter emitted from different angles from the point $ r_O = 3M $, $ \theta_O = \pi / 2 $ (\ref{ZVNS2}, \ref{ZVNSI}) and the point $ r_O = 3M $, $ \theta_O = \pi / 4 $ (\ref{ZVNS2NE}, \ref{ZVNSINE}). Yellow - geodetic falling on the horizon / singularity, forming a shadow. Blue - geodesic going to infinity. Red - geodetic wound on the TTS${}_\theta$ and forming an infinite set of relativistic images. Dotted lines correspond to the tetrad angle of emitting red geodesic $ \vartheta $ on the celestial sphere and determine the vertical size of the shadow.}
\label{Kerr2}
\end{figure}

If one decomposes the resulting  TTS${}_\theta$ in a Fourier series, one can build TTS${}_\theta$ filling of the spatial section of the entire spacetime:
\begin{equation} 
f_a(\theta)=a+f(\theta), 
\label{f3}
\end{equation}
where $a$  is a family parameter. By calculating the TTS, ATTS conditions for such surfaces, one can easily obtain the images shown in the Figs. \ref{Kerr2}. Note that the capture regions of tangent geodesics with a zero value of the impact parameter coincide trivially with TTR${}_\theta$ as it was in the Schwarzschild metric \cite{Synge}. 

These figures also depict a set of geodesics with a zero value of the impact parameter $\rho =0$  from the observation point $ r_O = 3M $ and various angles $\theta_O=\pi/2,\pi/ 4 $. It can be seen that there are geodetic wound around TTS${}_\theta$ and defining the shadow boundary (or rather, the angular size of its cross section $\Delta Y$ at $ X = 0 $ in stereographic coordinates on the celestial sphere), while the angle of deviation of light rays for such geodesics, strive to infinity \cite{Bozza:2002zj}, and accordingly a set of relativistic images \cite{Virbhadra:2008ws,Virbhadra:2002ju} arises. In addition, it is clear from Figs. \ref{ZVNS2NE}, \ref{ZVNSINE} that the angles corresponding to the vertical size $Y$ and $-Y'$ above and below the axis $Y=0$ of the shadow are asymmetric for a non-equatorial observer (see also section IX).

Thus, the notion of TTS${}_\theta$ really justifies itself, and allows, without explicitly solving the geodesic equations, to predict the behavior of isotropic geodesic flows, determining the possibility of the existence of relativistic images and analyzing the overall optical structure of the solution.

The solution ZV generates a non-integrable dynamic system \cite{Lukes,Dolan:2019gsr} which does not allow the separation of variables and, as a consequence, explicit analytical description of the geodesic motion. Therefore, it is especially useful to try to obtain a series of analytical formulas characterizing the structure of TTS${}_\theta$. We first consider the case of weakly deformed spaces with $ \delta = 1 + \epsilon $, where $ \epsilon $ is a small expansion parameter. This case is the most interesting because such solutions are hardly distinguishable from the classical Kerr-like solutions \cite{Toshmatov:2019qih,Abdikamalov:2019ztb}. We will look for the TTS${}_\theta$ surface as (see \cite{Glampedakis:2018blj} for comparison):
 \begin{equation} 
f(\theta)=3m+\epsilon m \sum^{\infty}_{n=0}a_n\cos^{2n}\theta+O(\epsilon^2).
\label{f4}
\end{equation} 
Moreover, the boundary conditions (\ref{f2}) are obviously fulfilled automatically, and the case $ \epsilon = 0 $ directly corresponds to the classical photon sphere $ r = 3m $, which is obviously a special case of the TTS${}_\theta$ surface \cite{Claudel:2000yi}. After substitution (\ref{f4}) in (\ref{f1}), the first component in the expansion of $ \epsilon $ automatically vanishes, and equalizing the nontrivial coefficient for the first power of $ \epsilon $ to zero leads to an infinite system of equations (as for spheroidal orbits \cite{Glampedakis:2018blj}):
\begin{align} 
&1-a_0+2a_1=0, \quad  2+a_0-22a_1+48a_2=0,\\
&-(1+4n^2)a_n+(22+38n+20n^2)a_{n+1}-8(2+n)(3+2n)a_{n+2}=0, \quad n>0.
\label{f6}
\end{align}  
These recurrence relations are explicitly resolved via $ a_0 $ using hypergeometric functions. In order for the series to converge, one needs to require $ a_n = 0 $ for $ n \rightarrow \infty $. Then one can get the following explicit expressions:
\begin{align} 
&a_0=1+\frac{3774F^1_1+333F^1_2}{12580 F^1_1-2220F^2_1+1110F^1_2-150F^2_2},\\
&a_n=\frac{1887\cdot 4^{-(n+2)}\Gamma\left(n-\frac{i}{2}\right)\Gamma\left(n+\frac{i}{2}\right)\left(8 \tilde{F}^n_1+(2n+1)\tilde{F}^n_2\right)}{37\left(34F^1_1-6F^2_1+3F^1_2\right)-15F^2_2},
\label{f6}
\end{align} 
where the following notation is introduced for hypergeometric functions
\begin{align} 
F^m_1={}_2\mathrm F_1\left(1,m,\frac{1}{2}+m;1+m-\frac{i}{2},1+m+\frac{i}{2};\frac{1}{4}\right),\\
F^m_2={}_2 \mathrm F_1\left(2,1+m,\frac{3}{2}+m;2+m-\frac{i}{2},2+m+\frac{i}{2};\frac{1}{4}\right),
\end{align}  
and $ \tilde{F} $ corresponds to regularized hypergeometric functions. The numerical values for the first coefficients are given in the Table \ref{STTS}.
\begin{table}[h]
\caption{Weakly deformed TTS${}_\theta$}
\begin{center}
\begin{tabular}{|c|c|c|c|c|c|c|c|c|}
\hline
$a_0$ & $a_1$ & $a_2$ &  $a_3$ & $a_4$ & $a_5$ & $a_6$\\
\hline
$1.366651$ & $0.1833256$ & $0.013885$ & $0.001618$ & $0.000232$  & $0.000037$ & $0.000006$\\
\hline
\end{tabular}
\end{center}
\label{STTS}
\end{table}

Similarly, we can study the case of extremely large $ \delta $, namely, the ZVI or Curzon-Chazy solution. The equation (\ref{f2}) for this metric has the following relatively simple form:
\begin{align} 
-f^4f''+ M^2 \cos\theta\sin\theta f^2 f'+f(2f^2-2Mf+M^2\sin^2\theta)f'^2 \nonumber\\ 
+M^2\cos\theta\sin\theta f'^3+ f^3(f^2-2Mf+M^2\sin^2\theta)=0.
\label{f9}
\end{align}
Based on the initial boundary conditions, we will look for its solution in the form of a series: 
\begin{align} 
f(\theta)=M\sum^{\infty}_{n=0}a_n\cos^{2n}\theta, \quad f'(\theta)=-2M\sin\theta\sum^{\infty}_{n=0}na_n\cos^{2n-1}\theta, \nonumber \\
f''(\theta)=M\sum^{\infty}_{n=0}(2(n+1)(2n+1) a_{n+1}-4n^2 a_n)\cos^{2n}\theta.
\label{f10}
\end{align}
Now we will nullify the coefficients for each degree $ \cos^{2m} \theta $ in the expression obtained by substituting (\ref{f10}) into (\ref{f9}) and replacing all $ \sin^2 \theta $ with $ 1 - \cos^{2} \theta $. It is clear that all $ a_n $ with $ n\leq m $ will be included in this expression, and once $a_{m + 1} $ with a linear image. Indeed, $ a_{m + 1} $ can only appear from the expansion of $ f '' (\theta) $, where it appears linearly. Thus, we can obtain a series of recurrence relations, of which $ a_{m + 1} $ is expressed as a solution to an equation of the first degree. Explicitly:
\begin{align} 
a_{n+1}=\frac{P_{n}(a)}{(2n+2)!!(2n+1)!!a^{4n+1}}, \quad P_{n}(a)=\sum^{4n+2}_{p=0}C_p a^{p}, \quad a_0=a,
\label{f11}
\end{align}    
where
\begin{align} 
&P_{0}=(a-1)^2, \nonumber \\
&P_{1}=2-12a+29a^2-42a^3+42a^4-30a^5+9a^6, \nonumber \\
&P_{2}=16 - 160 a + 658 a^2 - 1564 a^3 + 2571 a^4 - 
 3330 a^5  + 3615 a^6 - 3180 a^7 + 2115 a^8 \nonumber \\ &- 990 a^9 + 
 225 a^{10}, \nonumber\\
&P_{3}=272 - 3808 a + 22468 a^2 - 76024 a^3 + 171676 a^4 - 
 291232 a^5 + 414593 a^6 - 530086 a^7  \nonumber \\  &+ 604536 a^8 - 
 591030 a^9 + 481350 a^{10} - 317550 a^{11} + 161460 a^{12} - 
 58590 a^{13} + 11025 a^{14}.  \nonumber 
\end{align}  
By fixing some order $ m \gg 1 $ we can easily get an approximate expression for TTS${}_\theta$ with any accuracy needed. To do this, we determine $ a $ from the condition $ P_{m}(a) = 0 $, and of course, the solution of the algebraic equation is found numerically, and then we substitute the value found in (\ref{f11}). We present the calculations in the table (\ref{TTS}). In this case, the coefficients coincide with the decomposition of the numerical solution. 

Note that the existence of spherically asymmetrical TTS${}_\theta$ or spheroidal photon orbits in Curzon-Chazy solution may indicate the non-integrability of the corresponding geodesic system \cite{Glampedakis:2018blj}, which is confirmed by the study of the Poincar\'e sections in \cite{Dolan:2019gsr}.

\begin{table}[h]
\caption{TTS${}_\theta$ in ZVI}
\begin{center}
\begin{tabular}{|c|c|c|c|c|c|c|c|c|}
\hline
$a_0$ & $a_1$ & $a_2$ &  $a_3$ & $a_4$ & $a_5$ & $a_6$\\
\hline
$1.646576$ & $0.126948$ & $-0.005655$ & $0.000028$ & $0.000098$  & $-0.000026$ & $0.000004$\\
\hline
\end{tabular}
\end{center}
\label{TTS}
\end{table} 

It is easy to verify that the expressions obtained do indeed give a good approximation of an explicit numerical solution. However, a rough but very simple analytical approximation can be obtained for TTS${}_\theta$ and in the case of an arbitrary $ \delta $. To do this, consider the first and second terms in the expansion (\ref{f10}):
\begin{equation} 
f(\theta)=a+b\cos^2\theta.
\label{f21}
\end{equation} 
To determine the value of the parameter $ b $ we substitute
(\ref{f21}) into (\ref{a3a}, \ref{a3b}) and  consider enequalities at $\theta=0,\pi/2$:
\begin{align} 
&a-m (1+2\delta)\leq0, \quad (a+3b-m (1+2\delta)\leq0, \label{f22a}\\
&a^2-2a(b+m(1+\delta))+m(2b+m\delta(2+\delta))\leq0. 
\label{f22b}
\end{align}            
Replacing the second two inequalities with equalities, it is easy to get the value of the parameters $ a $ and $ b $ for the approximate TTS${}_\theta$. In this case, of course, the inequalities (\ref{f22a}, \ref{f22b}) at arbitrary angles should be verified, which is indeed the case. So we get:
 \begin{align} 
&a=m\left(1+\delta+\frac{\sqrt{3+2\delta^2}}{\sqrt{5}}\right), \quad b=\frac{m}{3}\left(\delta-\frac{\sqrt{3+2\delta^2}}{\sqrt{5}}\right).
\end{align} 

An important feature of the images we obtained Figs.\ref{Kerr2} is the presence of white spots corresponding to a sign-indetermined quadratic form. In the case of Kerr solutions, these   corresponded to the photon region, this is also so in the Zipoy-Voorhees case, but in a certain approximation.  

\begin{figure}[tb!]
\centering
 \subfloat[][ZV2]{
  		\includegraphics[scale=0.41]{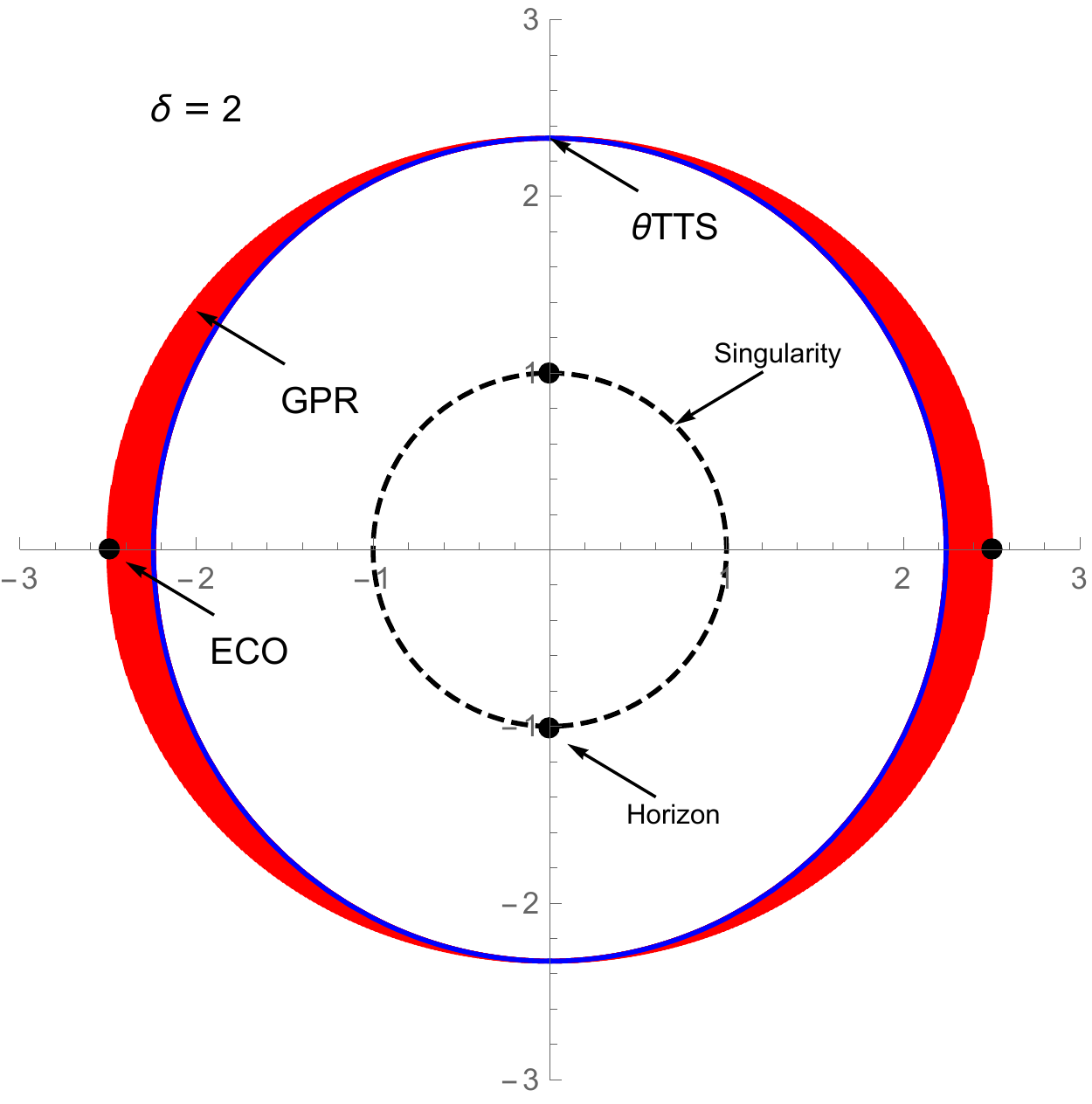}
		\label{PR2}
 }
 \subfloat[][ZV2]{
  		\includegraphics[scale=0.41]{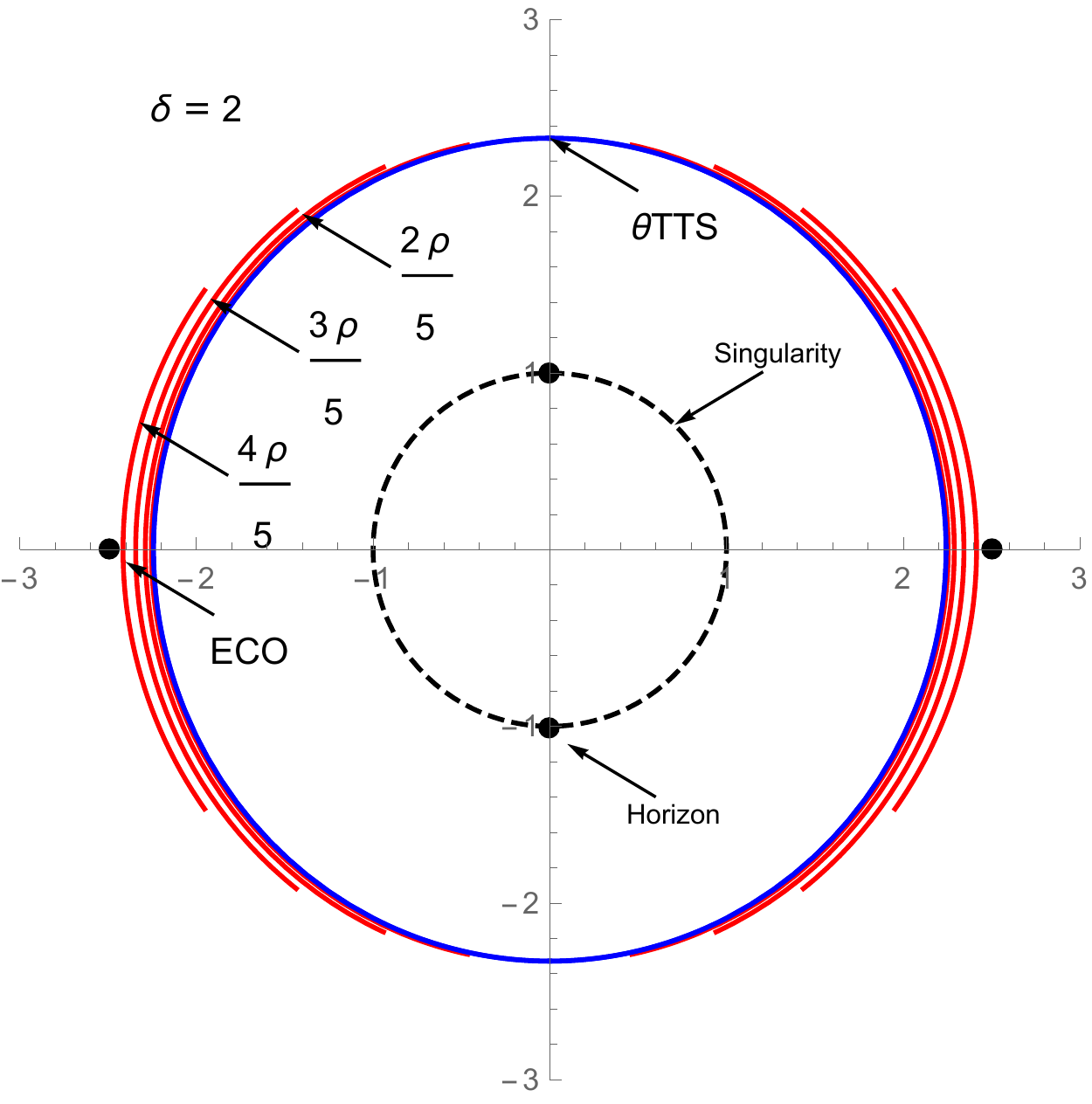}
		\label{PR2a}
 }
 \subfloat[][ZV2]{
  		\includegraphics[scale=0.41]{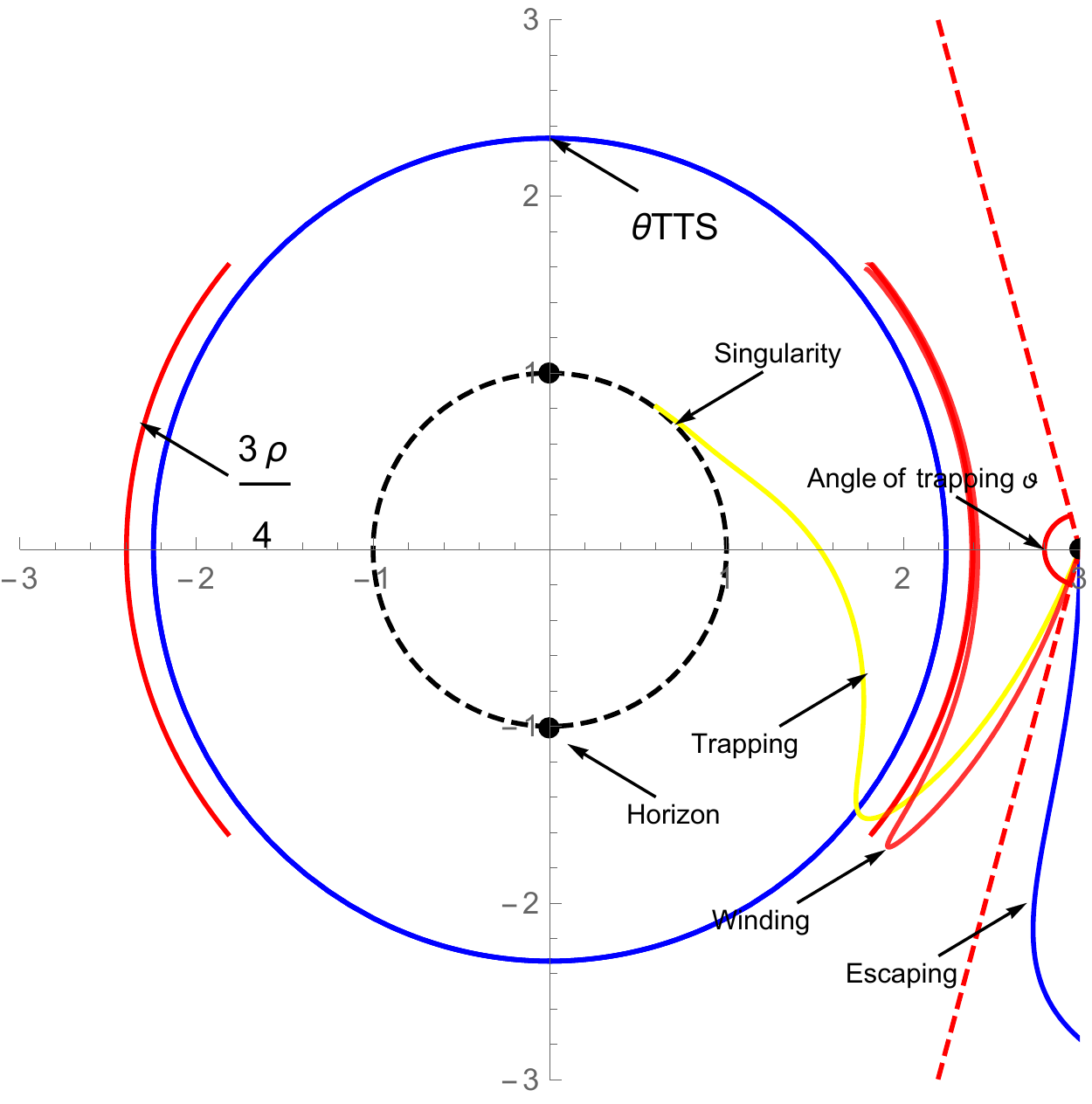}
		\label{PR2b}
 }
\\
 \subfloat[][ZVI]{
  		\includegraphics[scale=0.41]{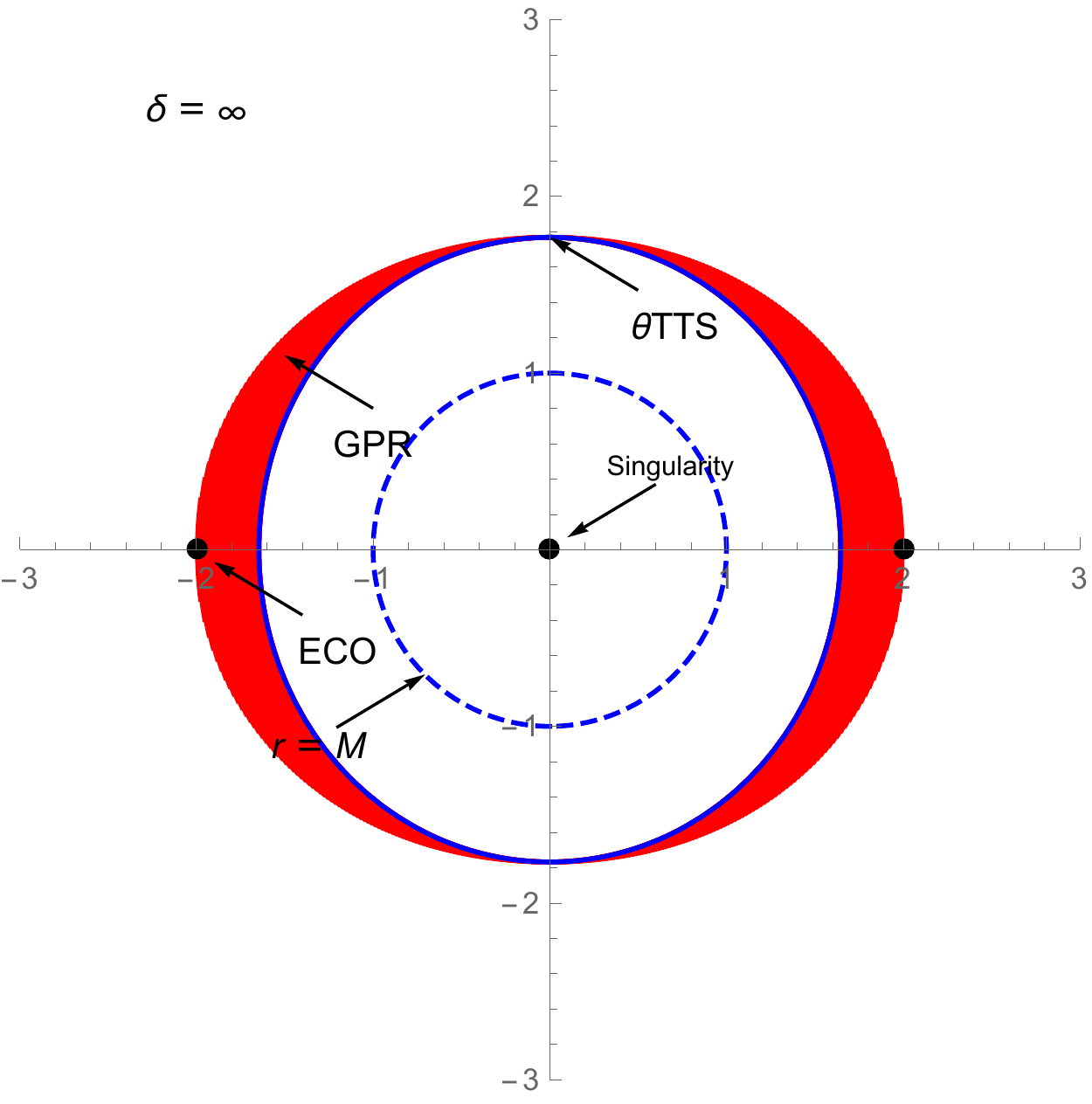}
		\label{PRG}
 }
\subfloat[][ZVI]{
  	\includegraphics[scale=0.41]{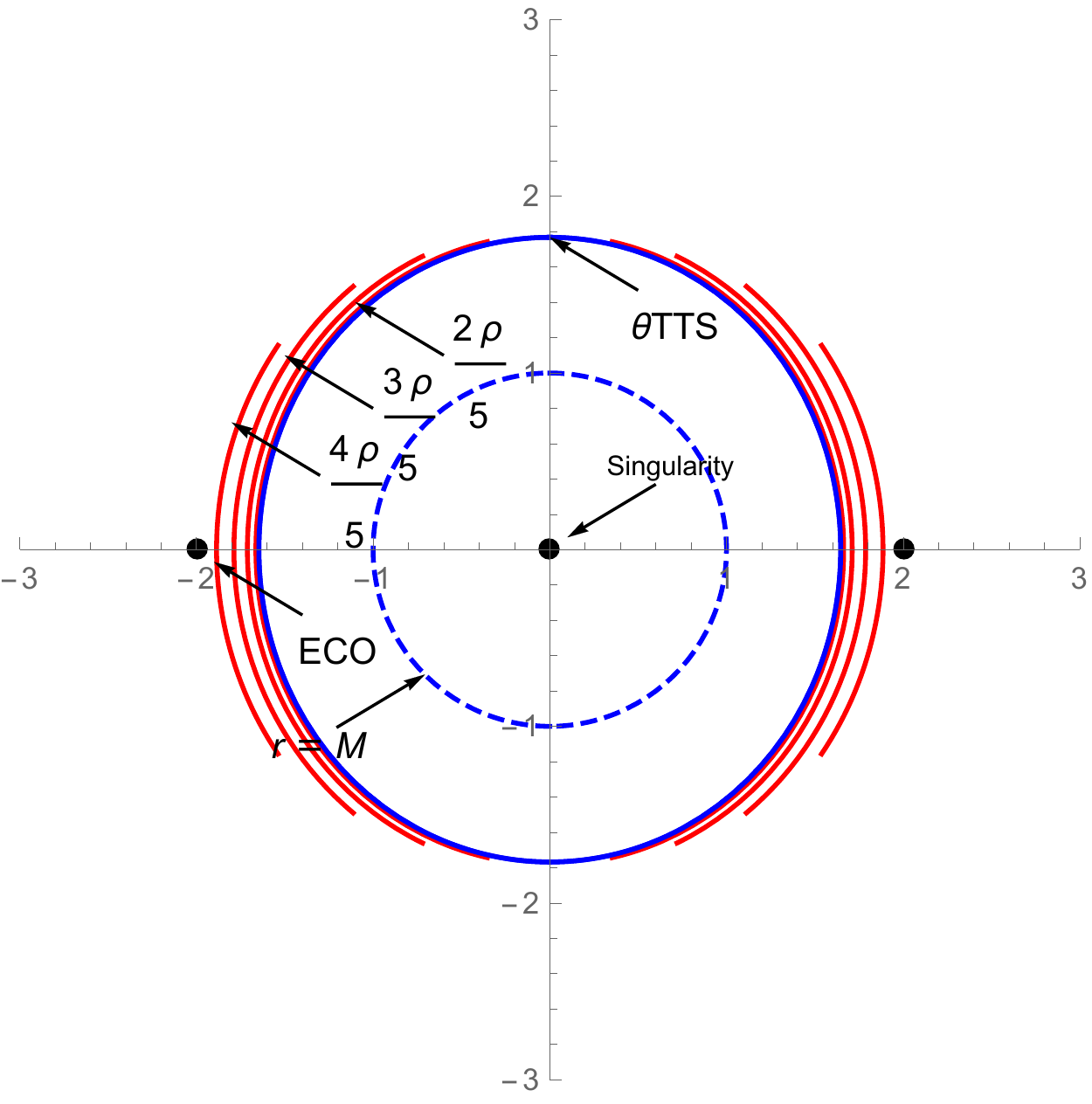}
		\label{PR10}
 }
 \subfloat[][ZVI]{
  		\includegraphics[scale=0.41]{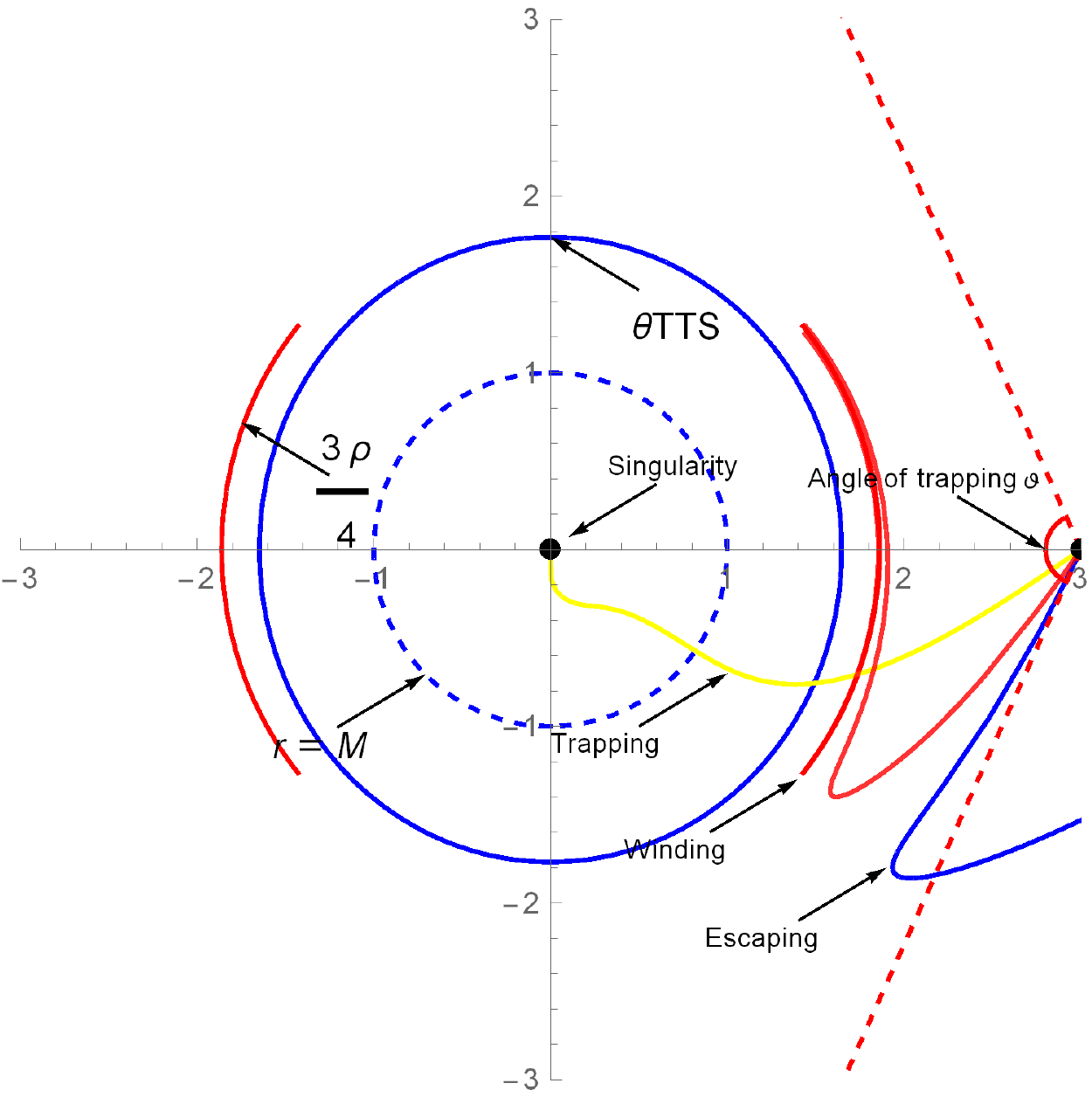}
		\label{PR10G}
 }
\caption{The photon region and the scattering of geodesics in the space ZV with $ M = 1 $ for $ \delta> 1 $. Figures (Fig. \ref{PR2}, \ref{PRG}) represent the image of a generalized photon region containing an infinite set of photon surfaces for each value of the impact parameter $ 0 <\rho <\rho_{\rm max} $, while the figures (Fig. \ref{PR2a}, \ref{PR10}) illustrate several separate surfaces from the photon region. The figures (Fig. \ref{PR2b}, \ref{PR10G}) illustrate the set of geodesics with a nonzero impact parameter from a certain observation point scattered on the surface of the photon region corresponding to this parameter. Red geodesics, as before, are responsible for the emergence of relativistic images and determine the border of the shadow.}
\label{Kerr3}
\end{figure}

\setcounter{equation}{0}

\section{Generalized photon region}

We have seen that TTS${}_\theta$ defines the behavior of geodesics with zero orbital angular momentum. As we showed in our work \cite{Galtsov:2019bty}, to analyze geodesics with other values ​​of the moment, it is necessary to construct a generalization of the photon region using the same formalism as in the case of TTS. A photon region in Kerr-like solutions is formed by a set of open surfaces $ r = {\rm const} $ on which there are closed isotropic geodesics, with each valid value of the impact parameter \cite{Teo}. 

From the point of view of the formalism \cite{Galtsov:2019bty} that we are considering, we must consider the unclosed surfaces $ f $, such that the values ​​of the second fundamental form on some isotropic tangent vectors (with a fixed ratio of orbital angular momentum and energy) were equal to zero $ \mathrm H (\dot{\gamma}, \dot{\gamma}) = 0 $. We will act as before, but now exclude the $  \dot{\theta}$ component:
\begin{align} 
\mathrm H(\dot{\gamma},\dot{\gamma})=\tilde{H}_{tt}\dot{t}^2+\tilde{H}_{\phi\phi}\dot{\phi}^2=0, \quad \tilde{H}_{tt}=H_{tt}-\frac{g_{tt}}{g_{\theta\theta}} H_{\theta\theta}, \quad \tilde{H}_{\phi\phi}=H_{\phi\phi}-\frac{g_{\phi\phi}}{g_{\theta\theta}} H_{\theta\theta}.
\label{g0}
\end{align}  
These expressions are conveniently written in terms of energy and impact parameter (\ref{c1}):
\begin{align} 
\mathrm H(\dot{\gamma},\dot{\gamma})=E^2( \tilde{\tilde{H}}_{tt}+\rho^2\tilde{\tilde{H}}_{\phi\phi})=0, \quad \tilde{\tilde{H}}_{tt}=\tilde{H}_{tt}/g^2_{tt}, \quad \tilde{\tilde{H}}_{\phi\phi}=\tilde{H}_{\phi\phi}/g^2_{\phi\phi}.
\label{g01} 
\end{align}   
It should be borne in mind that in this formula $ \rho^2 $ no longer takes arbitrary values, since the isotropy condition imposes the restriction $ g_{\phi \phi}\geq \rho^2| g_{tt} |  $. In particular for each $ \rho $ surfaces $f$ are limited by turning or boundary angles $\theta_b$ determined by equality $ g_{\phi \phi}= \rho^2| g_{tt} | $. At the same time, the condition (\ref{g01}) means that geodesics with a fixed $ \rho $ will remain on these surfaces, which is exactly a property of the photon region of Kerr-like solutions, as well as the general property of spheroidal photon orbits \cite{Glampedakis:2018blj}, representing a special case of fundamental photon orbits \cite{Cunha:2017eoe}.  

In terms of the principal curvatures, the equation (\ref{g01}) means:
\begin{align} 
\mathrm H(\dot{\gamma},\dot{\gamma})=E^2( (\lambda_\theta-\lambda_t)/|g_{tt}|+\rho^2(\lambda_\phi-\lambda_\theta)/g_{\phi \phi})=0, 
\end{align}  
or otherwise
\begin{align} 
 (\lambda_\theta-\lambda_t)+\xi_\rho(\theta)(\lambda_\phi-\lambda_\theta)=0, \quad \xi_\rho(\theta)=|g_{tt}|\rho^2/g_{\phi \phi}, \quad 0\leq\xi_\rho(\theta)\leq1.
\end{align} 
Thus, the photon region is a one-parameter $\rho $-family of the surfaces containing both TTS${}_\theta$($\lambda_\theta=\lambda_t$) and TTS${}_\phi$($\lambda_\phi=\lambda_t$) surfaces(in particular equatorial circular photon orbits, if it exists). In fact, $\xi_\rho(\theta_b)=1$ for the turning or boundary angle $\theta_b$. Accordingly, TTS${}_\phi$ are two-dimensional and contained in the boundary of each photon region surfaces. Note that $\xi_\rho(\theta)$ itself is a function and not a numeric parameter of the family but it depends linearly on such a parameter $\rho^2$.

\begin{figure}[tb!]
\centering
 \subfloat[][$\delta=2$]{
  		\includegraphics[scale=0.45]{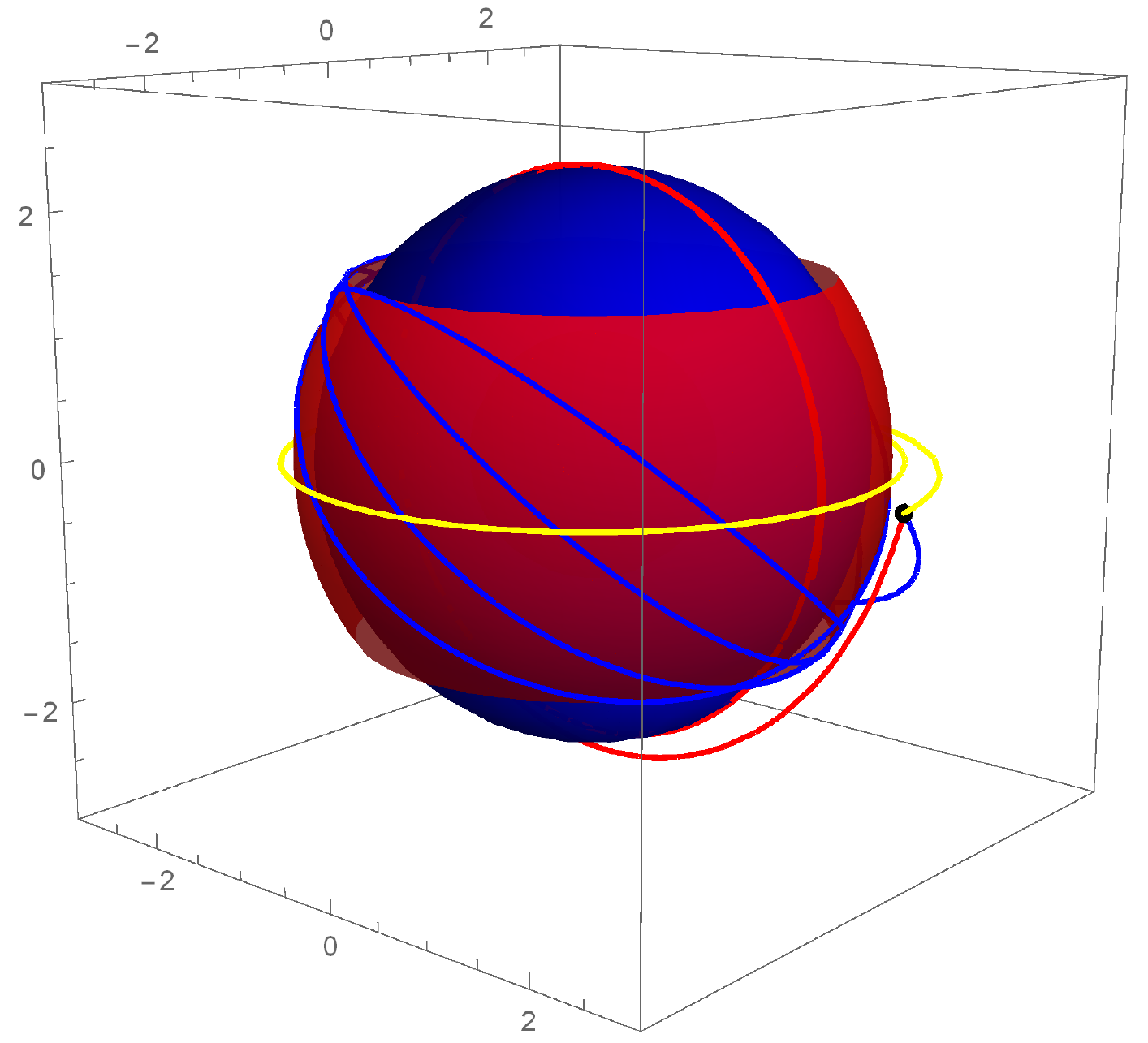}
		\label{ZVIPR3D}
 }
 \subfloat[][$\delta=\infty$]{
  		\includegraphics[scale=0.45]{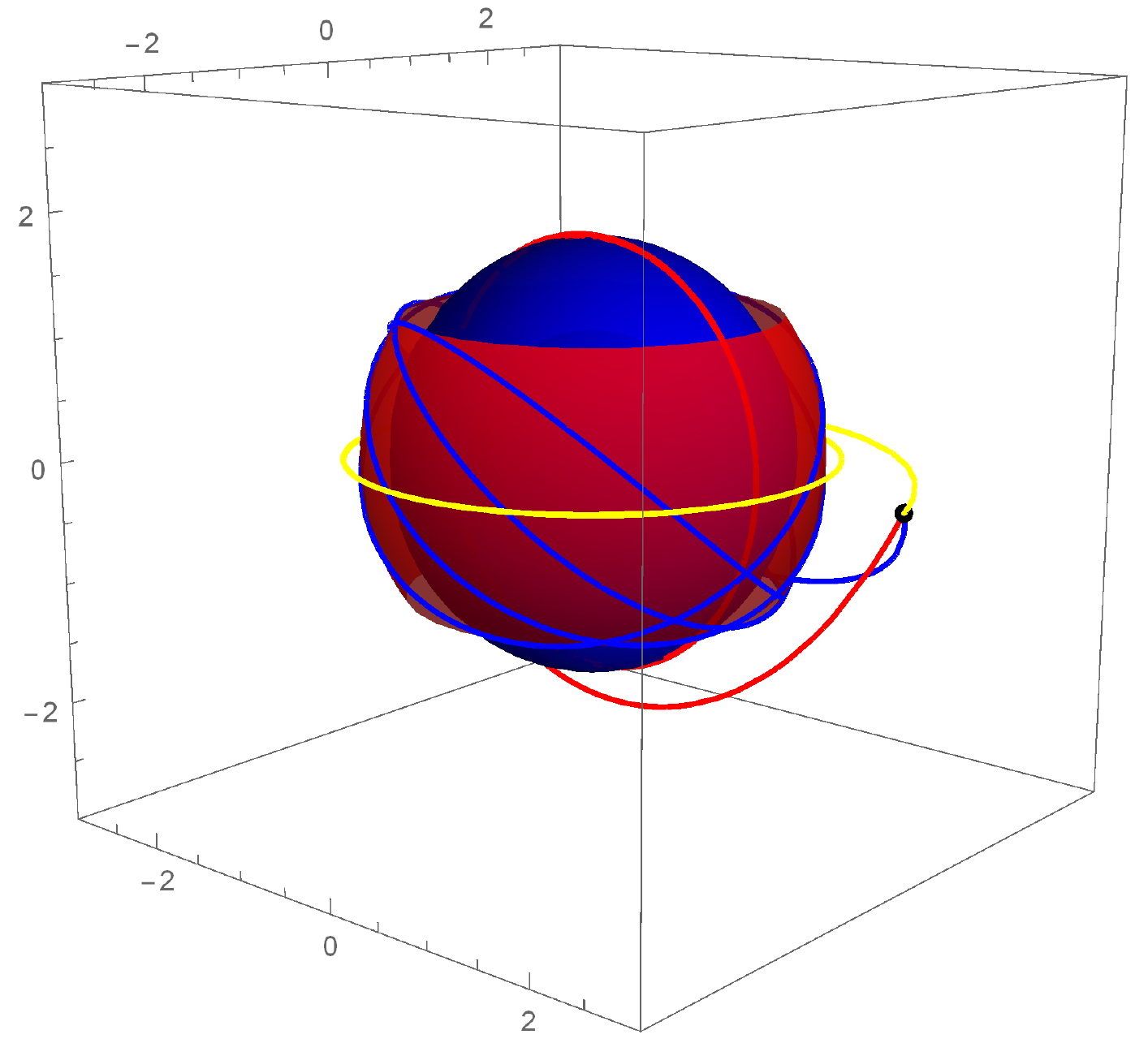}
		\label{ZVIPR3D1}
 }
\caption{Photon region, TTS${}_\theta$ and ECO in the ZV spacetime with $ M = 1 $ for $ \delta>1$. Three-dimensional images of individual surfaces $\rho=0$, $\rho=4/5\rho_{\rm max}$ and $\rho=\rho_{\rm max}$ from the photon region and the wound null geodesics emitted from a certain observation point. }
\label{Kerr3a}
\end{figure}

We can obtain the following explicit form of equation (\ref{g01}): 
\begin{align} 
[e^{-\alpha}(2f''-f'(\partial_\theta\alpha_\lambda+\partial_\theta\beta_\lambda+e^{\lambda_\beta}f'^2\partial_\theta \alpha_\lambda)+(f'^2(\partial_r\alpha_\beta+\partial_r\lambda_\beta)+e^{\beta_\lambda}\partial_r\alpha_\beta))] \nonumber \\
-\rho^2[\alpha\rightarrow\gamma]=0, \label{g1a} \\
e^{\gamma-\alpha}\geq\rho^2,
\label{g1}
\end{align}  
where $[\alpha\rightarrow\gamma]$ 
means the expression for the first line with the replacement of $ \alpha $ by $ \gamma $. It is clear that in the case of $ \rho = 0 $ we get exactly the equation for TTS${}_\theta$.
We also need the boundary condition, namely $\tilde{H}_{\phi\phi}=0$ (\ref{a3b}):
\begin{align} 
\partial_r \alpha_\gamma=e^{\lambda_\beta}f'\partial_\theta \alpha_\gamma,
\label{g2}
\end{align}  
at the $ \theta_b$ boundary of a generalized photon region. Now the equation is easy to solve numerically using the shooting method. Set the initial conditions out of symmetry:
\begin{align} 
f'(\pi/2)=0, 
\label{g3}
\end{align}     
and require that for the boundary $\theta_b$ defined by equality (\ref{g1}) on the solution $ f $, the condition (\ref{g2}) is satisfied. Acting as in the case of calculating TTS${}_\theta$, we obtain a family of photon region surfaces for each value of $ \rho $ forming a generalized photon region Figs. \ref{Kerr3}. These figures also depict a set of geodesics with a non-zero value of the impact parameter $\rho\neq0$  from the observation point $ r_O = 3M $ and angles $\theta_O=\pi/2 $. It can be seen that there are geodetic wound around photon region surface (see also Figs. \ref{Kerr3a}) and defining the shadow boundary and a set of relativistic images \cite{Virbhadra:2008ws,Virbhadra:2002ju} just like it was for spherical photon orbits in the Kerr metric \cite{Grenzebach,Grenzebach:2015oea,Teo}.  Thus, these properties of isotropic geodesics leading to the characteristic structures of the photon region are not a unique effect generated by rotation and even non-static space \cite{Cunha:2017eoe}.

 It is not difficult to see that the generalized photon region fills just the white external region from the TTS${}_\theta$ filling, complementing it in this way, however, only approximately as the surfaces in the photon region, though not significantly, differ from the surfaces in TTS${}_\theta$ filling. To build the maximum filling it is necessary to build the expansion of the surfaces of the photon region, adding to them the corresponding PATTS covers. This can be done for example by decomposing the surfaces of the photon region in a Fourier series in the domain of their definition, and extending their values ​​to the remaining $\theta$. At the same time, the inner white areas are still not classified.

Note that, we can easily find an explicit analytical expression for the boundary of the photon region in the equatorial plane, as well as the maximum value of the impact parameter $ \rho_{\rm max} $. Indeed, boundaries of the photon region surfaces are two-dimensional and presented by (A)TTS${}_\phi$. It is really clear that for the boundary in the equatorial plane mast be $\theta_b=\pi/2$ and so $ e^{\gamma- \alpha} = \rho^2_{\rm max} $ only when $\theta_b=\pi/2$. For ZV we get
\begin{align}  
r_{\rm max}=m(1+2\delta), \quad \rho_{\rm max}=m(2\delta-1)^{-\delta+1/2}(2\delta+1)^{\delta+1/2}.
\label{g4}
\end{align}    
 In particular, when $\delta\rightarrow\infty$ we get $\rho_{\rm max}=2 e M$. At the same time, due to the fact that $\theta_b=\pi/2$ is also a turning point, geodesics cannot leave this two-dimensional (A)TTS${}_\phi$ surface and form equatorial circular orbit which only exists when $\delta>1/2$ \cite{Toshmatov:2019qih,Herrera:1998rj}.  The photon region itself then in some sense interpolates between (A)TTS${}_\theta$ ($\rho=0$) and this circular equatorial orbit ($\rho=\rho_{\rm max}$) Figs. \ref{Kerr3}. When $\delta<1/2$, the photon region has a singularity as a boundary and, therefore, the optical properties of the solution changes(see section VIII below and \cite{Abdikamalov:2019ztb}).  

As was demonstrated in section IV, knowing the value of $\rho_{\rm max}$ you can determine the equatorial size of the shadow for the equatorial observer, determined by the rays relating to the photon region. Indeed, from (\ref{c6}) we get:
 \begin{align}
\Delta X|_{Y=0}=4\tan \left(\frac{1}{2}\arcsin\left( m \left(\frac{2\delta-1}{r_O-2m}\right)^{-\delta+1/2}\left(\frac{2\delta+1}{r_O}\right)^{\delta+1/2}\right)\right).
\label{g5}
\end{align}    
In the particular case of Curzon-Chazy metric:
 \begin{align}
\Delta X|_{Y=0}=4\tan \left(\frac{1}{2}\arcsin \left( \frac{2Me^{\frac{r_O-2M}{r_O}}}{r_O}\right)\right).
\label{g6}
\end{align}    

\setcounter{equation}{0}

\section{The case $\delta<1$} 

\begin{figure}[tb]
\centering
\subfloat[][ZV3/5, $\pi/2$]{
  		\includegraphics[scale=0.41]{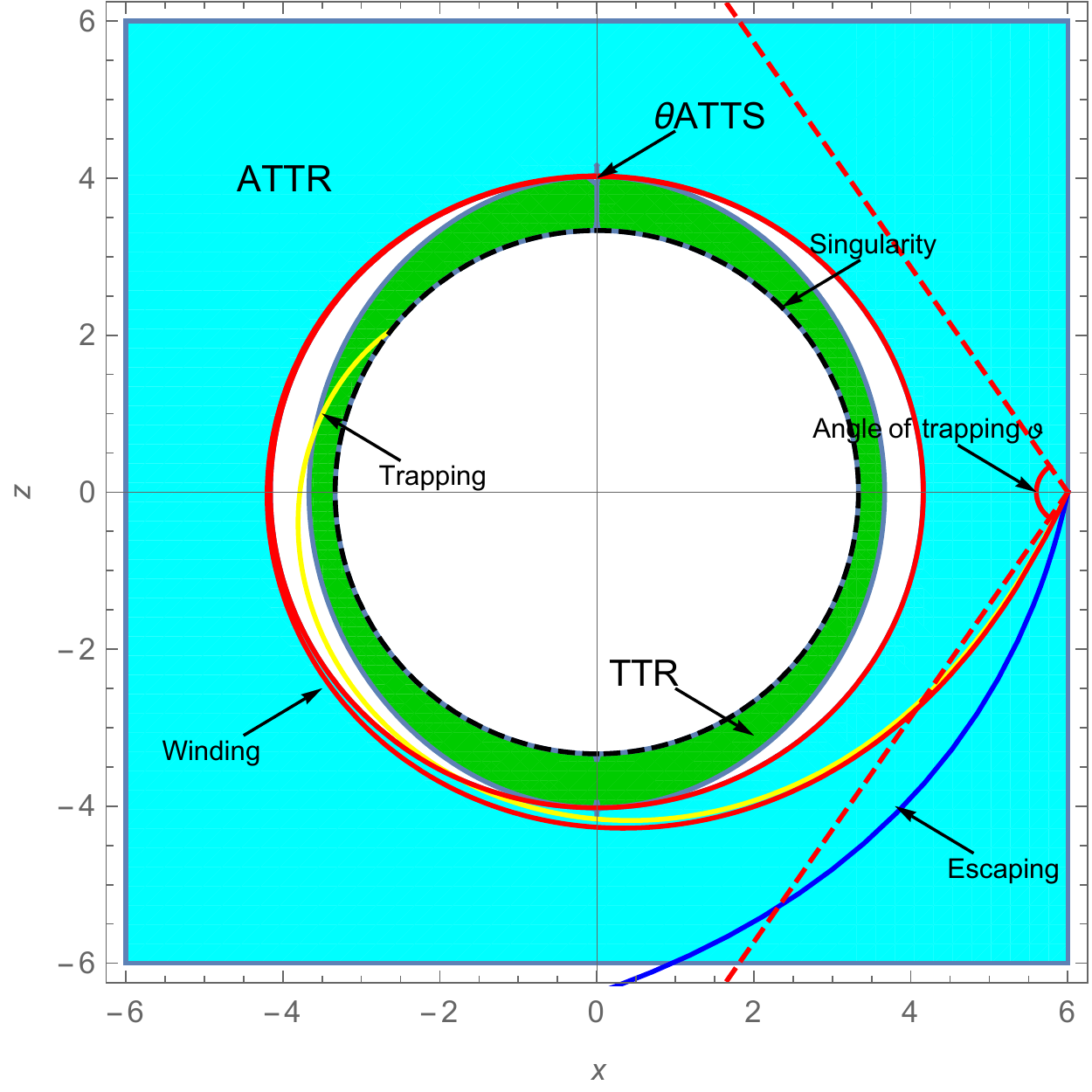}
		\label{ZVNS35}
 }
 \subfloat[][ZV3/5, $\pi/4$]{
  		\includegraphics[scale=0.41]{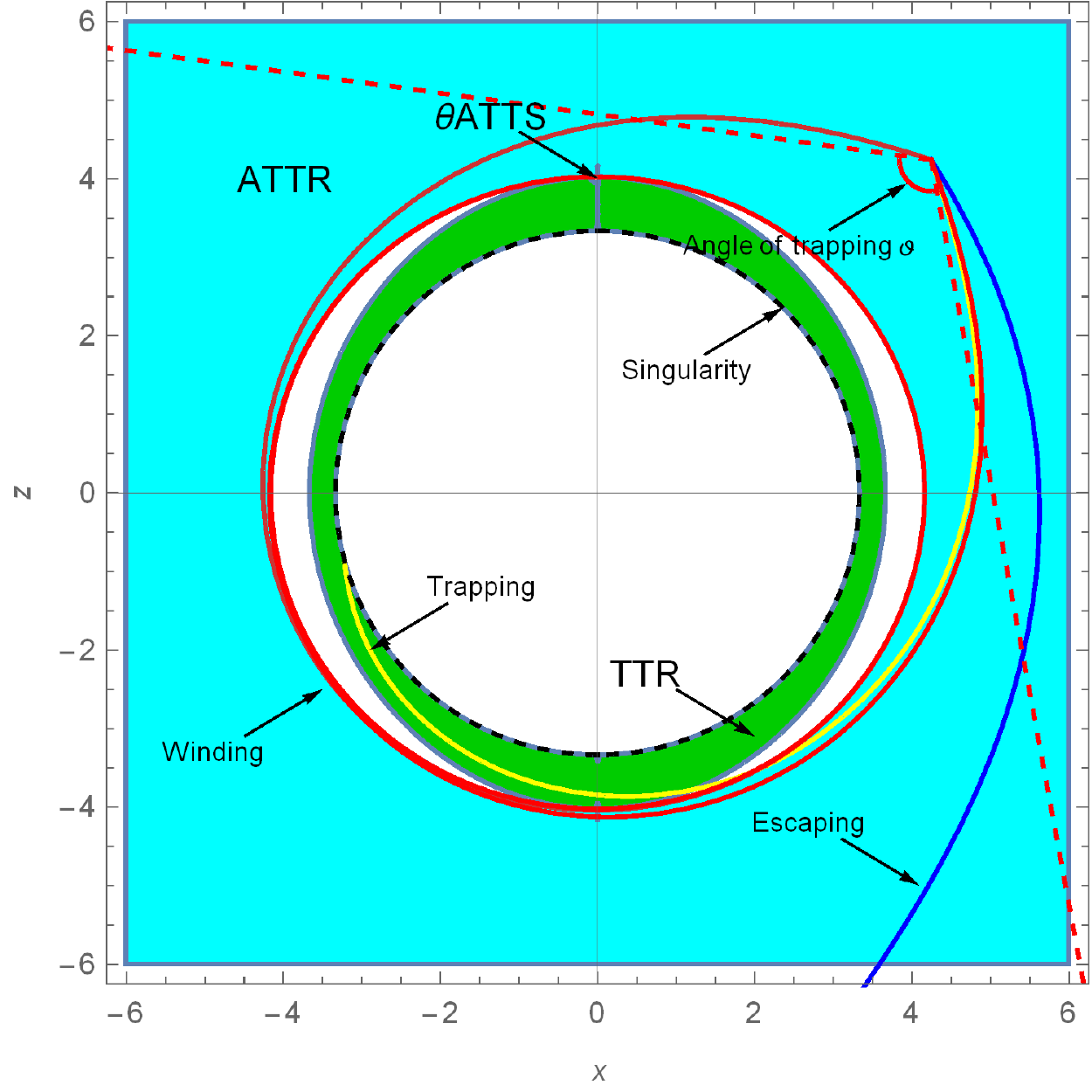}
		\label{ZVNS35NE} 
 }
\subfloat[][ZV$\delta$]{
  		\includegraphics[scale=0.41]{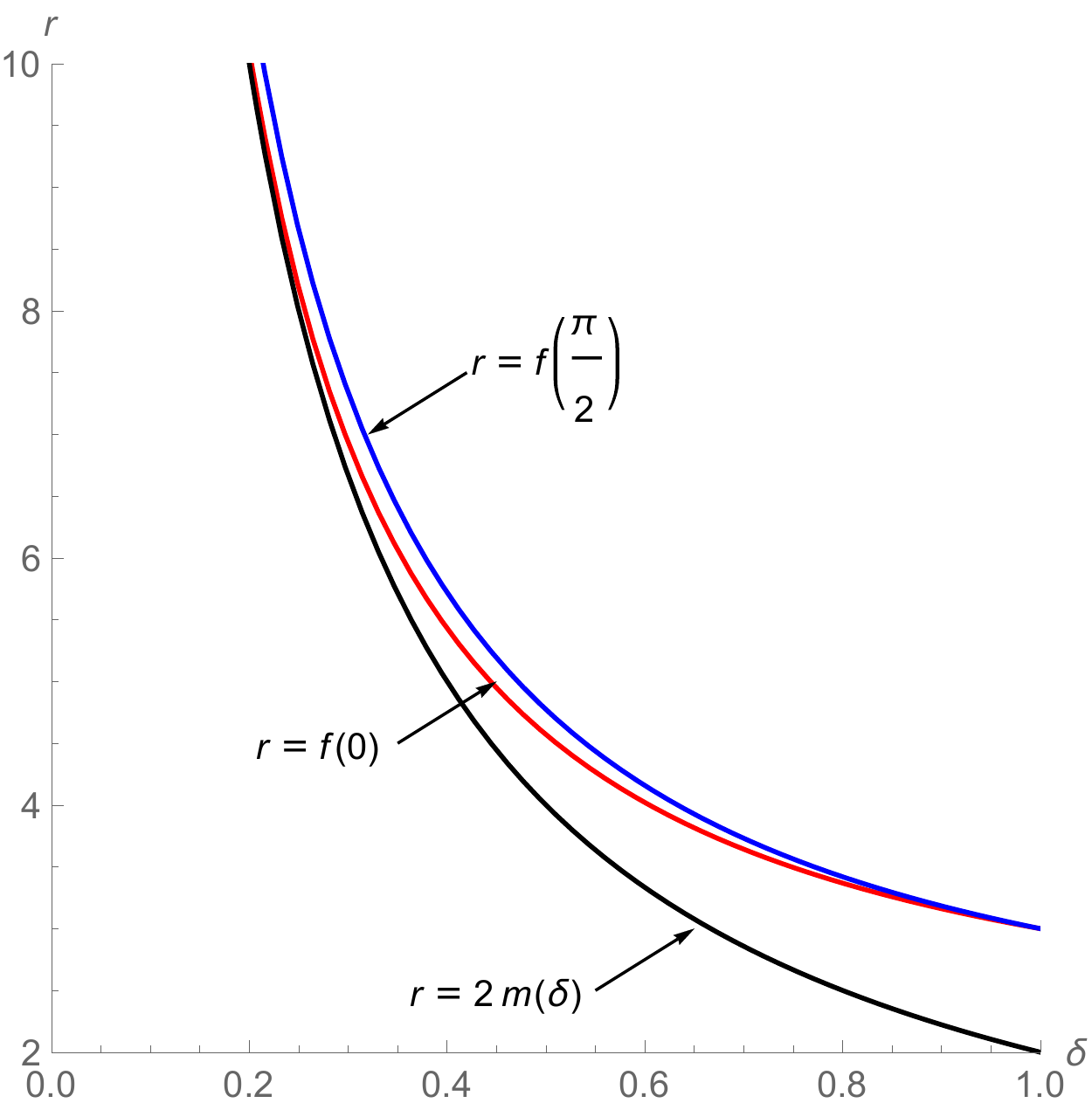}
		\label{ZVGRAN1}
 }
\\
 \subfloat[][ZV3/5]{
  		\includegraphics[scale=0.41]{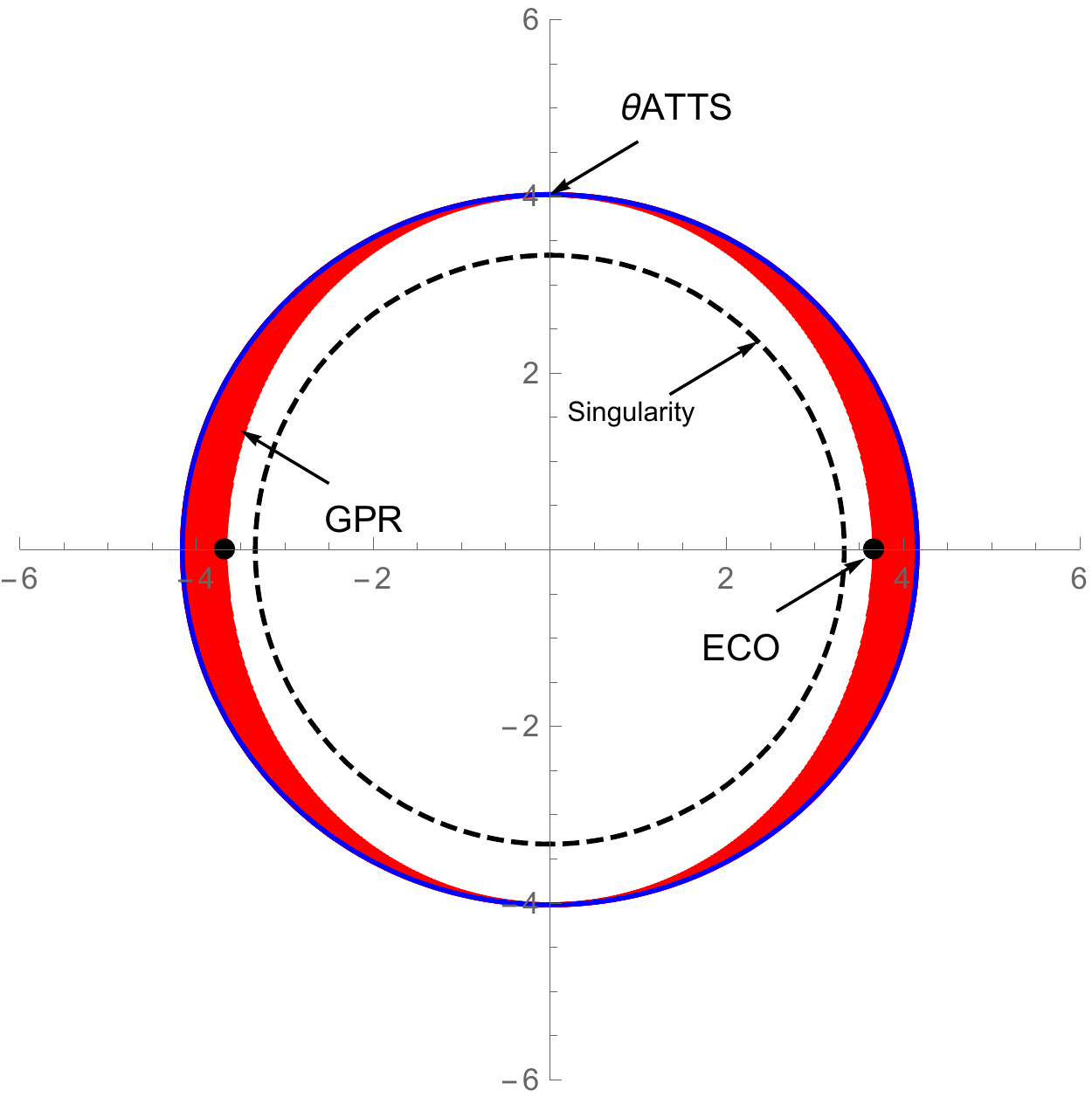}
		\label{PR35}
 } 
 \subfloat[][ZV3/5]{
  		\includegraphics[scale=0.41]{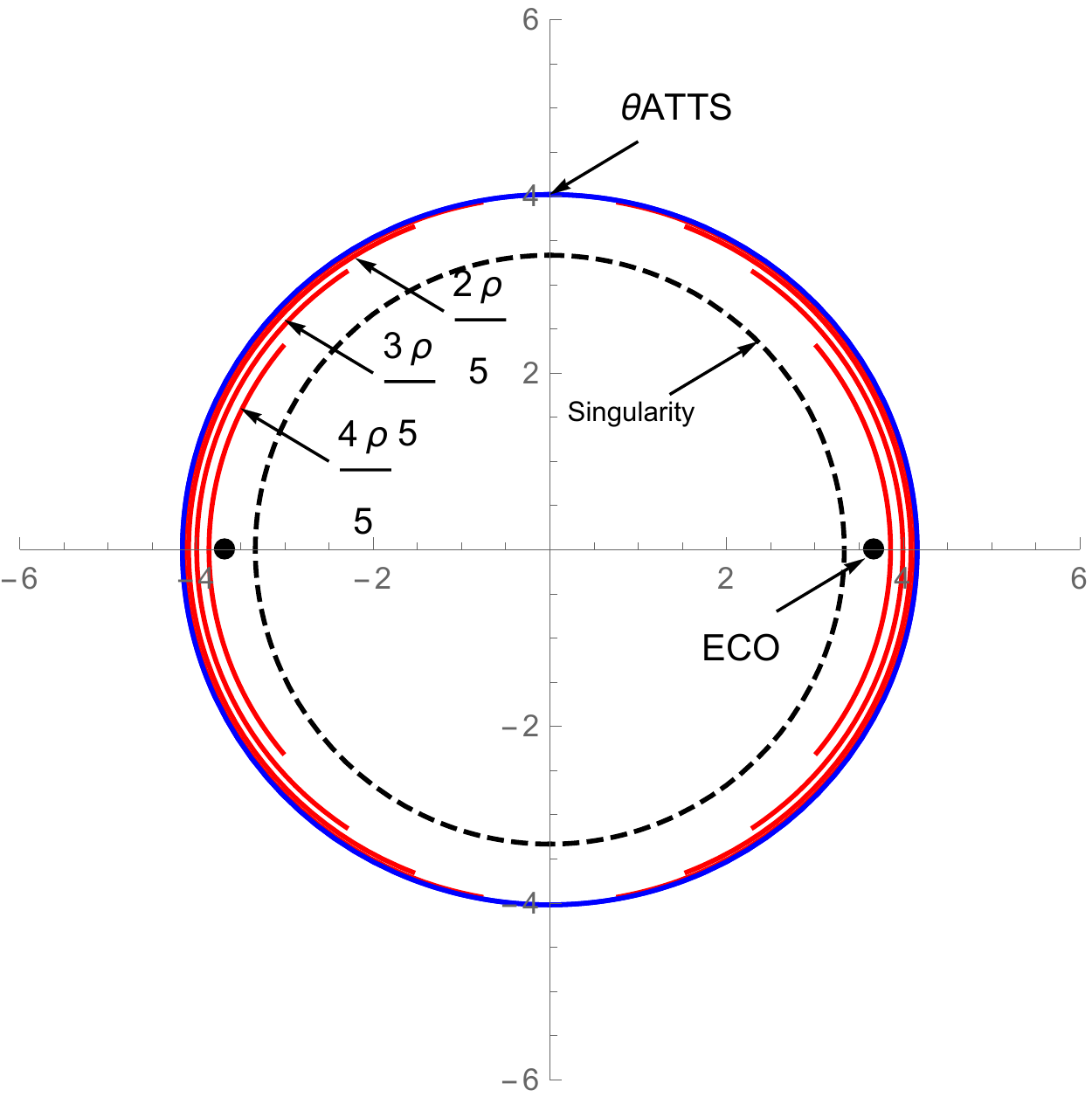}
		\label{PR35a}
 }
 \subfloat[][ZV3/5]{
  		\includegraphics[scale=0.41]{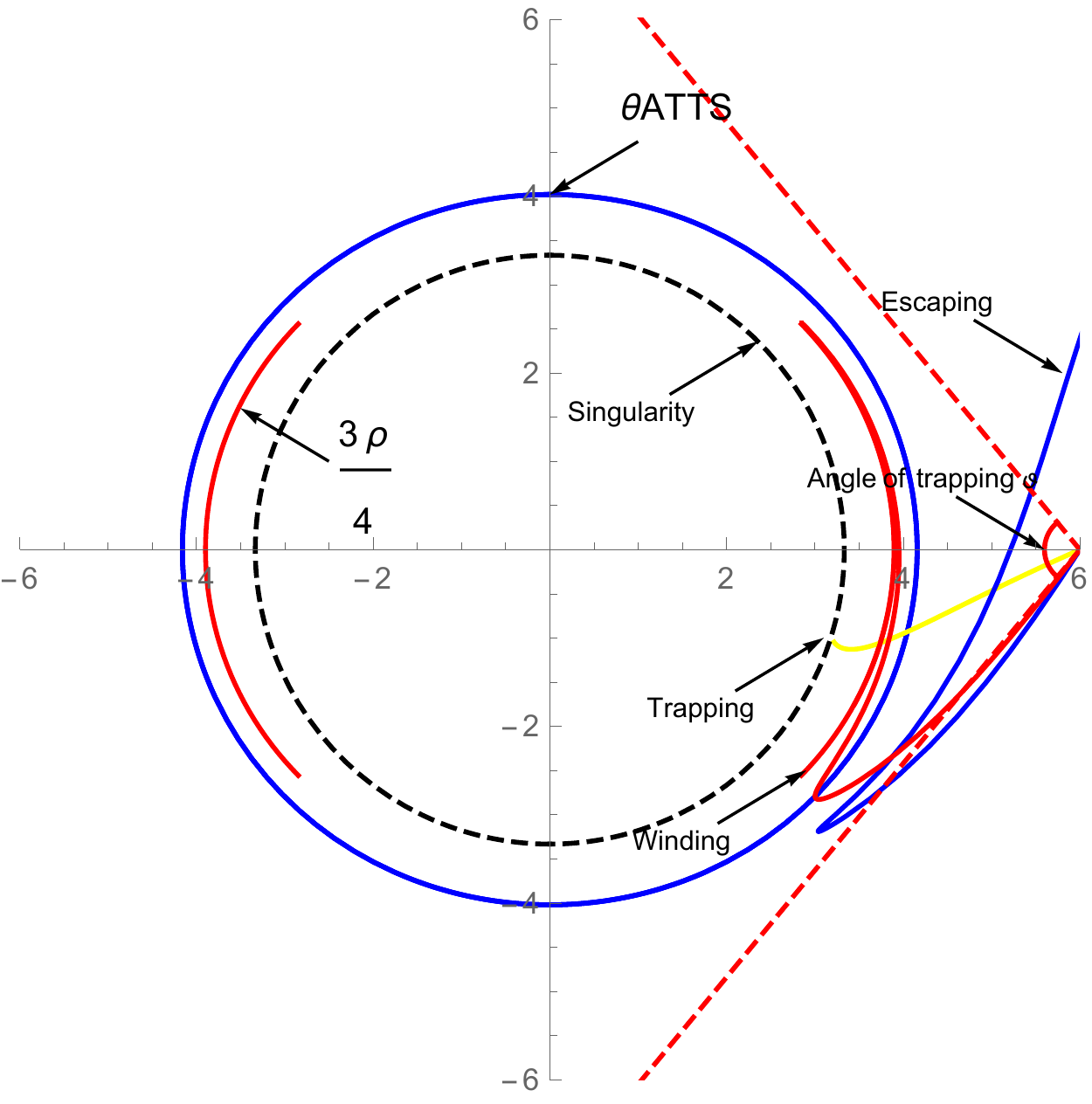}
		\label{PR35b}
 }
\caption{ATTS${}_\theta$, photon region and geodesic scattering in the space ZV for $\delta<1$. The photon region is contained inside ATTS${}_\theta$ (Fig.\ref{PR35},\ref{PR35b}).}
\label{Kerr2a}
\end{figure}

The case $\delta<1$ should be considered separately. The first important difference is that instead of TTS${}_\theta$ there is ATTS${}_\theta$ Figs. \ref{ZVNS35} - \ref{ZVNS35NE}.  The second is that the photon region is located inside the ATTS${}_\theta$ surface Figs. \ref{PR35} - \ref{PR35b}, which differs significantly from the case of Kerr-like solutions \cite{Grenzebach,Grenzebach:2015oea}. The value of $ r_{\rm max} $ becomes the minimum for the equatorial size of the surfaces of the photon region. As a result, instead of a slightly oblate solution, we get extended in the direction of the poles. At the same time, of course, there are still relativistic images \cite{Virbhadra:2008ws} created by geodetic wound on the surface in the photon region Fig. \ref{PR35b}.  However, a closed TTS will exist only for $ \delta> 1/2 $ values, since the photon region touches the singularity with $ \delta = 1/2 $ (\ref{g4}). 

\begin{figure}[tb]
\centering
\subfloat[][ZV2/5, $\pi/2$]{
  		\includegraphics[scale=0.41]{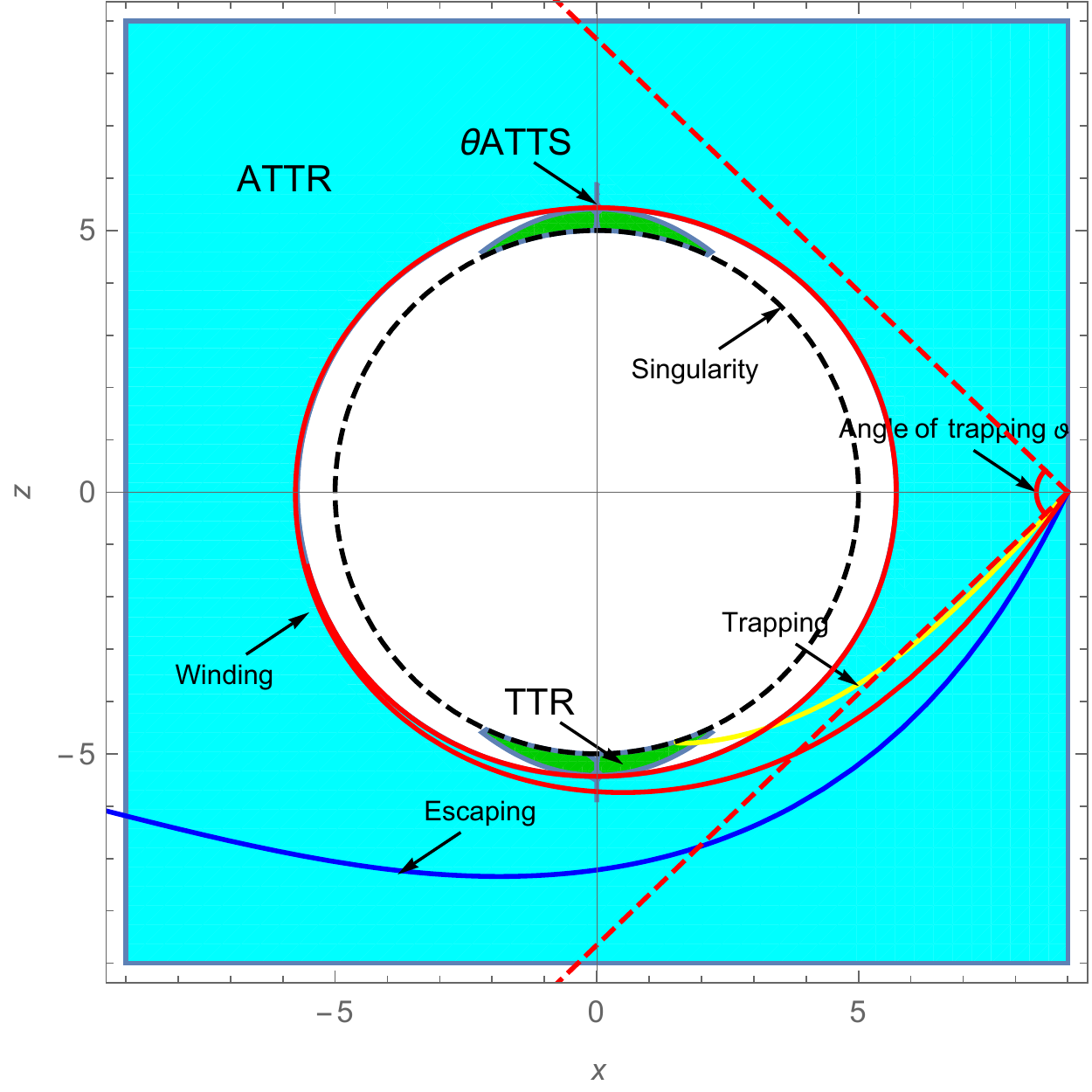}
		\label{ZVNS25}
 }
\subfloat[][ZV2/5, $\pi/2$]{
  		\includegraphics[scale=0.37]{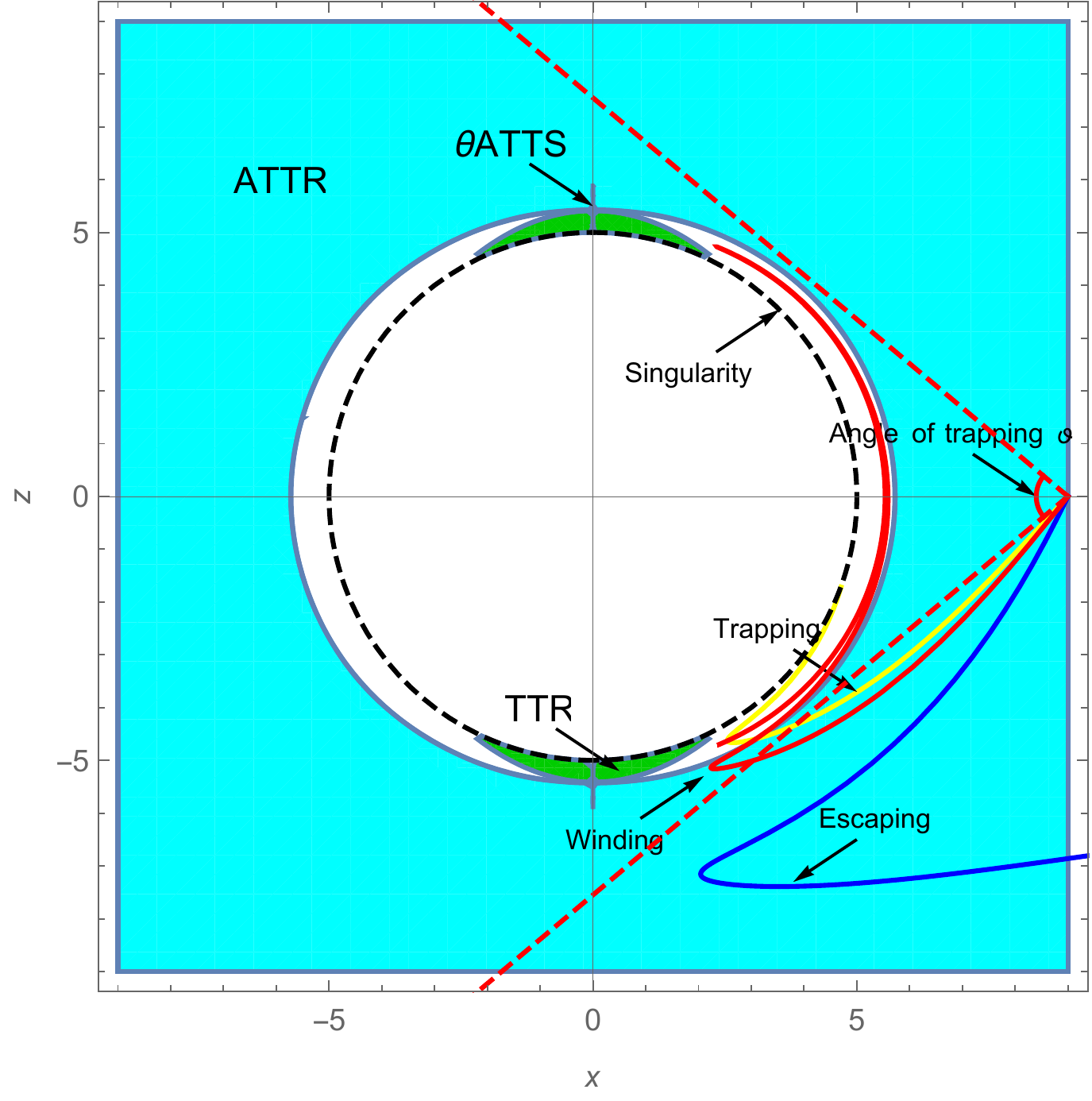}
		\label{ZVNS25a}
 }
\subfloat[][ZV2/5, $\pi/2$]{
  		\includegraphics[scale=0.37]{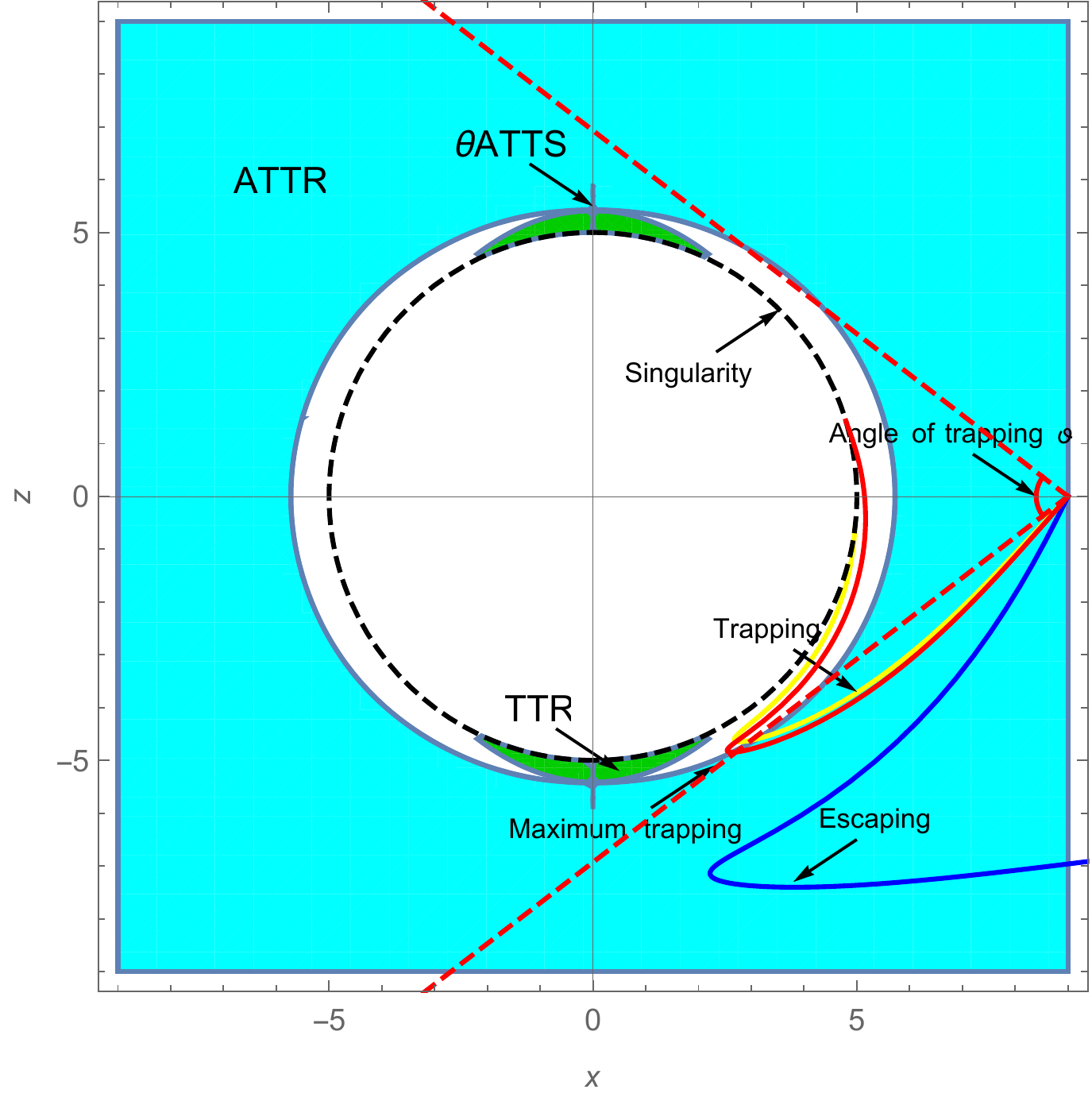}
		\label{ZVNS25b}
 }
\caption{ATTS${}_\theta$, and geodesic scattering in the space ZV for $\delta<1/2$. In the case $\delta<1/2$ there are no closed TTS, and the photon region touches the singularity, as a result the relativistic images disappear in this region.}
\label{Kerr2b}
\end{figure}

With $ \delta <1/2 $ the solution belongs to the second type II of our classification \cite{Galtsov:2019bty} and will have a clearly different optical structure Figs. \ref{Kerr2b} (see also \cite{Abdikamalov:2019ztb}). In particular, relativistic images will not appear near the entire border of the shadow, although they will still exist in the vicinity of the poles. Indeed, for small values of the impact parameter, there exist photon region surfaces that are approximately closed on  PTTR${}_\theta$ Figs. \ref{ZVNS25}, \ref{ZVNS25a} thus in the vicinity of the poles there are relativistic images. For large values of the impact parameter, the surface of the photon region does not exist and the geodesics either fall on the singularity (created shadow) or go to infinity, but cannot wound Fig. \ref{ZVNS25b} and produce relativistic images.

As can be seen from the Fig. \ref{ZVGRAN1}, the region of the possible existence of the PTTS (between the singularity and ATTS${}_\theta$ near the poles) decreases with decreasing $\delta$, hence the region of existence of relativistic images decreases, and the solution itself asymptotically approaches the third type III.

\setcounter{equation}{0}

\section {Shadow}
The shadow of the ZV solution is constructed and investigated, for example, in the work \cite{Abdikamalov:2019ztb}. Here, we provide more details about the structure of these images as well as the existence of relativistic images, based on the structure of the characteristic surfaces. 

\begin{figure}[tb]
\centering
 \subfloat[][ZV2]{
  		\includegraphics[scale=0.4]{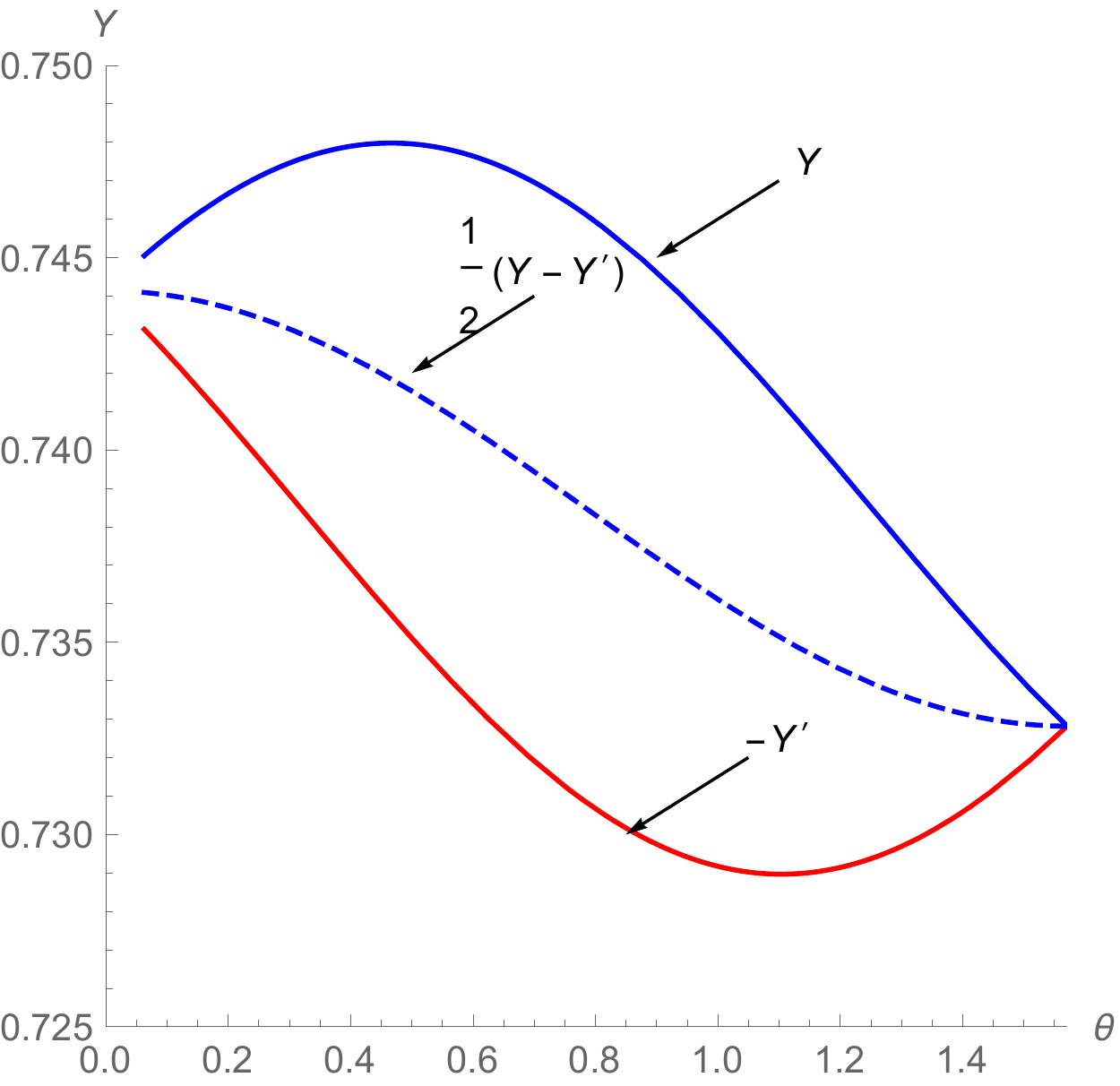}
		\label{ZVGAN}
 } 
\quad
 \subfloat[][ZVI]{
  		\includegraphics[scale=0.4]{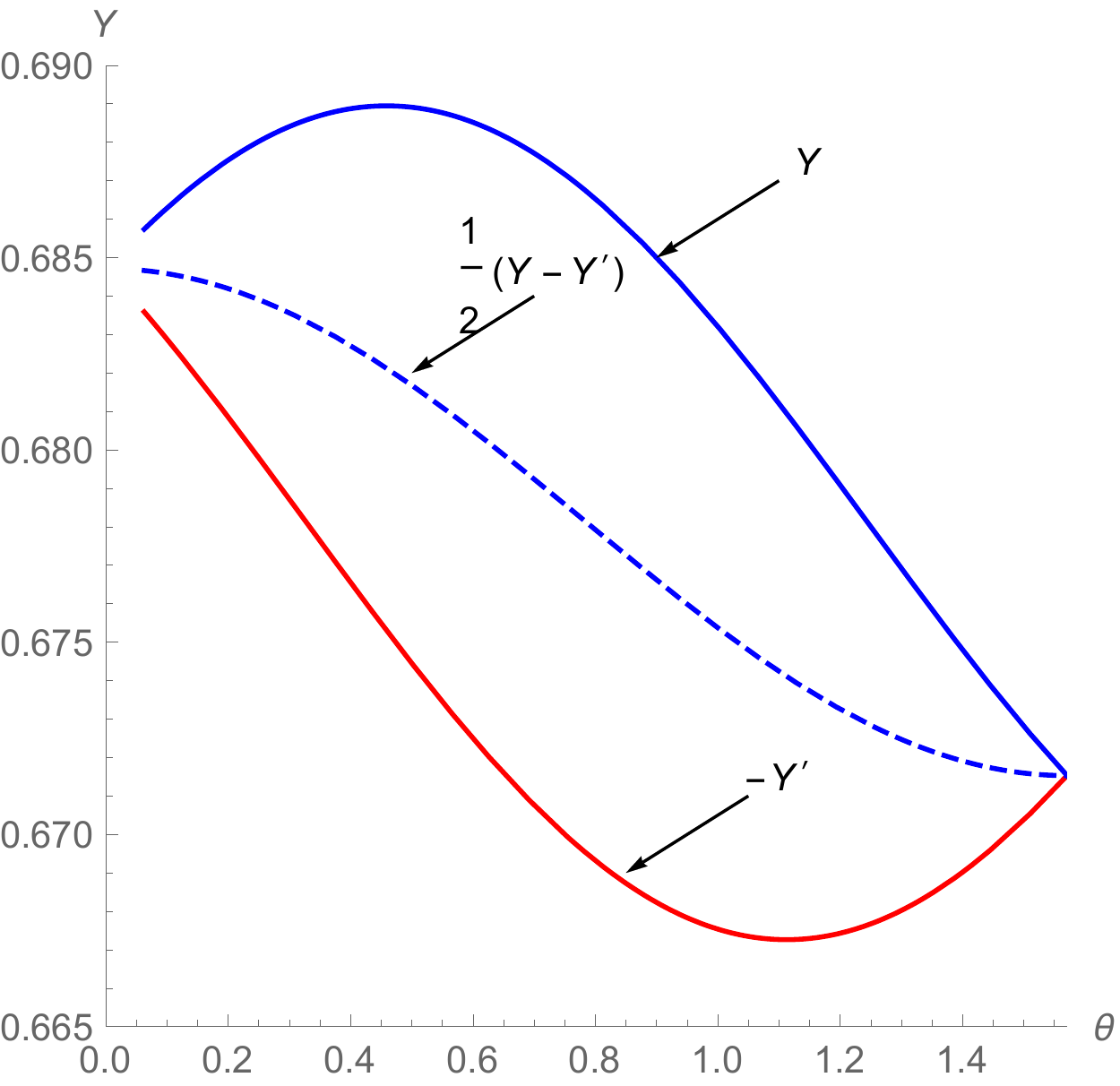} 
		\label{ZVGRAI}
 }
\quad
 \subfloat[][ZV$\delta$]{
  		\includegraphics[scale=0.4]{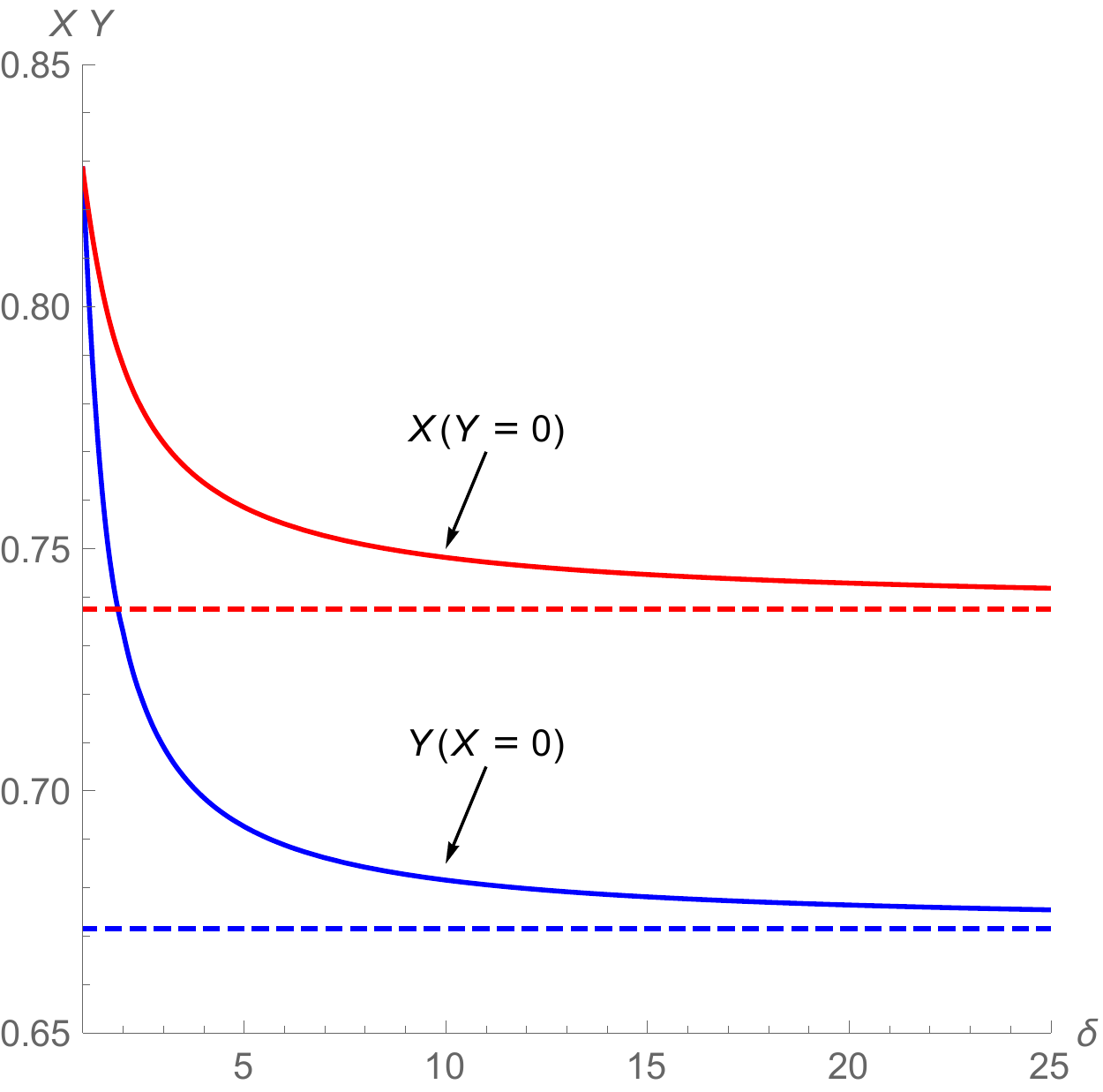}
		\label{ZVGRAN}
 }
 \caption{Positions  of shadow size maxima  $Y_{\rm max}$ (blue), $-Y'_{\rm max}$ (red) above and below the $Y=0$ respectively and vertical size of the shadow  $\Delta Y/2$ (dotted) at $X=0$ depending on the observation angle $\theta_O$ are shown in the figures (Fig. \ref{ZVGAN}, \ref{ZVGRAI}) for the case  $\delta=2,\delta=\infty$. The dependence of the size of the shadow $\Delta Y/2$ (red) at $X=0$ and $\Delta X/2$ (blue) at $Y=0$ on the value of $\delta$ при $\theta_O=\pi/2$ is shown in the figure (Fig.  \ref{ZVGRAN}), the dotted line corresponds to the limit solutions of ZVI.}
\label{Kerr2n}
\end{figure}

Obviously, the shadow of a non-spherical object like TTS${}_\theta$  will not be symmetric when viewed outside the equatorial plane.  To determine the vertical size of the shadow, one can use the numerical method of shooting. For this, a set of geodesics with a zero value of the impact parameter is emitted from the observation point, and the angle $ \vartheta $ on the celestial sphere is searched for the boundary between the geodesics going to infinity and falling on the horizon / singularity. Thus obtained, the dependence of the position of the maxima of the size of the shadow $ Y_{\rm max} $, $ -Y'_{\rm max} $ above and below the axis $ Y = 0 $, respectively, and the vertical size of the shadow $  \Delta Y / 2 $ at $ X = 0 $ from the observation angle $ \theta_O $ and $ r_O=6M$ is shown in the figures Fig. \ref{ZVGAN}, \ref{ZVGRAI} for the case of $ \delta = 2, \delta = \infty $, respectively, and clearly corresponds to non-spherical structure TTS${}_\theta$ . The dependence of the shadow size $ \Delta Y / 2 $ at $ X = 0 $ and $ \Delta X / 2 $ at $ Y = 0 $ (defined explicitly by the formula (\ref{g5}) for the equatorial observer) on the value of $ \delta $ with $ \theta_O = \pi / 2 $ is shown in the Fig. \ref{ZVGRAN} and obviously correlates with the general behavior (Figs. \ref{K}).

\begin{figure}[tb!]
\centering
		 \subfloat[][FS, $\pi/2$]{
  		\includegraphics[scale=0.42]{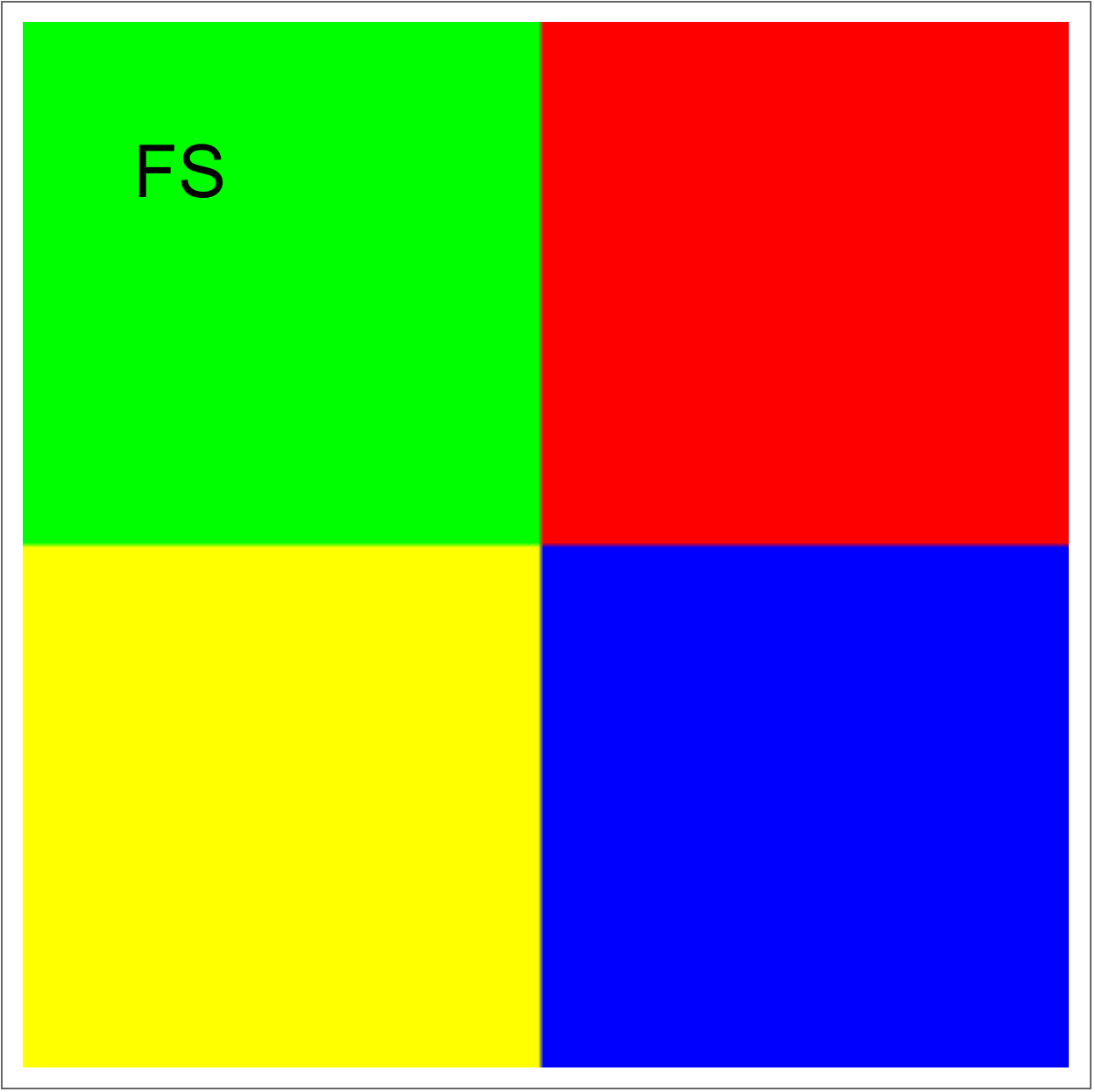}
		\label{SH0E}
		}
				 \subfloat[][FS, $\pi/4$]{
  		\includegraphics[scale=0.42]{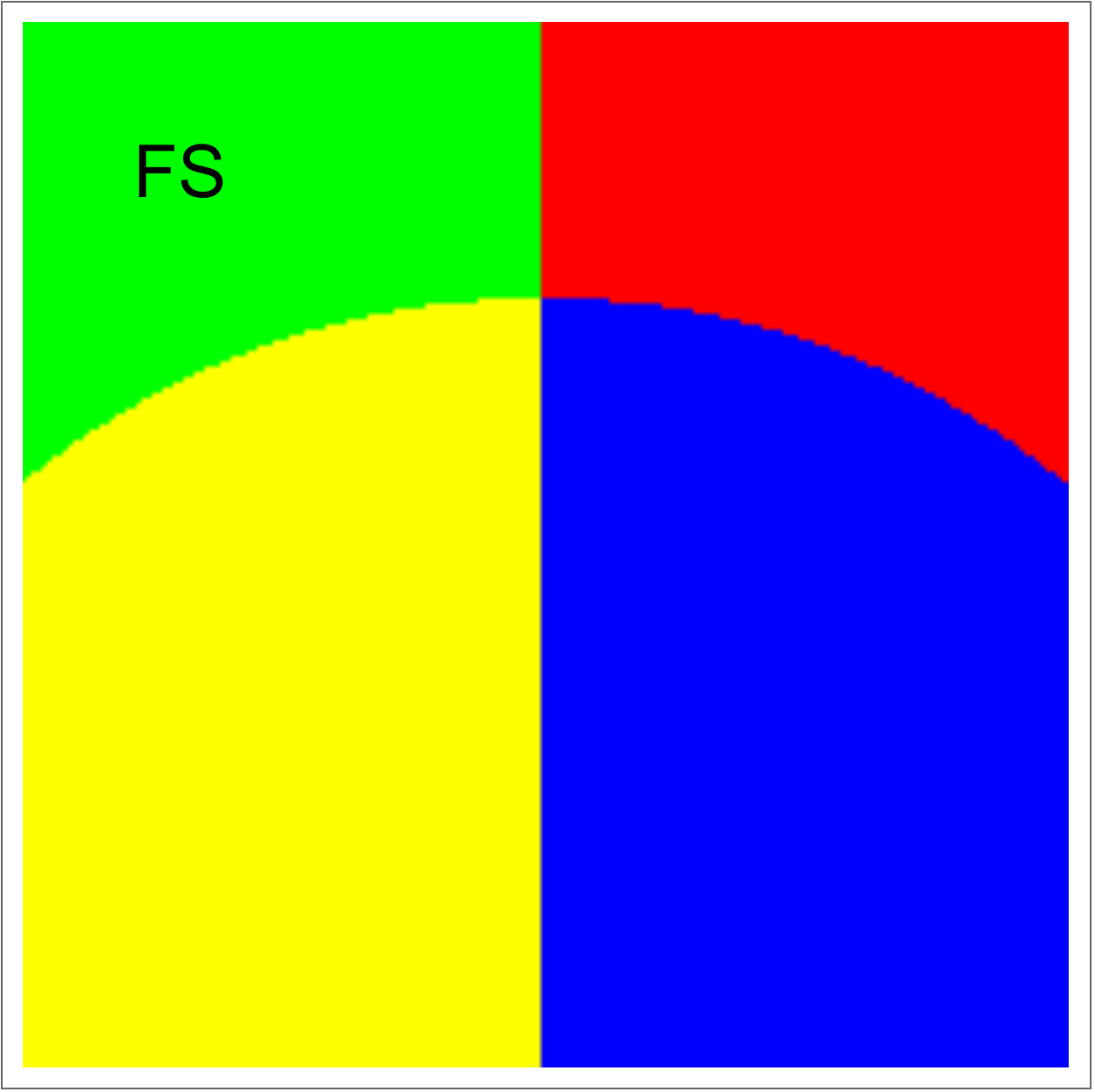}
		\label{SH0NE}
		}
				 \subfloat[][FS, $\pi/12$]{
  		\includegraphics[scale=0.42]{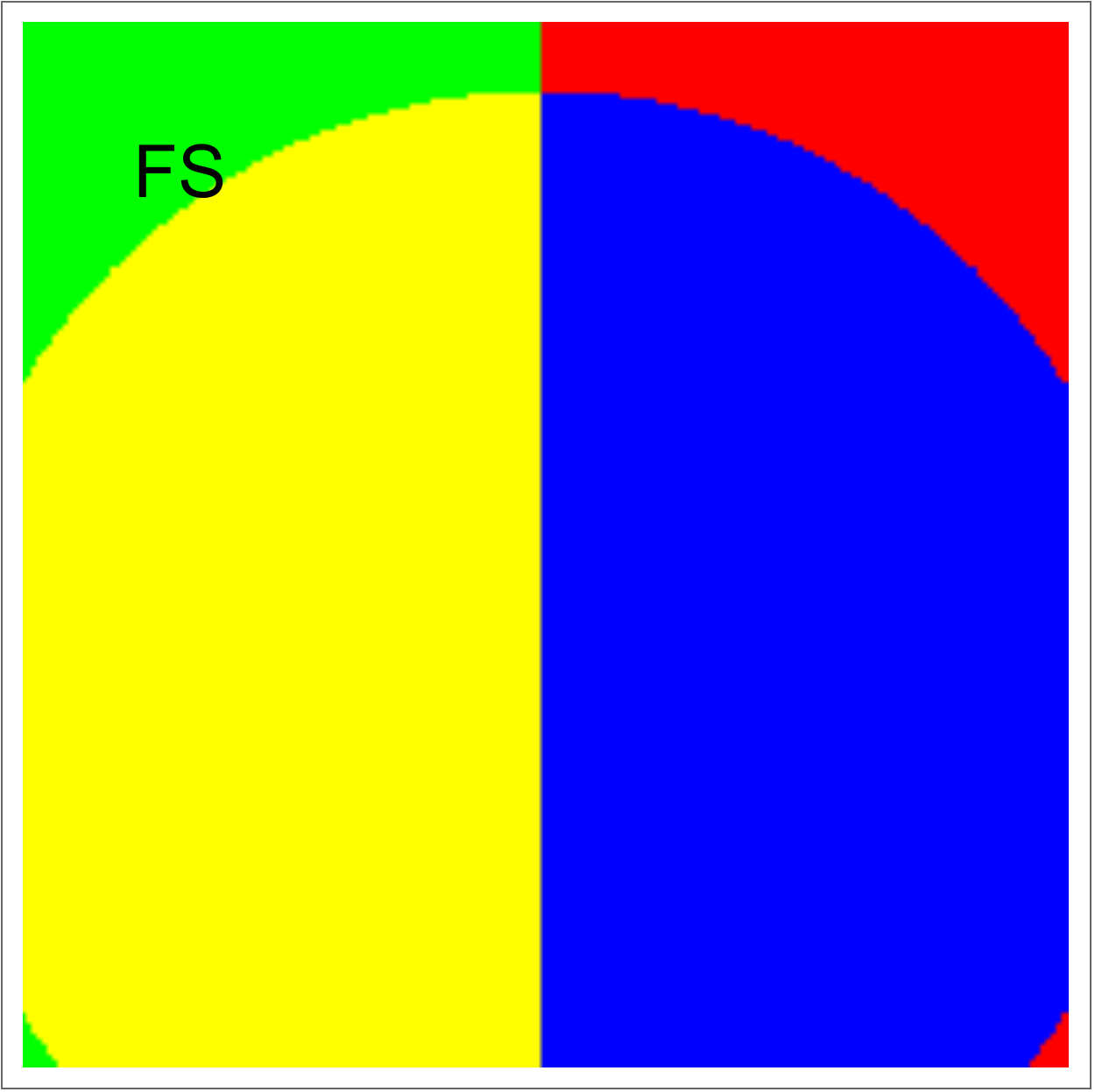}
		\label{SH0NE}
		}
\caption{The flat space. Observer with coordinate $r_O =5M $ and various $\theta_O$. The sky $r =30M $ is painted in 4 colors. Size $200\times200$ pixels.}
\label{KerrF}
\end{figure}

\begin{figure}[tb!]
\centering
		 \subfloat[][ZV1, $\pi/2$]{
  		\includegraphics[scale=0.42]{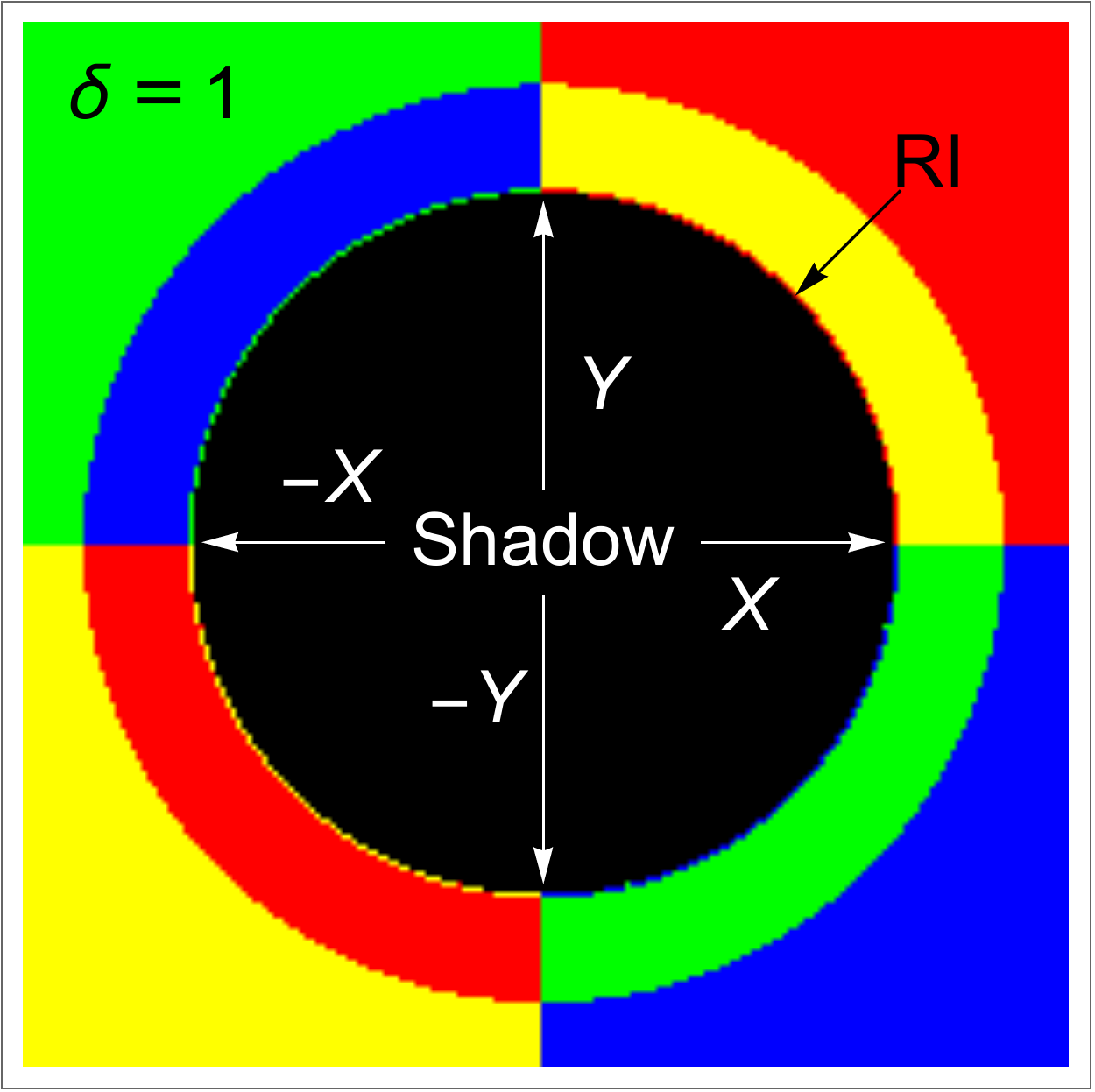}
		\label{SH1E}
		}
		 \subfloat[][ZV1, $\pi/4$]{
  		\includegraphics[scale=0.42]{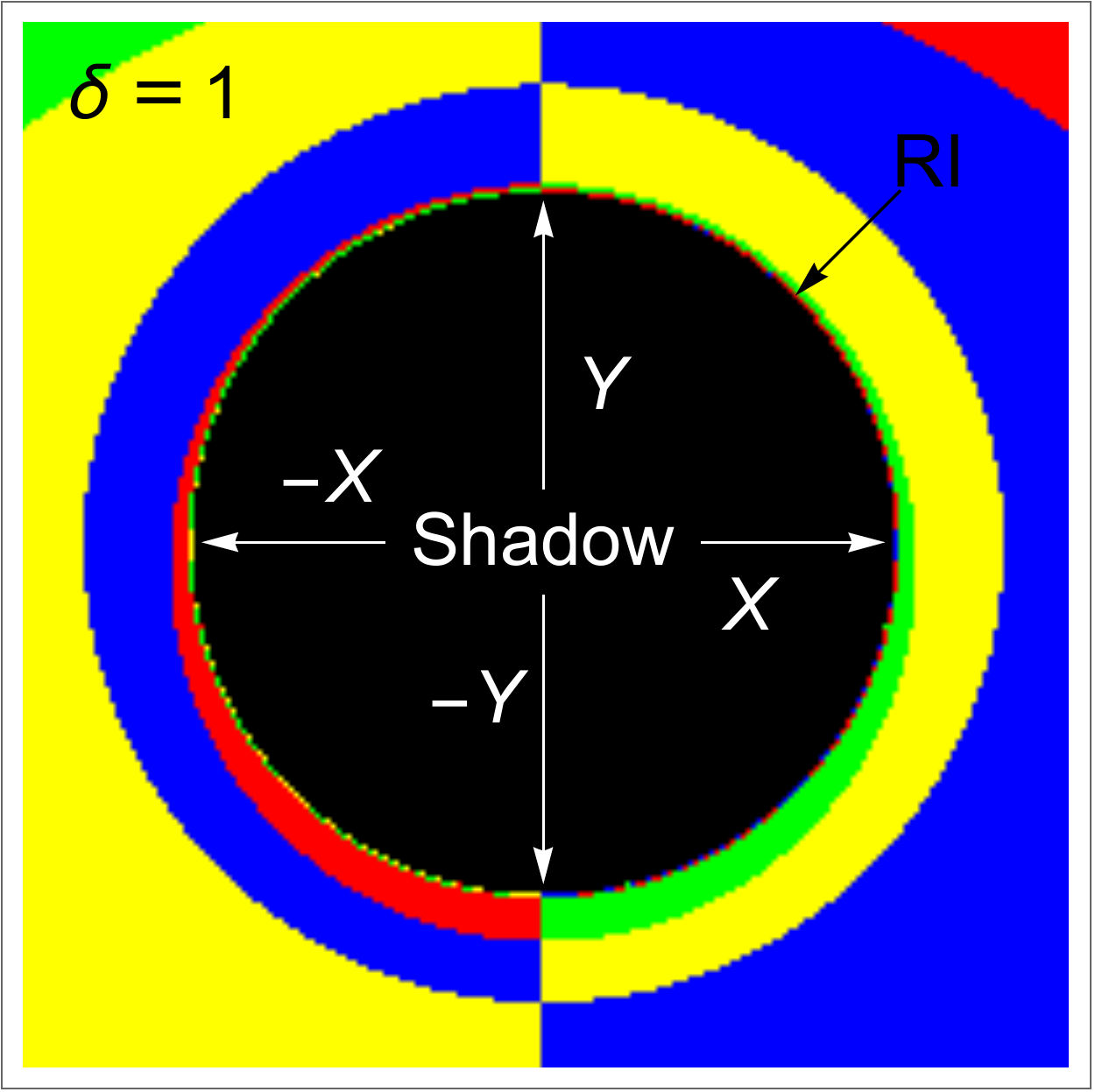}
		\label{SH1NE}
		}
		 \subfloat[][ZV1, $\pi/12$]{
  		\includegraphics[scale=0.42]{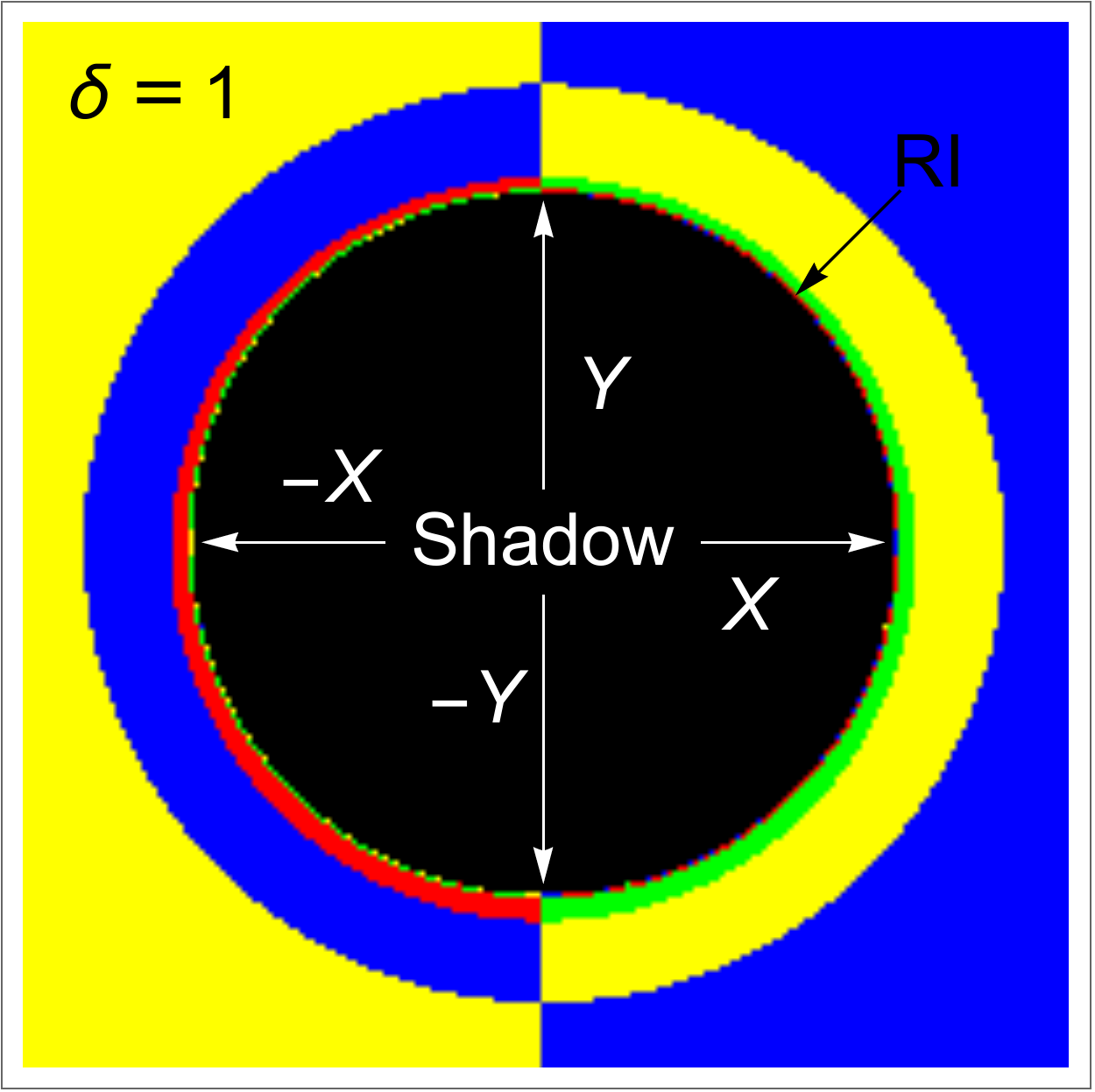}
		\label{SH1N}
		}
				\\
		 \subfloat[][ZV2, $\pi/2$]{
  		\includegraphics[scale=0.42]{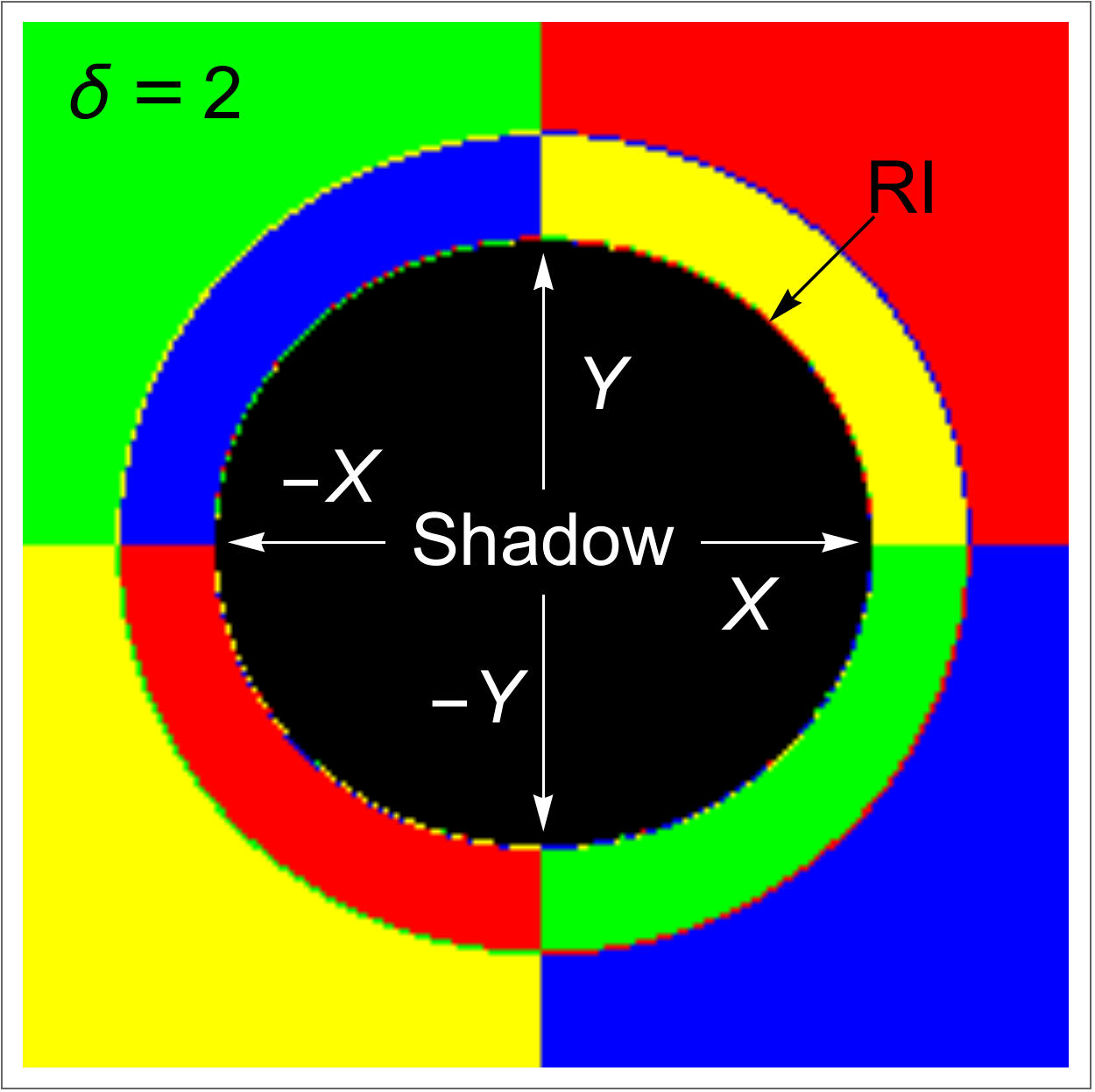}
		\label{SH2E}
		}
		 \subfloat[][ZV2, $\pi/4$]{
  		\includegraphics[scale=0.42]{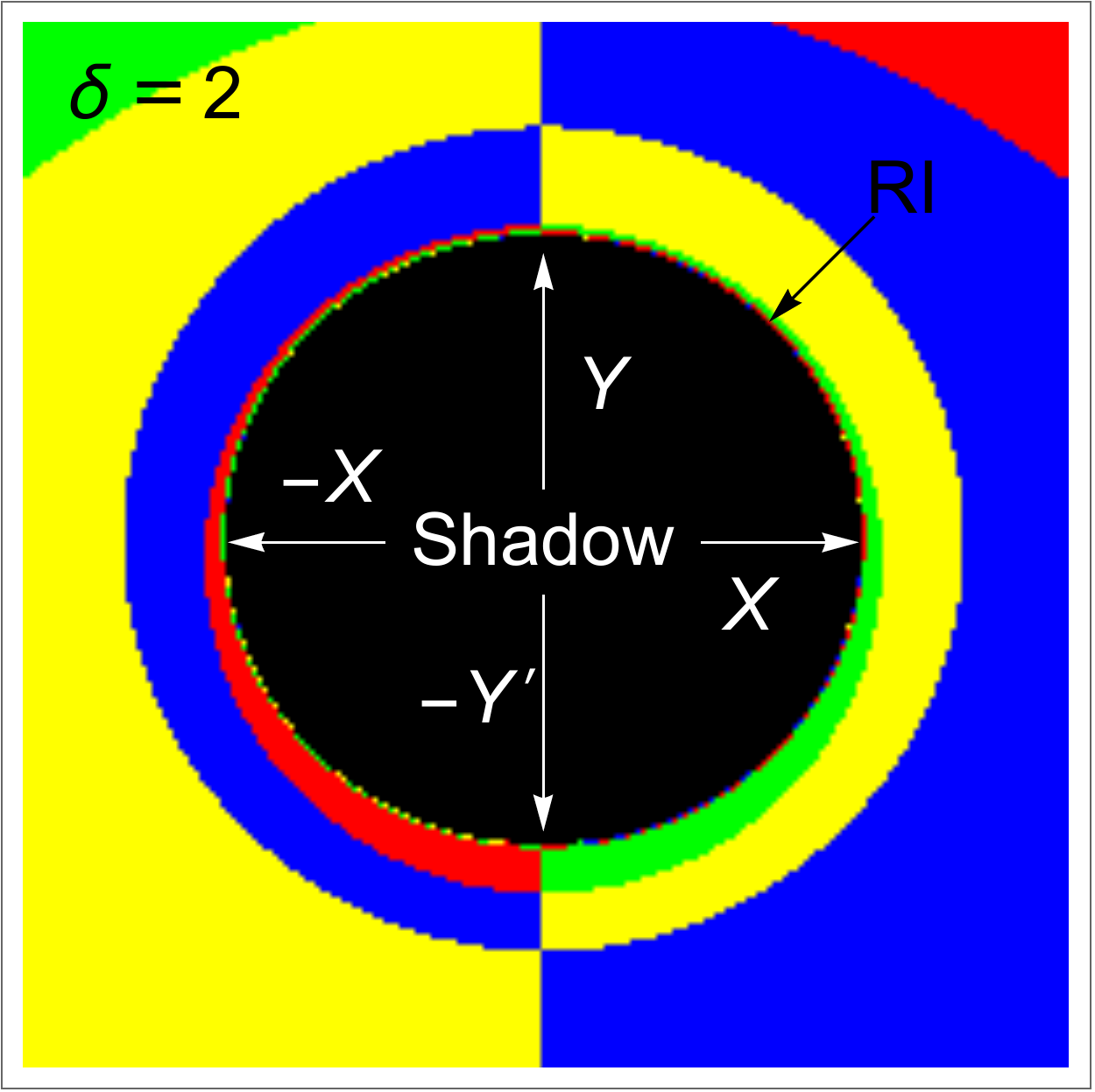}
		\label{SH2NE}
		}
		 \subfloat[][ZV2, $\pi/12$]{
  		\includegraphics[scale=0.42]{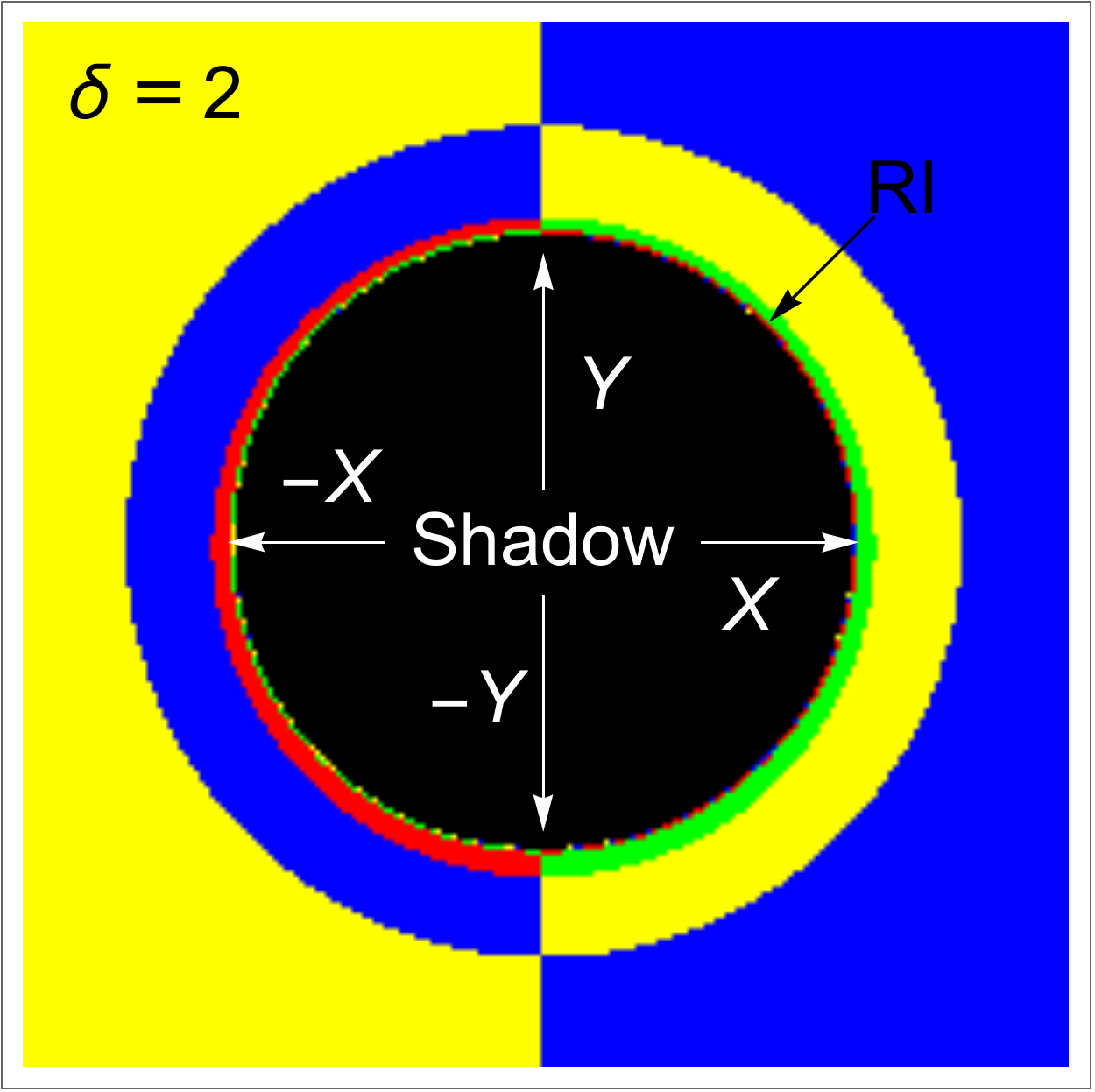}
		\label{SH2N}
		}
				\\
		 \subfloat[][ZVI, $\pi/2$]{
  		\includegraphics[scale=0.42]{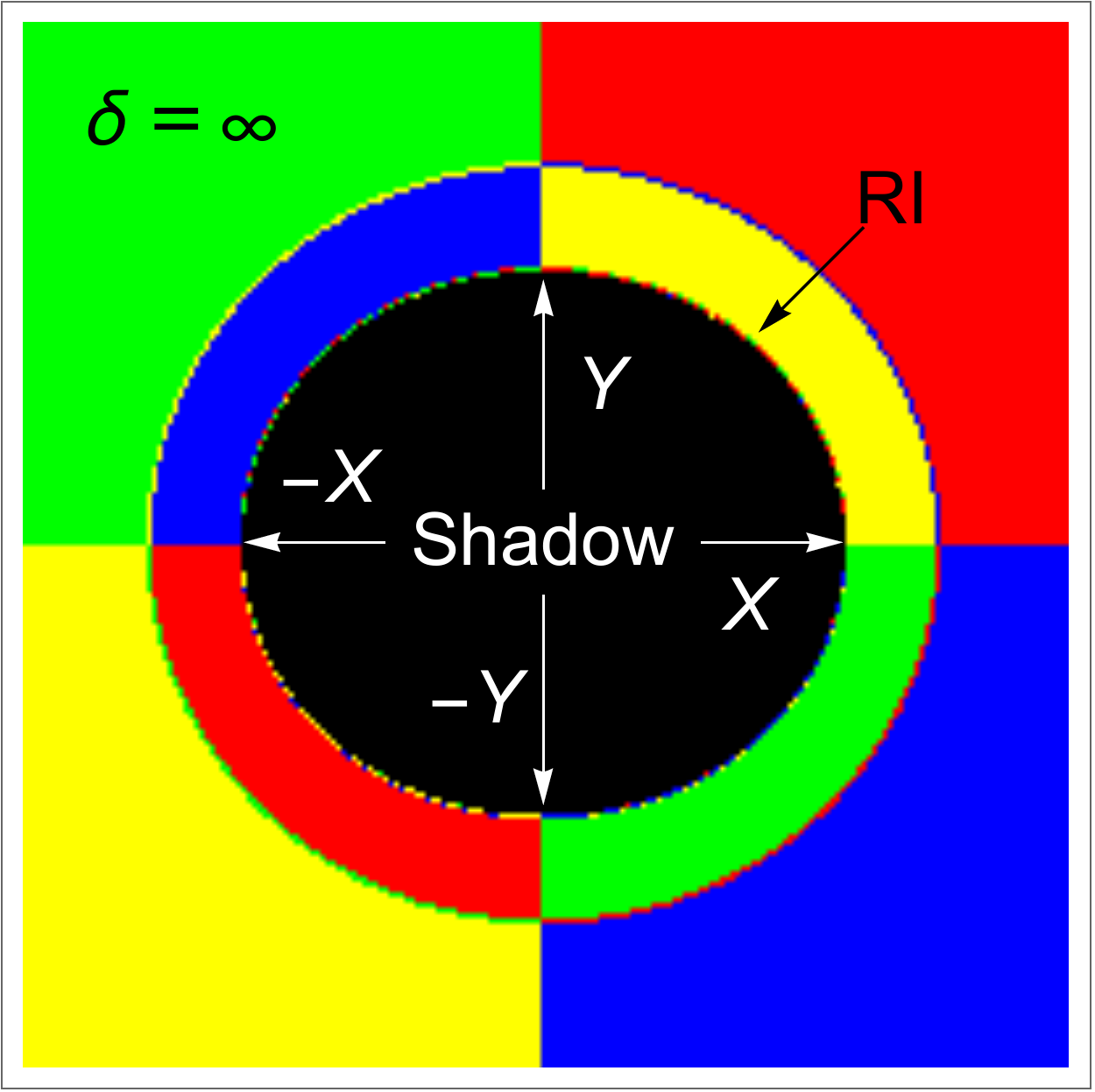}
		\label{SH20E}
		}
		 \subfloat[][ZVI, $\pi/4$]{
  		\includegraphics[scale=0.42]{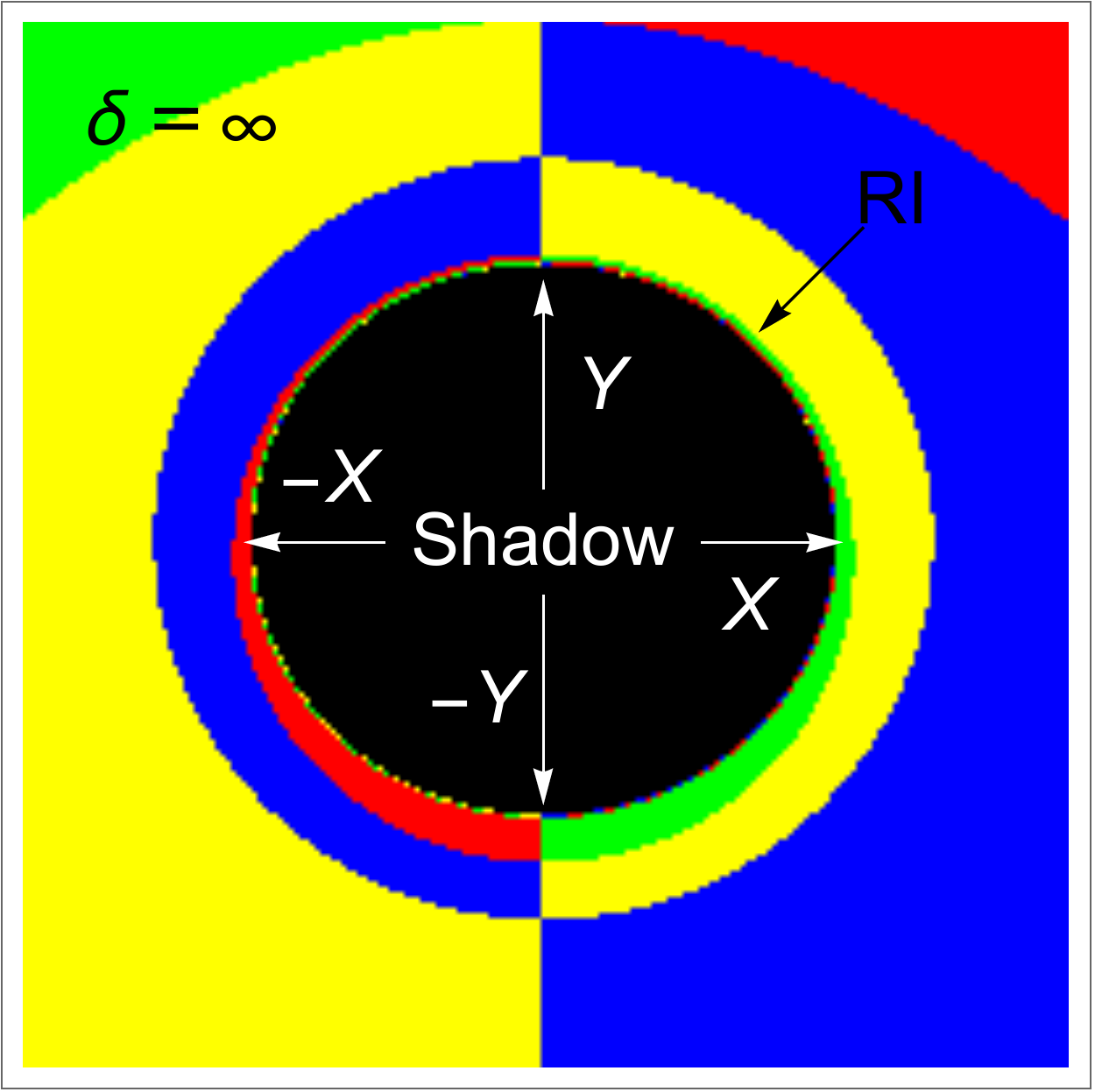}
		\label{SHINE}
		}
		 \subfloat[][ZVI, $\pi/12$]{
  		\includegraphics[scale=0.42]{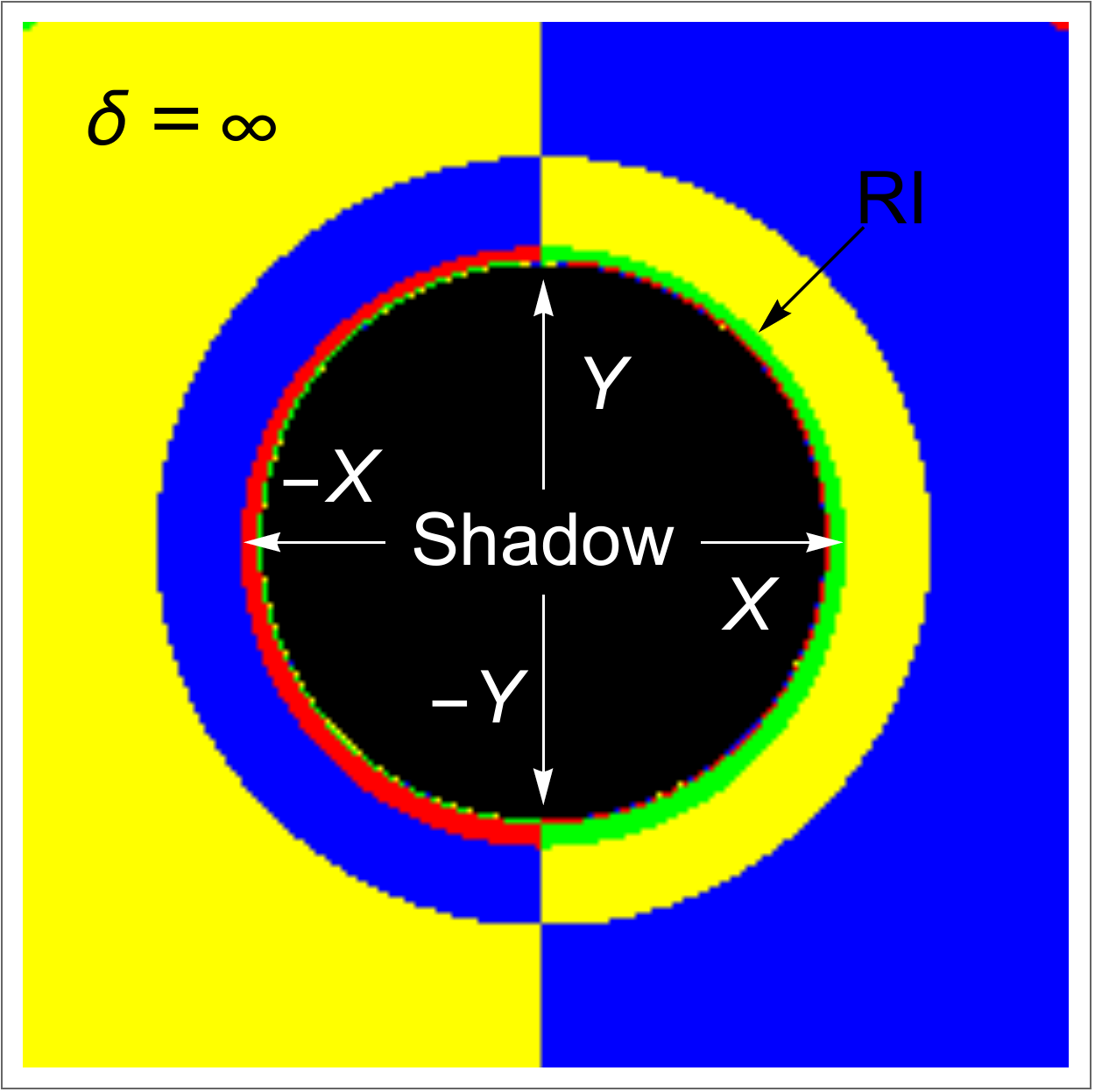}
		\label{SHIN}
		}
\caption{The shadow of the ZV space with $M=1 $ and $ \delta \geq1 $. Observer with coordinate $r_O =5M $ and various $\theta_O$. Black is a shadow. The sky $r =30M $ is painted in 4 colors. Size $200\times200$ pixels.}
\label{Kerr4}
\end{figure}

\begin{figure}[tb!]
\centering
		 \subfloat[][ZV3/5, $\pi/2$]{
  		\includegraphics[scale=0.42]{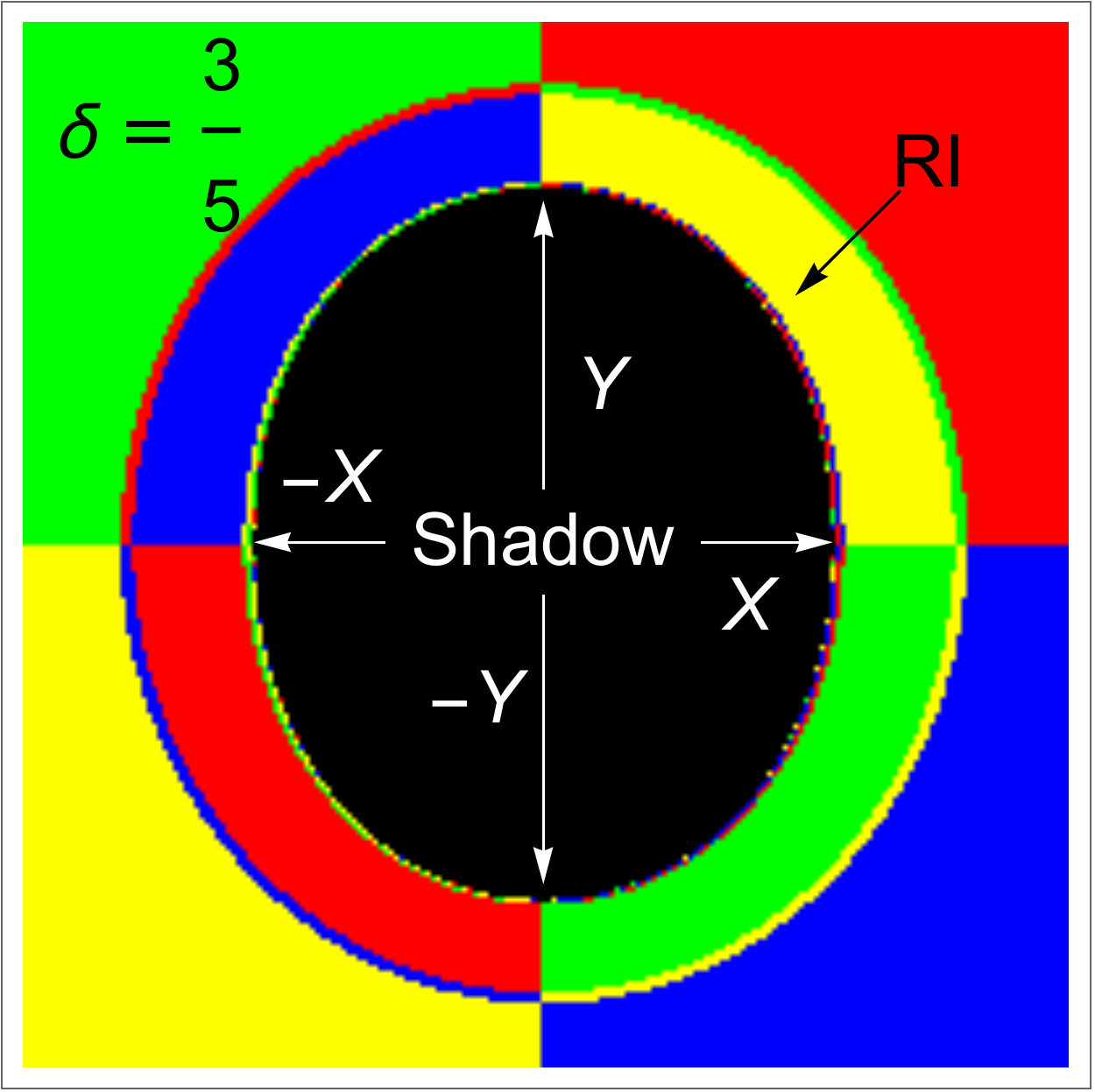}
		\label{NSH35E}
		}
			 \subfloat[][ZV3/5, $\pi/4$]{
  		\includegraphics[scale=0.42]{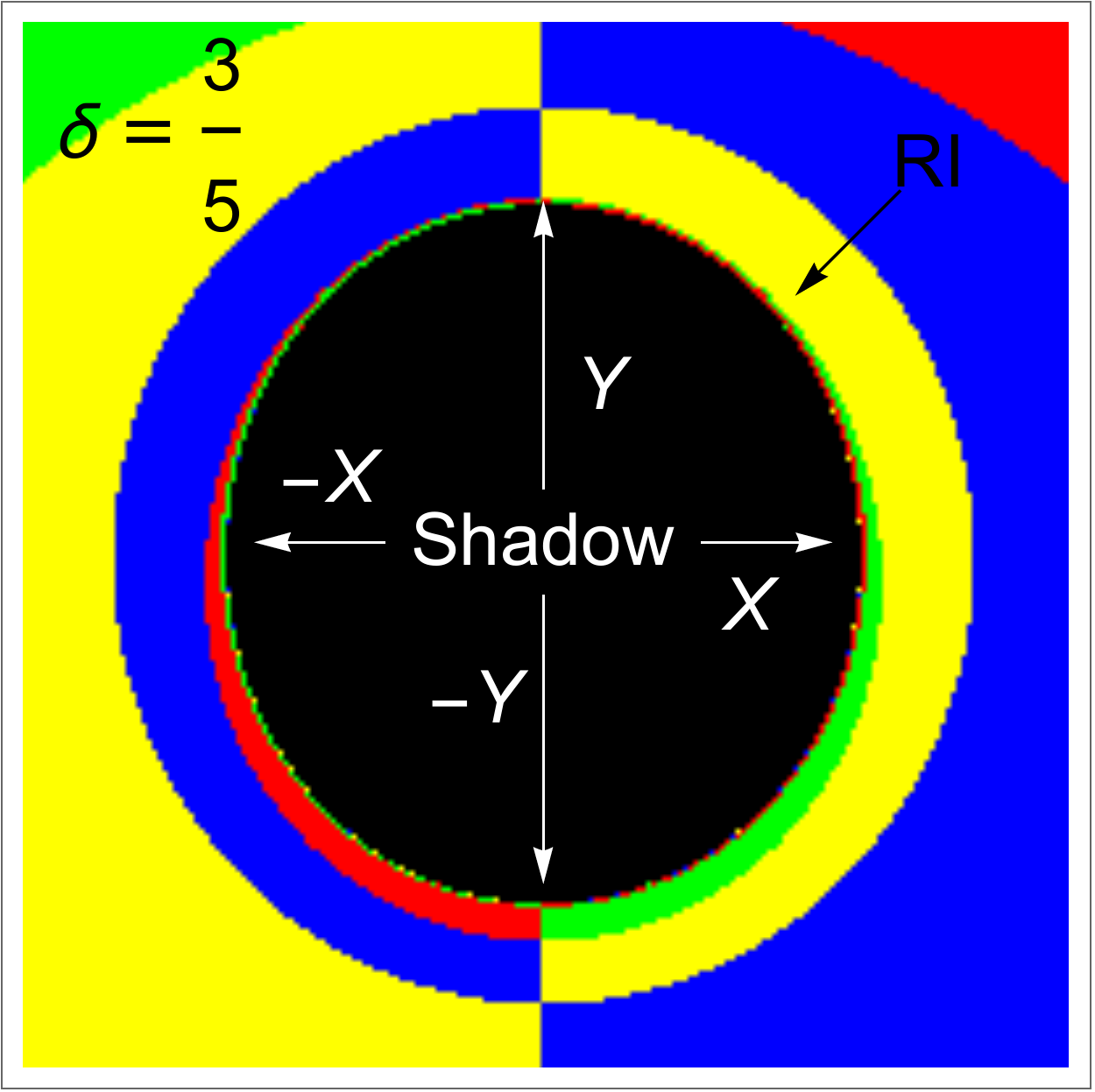}
		\label{NSH35NE}
		}
			 \subfloat[][ZV3/5, $\pi/12$]{
  		\includegraphics[scale=0.42]{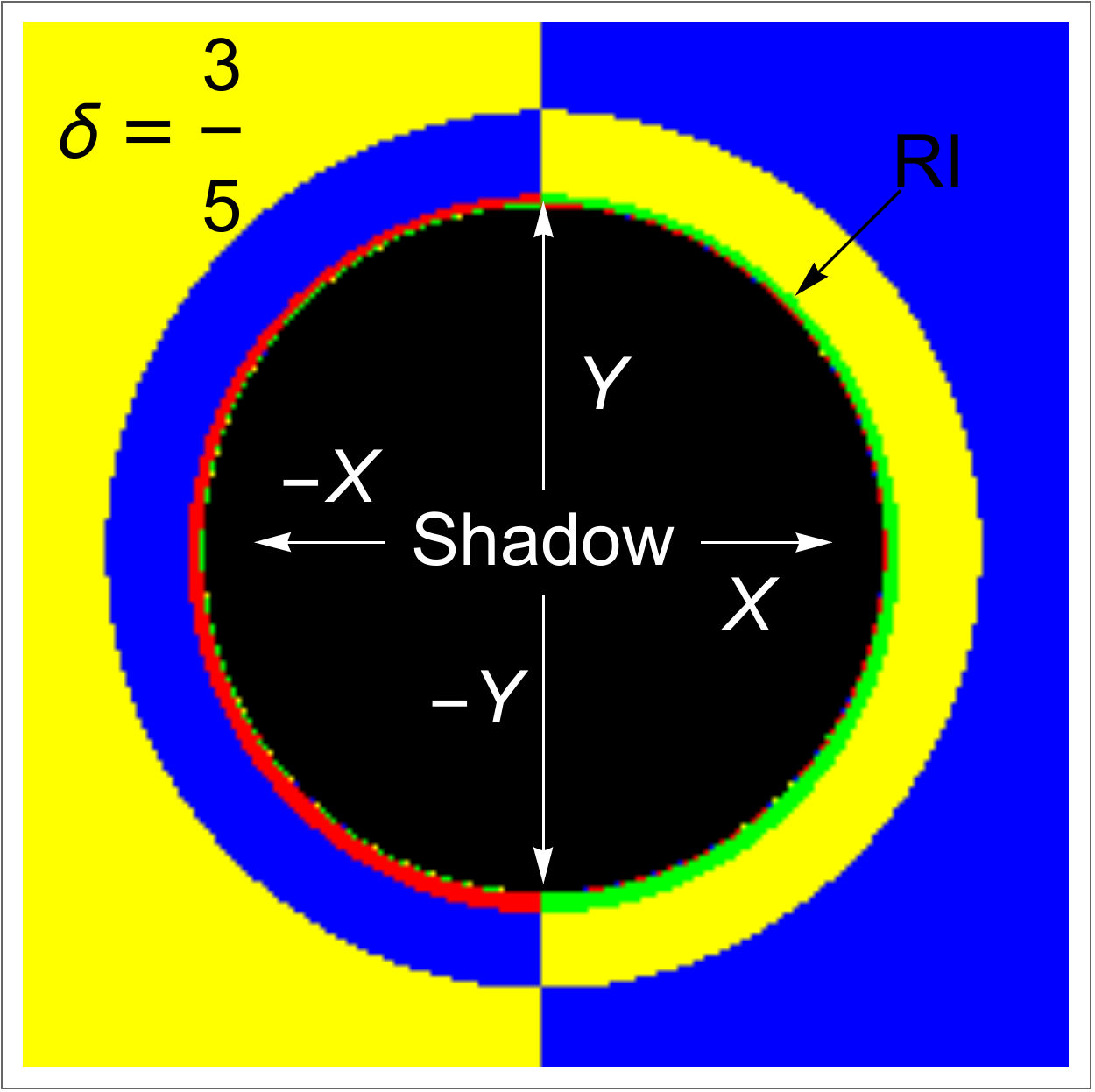}
		\label{NSH35N}
		}
			\\
		 \subfloat[][ZV2/5, $\pi/2$]{
  		\includegraphics[scale=0.42]{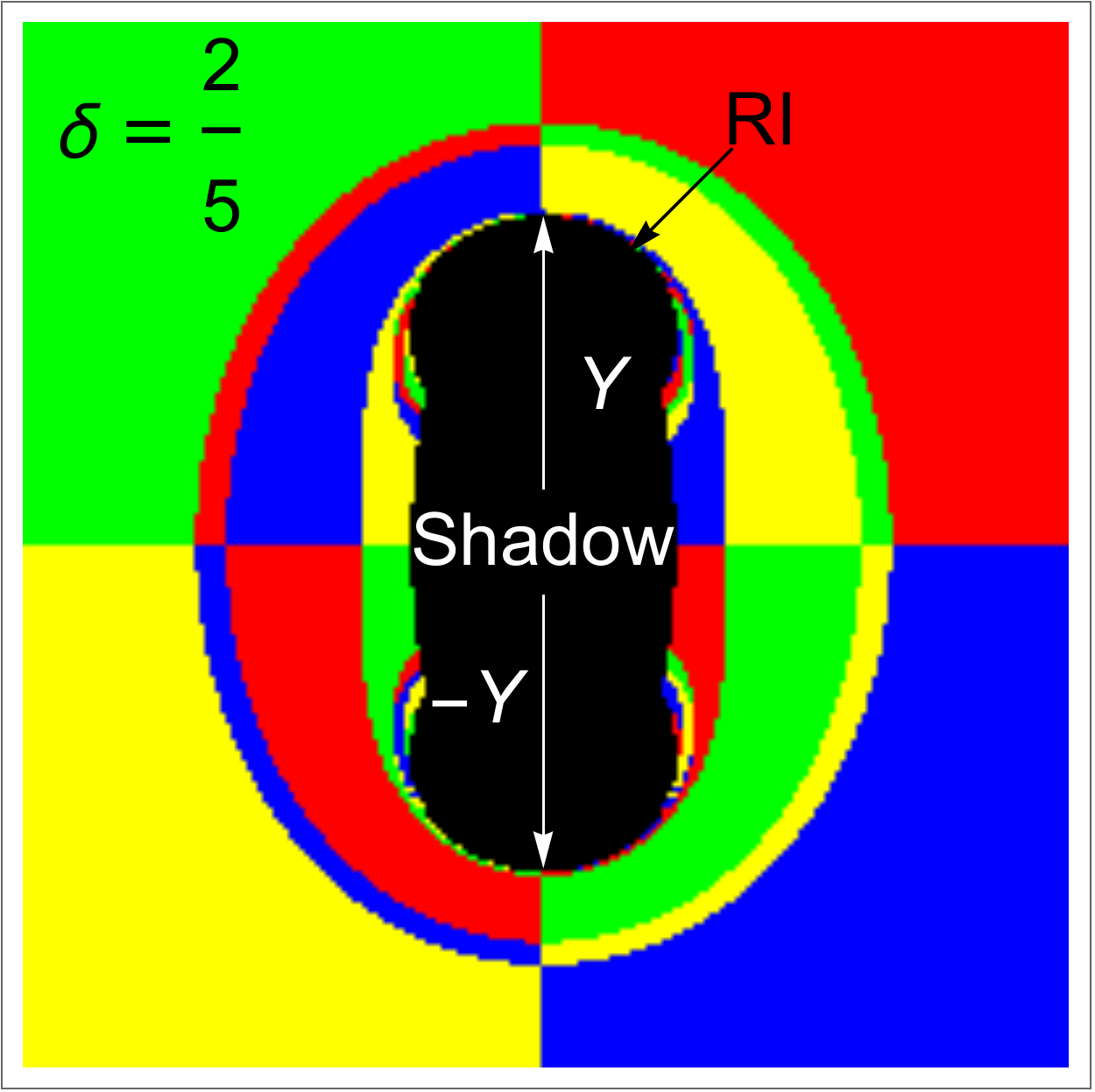}
		\label{NSH25E}
		}
			 \subfloat[][ZV2/5, $\pi/4$]{
  		\includegraphics[scale=0.42]{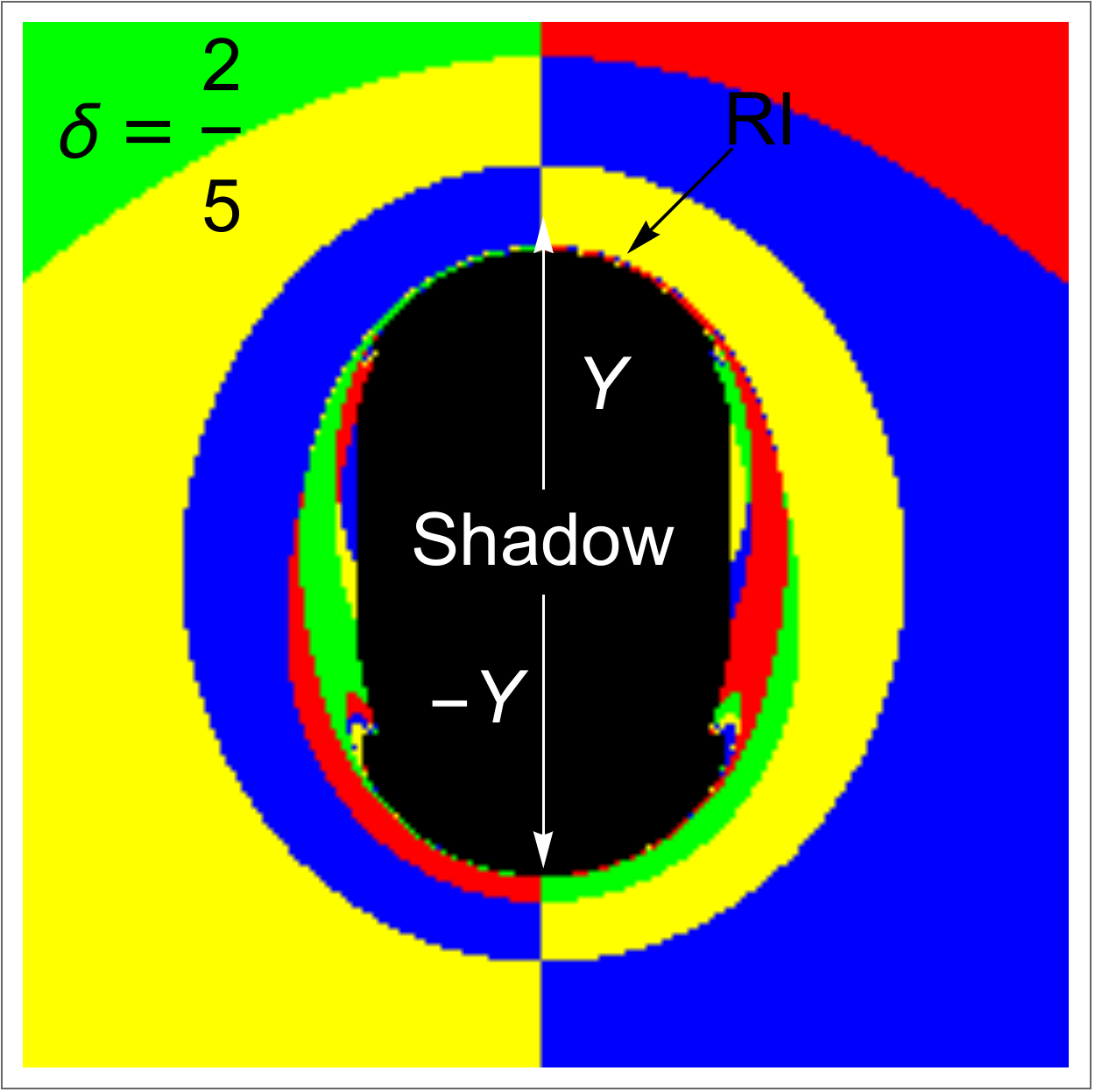}
		\label{NSH25NE}
		}
			 \subfloat[][ZV2/5, $\pi/12$]{
  		\includegraphics[scale=0.42]{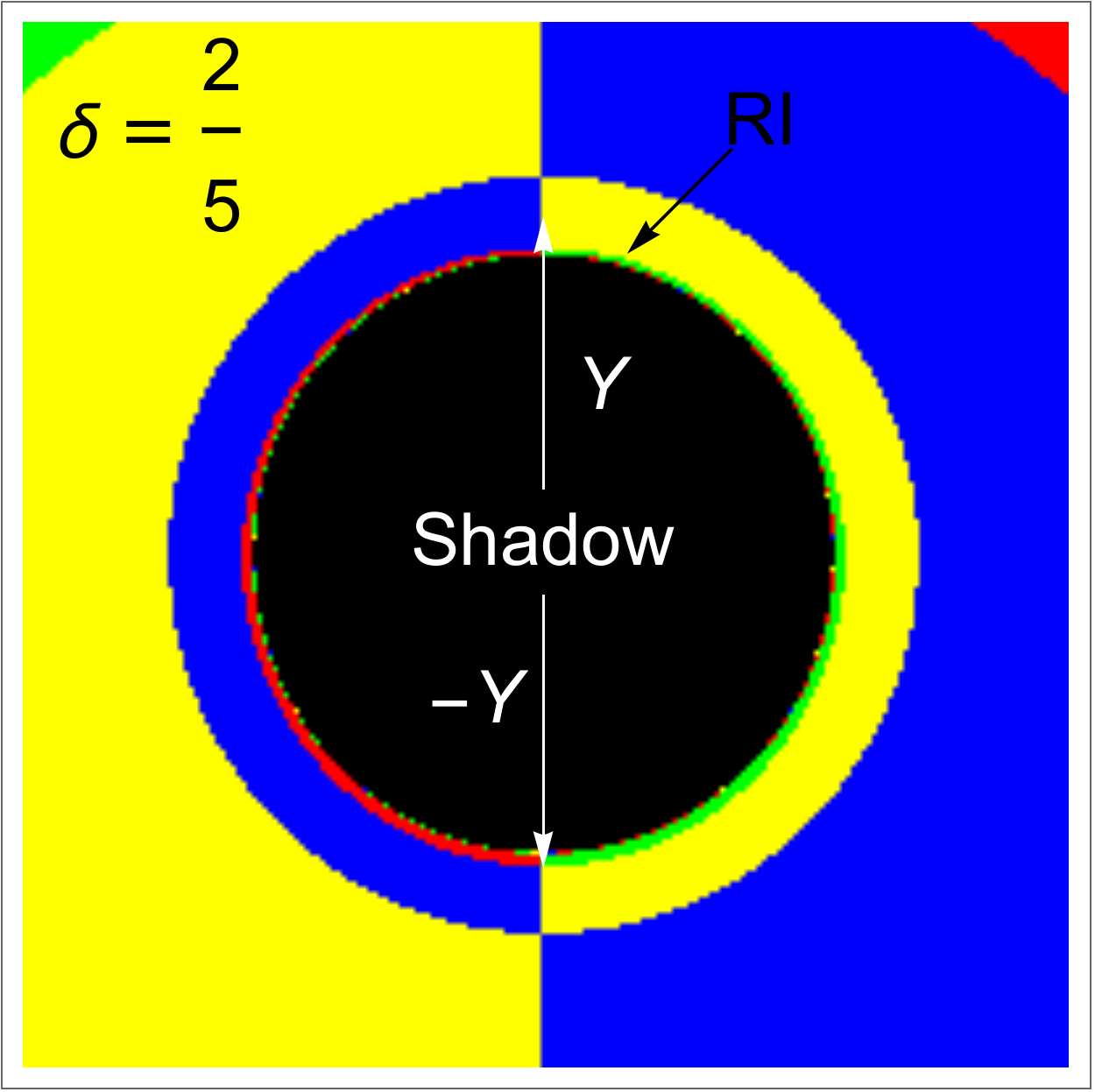}
		\label{NSH25N}
		}
		\\
		 \subfloat[][ZV1/5, $\pi/2$]{
  		\includegraphics[scale=0.42]{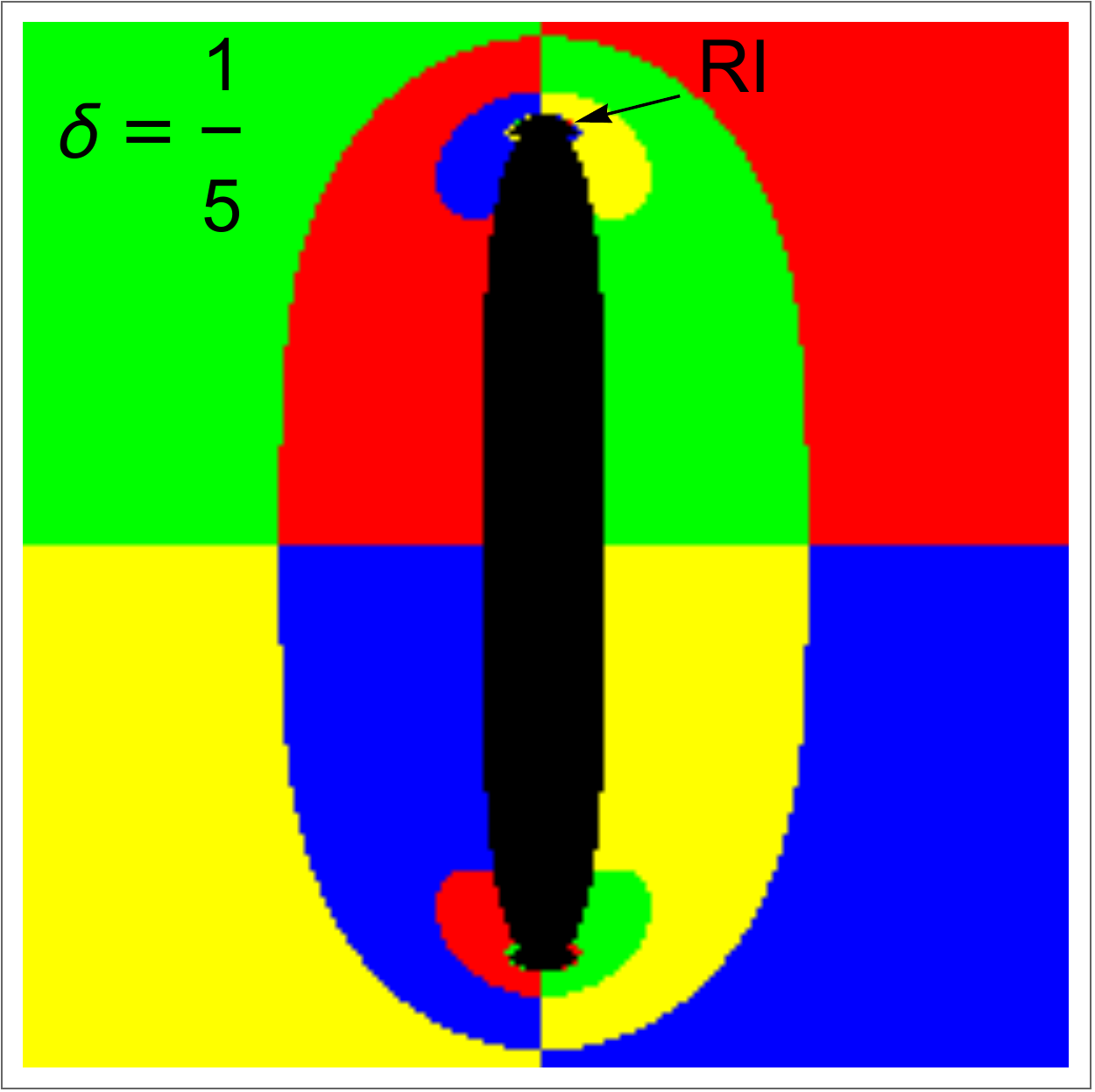}
		\label{NSH15E}
		}
			 \subfloat[][ZV1/5, $\pi/4$]{
  		\includegraphics[scale=0.42]{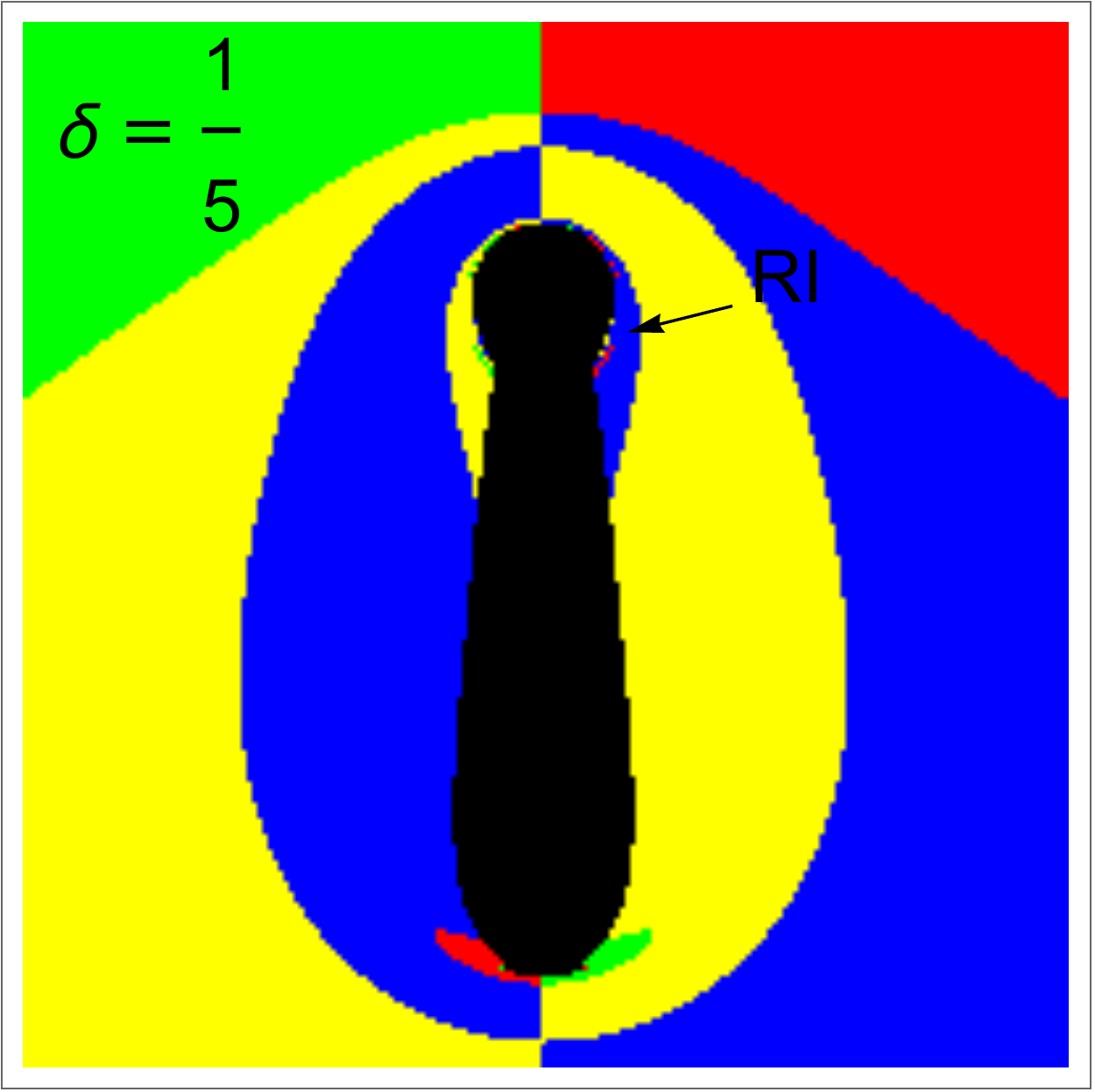}
		\label{NSH15NE}
		}
			 \subfloat[][ZV1/5, $\pi/12$]{
  		\includegraphics[scale=0.42]{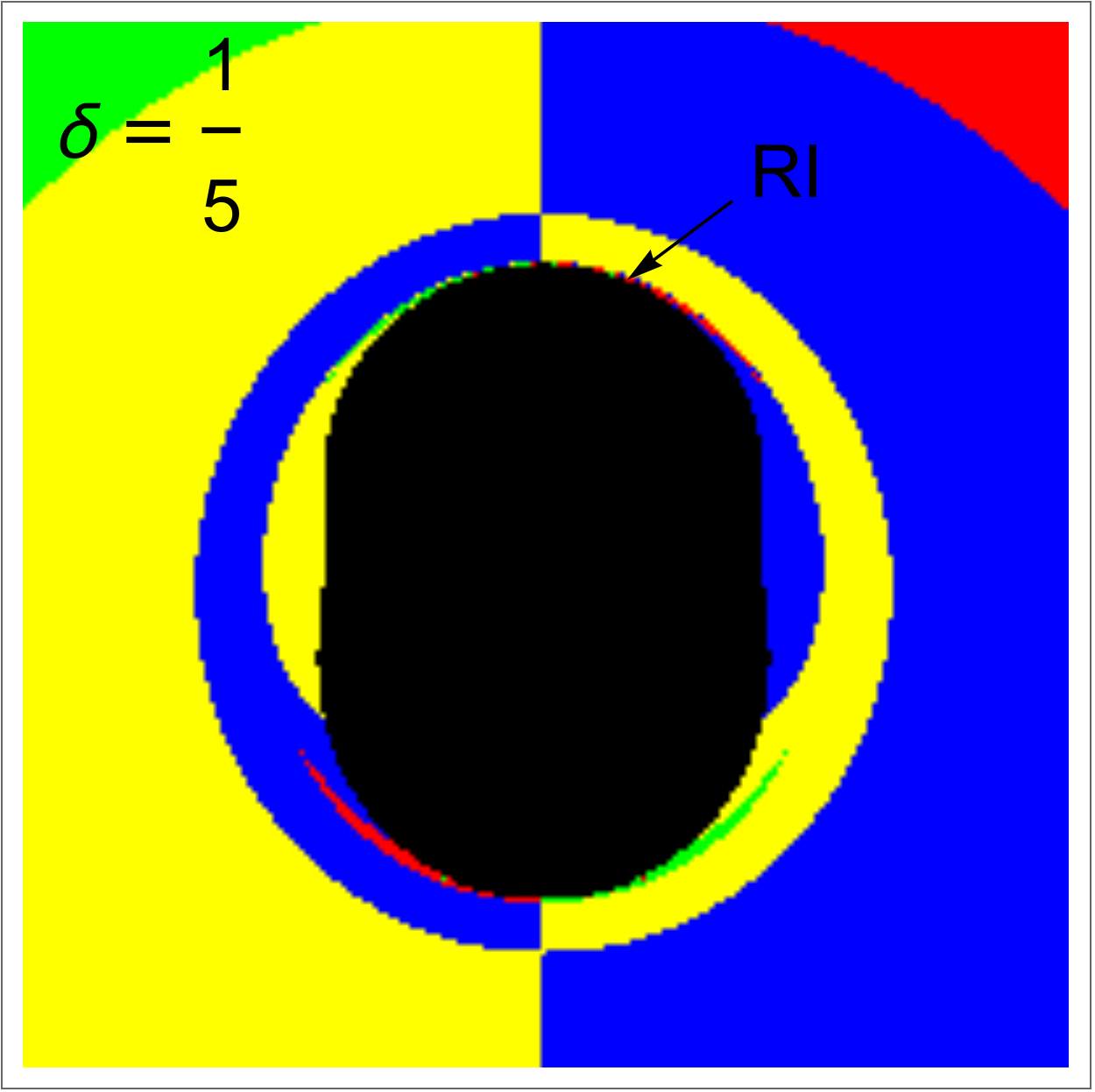}
		\label{NSH15N}
		}
\caption{The shadow of the ZV space $\delta<1$. Solutions with $\delta<1/2 $ belong to optical type II. Size $200\times200$ pixels.}
\label{Kerr5}
\end{figure}

Finally, we build the complete image of the shadow along with the relativistic images from the ZV solution Figs. \ref{KerrF} - \ref{Kerr5} just like it was done in \cite{Cunha:2015yba,Bohn:2014xxa,Cunha:2017wao,Cunha:2018acu}. From the observation point $ r_O = 5M $, a set of geodesics in different directions on the celestial sphere is launched in stereographic coordinates $ (X, Y) $ (\ref{c5}) and tracked to which part of the sphere $ r = 30M $ colored with four colors will fall into the geodesic. In this case, the initial conditions for the system of equations (\ref{c2}) are determined for each pixel from the formulas (\ref{c4}), where it is assumed $ E = 1 $. Geodesic falling on singularity form a shadow. In this case, the calculations also use the symmetry of $ X \rightarrow-X $.

The basic properties of the shadow were predicted by us earlier. In particular, when $ \delta> 1 $ we expect that the shadow is somewhat flattened (due to the presence of the photon region), with $ \delta <1 $ extended (the photon region inside ATTS${}_\theta$). When $ \delta <1/2 $, the optical structure differs significantly. In particular, some of the relativistic images disappear.

In the Figs. \ref{SH0E} - \ref{SH0NE}, an image of the celestial sphere of empty space is given as a trivial example for an observer located at points with different angles of $ \theta_O $. Figs. \ref{SH1E} - \ref{SH1N} represent the image of the Schwarzschild metric for different viewing angles. The shadow of a spherically symmetric object, as expected, is a circle, and does not depend on the angle of observation. The appearance of relativistic images can be observed on the border of the shadow.

The Figs. \ref{SH2E} - \ref{SHIN} represent the image of the deformed ZV metrics with $ \delta> 1 $ for different observer positions. At the border of the shadow, we can again observe the appearance of relativistic images, but this time the shadow is oblate and the their shape depends on the observer angle $ \theta_O $. In addition, its size is significantly reduced compared to the spherically symmetric metric, as was previously predicted Figs. \ref{K}.

Figs. \ref{Kerr5} represent the image of deformed ZV metrics with $ \delta <1 $. As expected, the shadow is extended in the direction of the poles. When $ \delta <1/2 $ (Figs. \ref{NSH25E} - \ref{NSH15N}), the solutions are significantly deformed, relativistic images disappear in the vicinity of the equatorial plane, the image differs significantly from type I Kerr-like solutions \cite{Abdikamalov:2019ztb}. Thus, the classification \cite{Galtsov:2019bty} introduced by us remains valid in the case of ZV metrics.

\section{Conclusion and discussion}
In this paper we developed further our formalism, intended to characterize the optical appearance of metrics associated with compact astrophysical objects without resorting to the integration of geodetic equations. In absence of spherical symmetry, crucial role in such an analysis is played by transversely trapping photon surfaces whose definition generalize that of the photon spheres. TTSs can be determined by examining the properties of the second quadratic forms of timelike hypersurfaces in space-time. Further specifying the TTSs in a static axially-symmetric spacetime according relations between their principal curvatures, one is able to define suitable generalization of the notion of the photon regions known in the Kerr metric to  situation when geodesic equations are non-integrable and consequently the fundamental photon orbits can not be described explicitly.

In terms of principal curvatures $\lambda_A,\;A=t,\,\theta,\phi $, the totally umbilic photon surfaces correspond to equal $\lambda_A$, while for the newly introduced TTS${}_\theta$, only two principal curvatures are equal ($\lambda_\theta=\lambda_t$). Similarly, one can define the surface TTS${}_\phi$. Using these definitions, we then described the generalized photon region as a one-parameter family of surfaces containing both three-dimensional (A)TTS${}_\theta$ and two-dimensional surfaces (A)TTS${}_\phi$ at each  boundary. In particular, if there exists an equatorial circular photon orbit (ECO), the photon region can be regarded as some one-parameter $\rho $-interpolation between the ECO ($\rho=\rho_{\rm max}$) and the (A)TTS${}_\theta$ ($\rho=0$). 

Applying this technique to Zipoy-Voorhees spacetime, we demonstrated that
the existence of a photon region and the TTSs is not a unique feature of rotating solutions. At the same time,  the presence of spheroidal photon orbits \cite{Glampedakis:2018blj} indicates the non-integrability of the corresponding dynamical system and the occurrence of geodesic chaos \cite{Glampedakis:2018blj,Lukes,Dolan:2019gsr}. 

We got a series of numerical and approximate analytical expressions   (\ref{f4}, \ref{f10}) and analyzed their relationship with the behavior of the scattered geodesics. We have found that geodesics with $\rho=0$ may wind on the (A)TTS${}_\theta$ surface creating relativistic images \cite{Virbhadra:2008ws,Virbhadra:2002ju} and determining the vertical size of the shadow in the coordinates of the stereographic projection of the observer’s celestial sphere. Instead, geodesic with nonzero impact parameter $\rho\neq0$ can wind on various photon region surfaces similarly to the Kerr case \cite{Teo}  determining the shape of the shadow and the relativistic images. We also managed to obtain an explicit analytical formula for the equatorial size of the shadow (\ref{g5}, \ref{g6}).

We find that  ZV family of solutions falls under the optical classification of Kerr-like solutions that we introduced previously \cite{Galtsov:2019bty} and has a somewhat similar to Kerr optical structure (this was also observed in \cite{Toshmatov:2019qih,Abdikamalov:2019ztb}). Namely, the solutions with $ \delta> 1/2 $, including Chazy-Curzon metric, belong to the type I, and with $ \delta <1/2 $  --- to the type II. For $ \delta>1 $ the shadow is somewhat flattened and  extended with respect to the case $ \delta <1 $. These features are confirmed by the explicit calculation of the shadow and the structure of the optical images in Figs. \ref{Kerr4} - \ref{Kerr5} (see also \cite{Abdikamalov:2019ztb}). We also obtained the explicit parameters of the shadow  in Figs. \ref{Kerr2n}, and analyzed their relationship with the behavior of the scattered geodesic and the structure of the characteristic surfaces.

We believe that the description of the optical properties of the  metrics in the geometric terms of principal curvatures without solving the geodesic equations can be  useful in further applications, in particular, in analysis of integrability, and is worth to be   developed further.
\section*{Acknowledgement}
We thank G\'erard Cl\'ement and Igor Bogush for  valuable comments. The work was supported by the Russian Foundation for Basic Research on the project 17-02-01299a, and by the Government program of competitive growth of the Kazan Federal University. The
authors would like to acknowledge the networking support
of the COST Action No. CA16104.


\begin{thebibliography}{lab}

\bibitem{Cardoso:2019rvt} 
  V.~Cardoso and P.~Pani,
  arXiv:1904.05363 [gr-qc].

\bibitem{Perlick:2004tq} 
  V.~Perlick,
  Living Rev.\ Rel.\  {\bf 7}, 9 (2004).

\bibitem{Cunha:2018acu} 
  P.~V.~P.~Cunha and C.~A.~R.~Herdeiro,
  Gen.\ Rel.\ Grav.\  {\bf 50}, no. 4, 42 (2018)
  doi:10.1007/s10714-018-2361-9
  [arXiv:1801.00860 [gr-qc]]. 
  
\bibitem{Bohn:2014xxa} 
  A.~Bohn, W.~Throwe, F.~Hébert, K.~Henriksson, D.~Bunandar, M.~A.~Scheel and N.~W.~Taylor,
  Class.\ Quant.\ Grav.\  {\bf 32}, no. 6, 065002 (2015)
  doi:10.1088/0264-9381/32/6/065002
  [arXiv:1410.7775 [gr-qc]].
  
\bibitem{Cunha:2015yba} 
  P.~V.~P.~Cunha, C.~A.~R.~Herdeiro, E.~Radu and H.~F.~Runarsson,
  Phys.\ Rev.\ Lett.\  {\bf 115}, no. 21, 211102 (2015)
  doi:10.1103/PhysRevLett.115.211102
  [arXiv:1509.00021 [gr-qc]].

\bibitem{Cunha:2017wao} 
  P.~V.~P.~Cunha, J.~A.~Font, C.~Herdeiro, E.~Radu, N.~Sanchis-Gual and M.~Zilhão,
  Phys.\ Rev.\ D {\bf 96}, no. 10, 104040 (2017)
  doi:10.1103/PhysRevD.96.104040
  [arXiv:1709.06118 [gr-qc]].
  
\bibitem{Abdikamalov:2019ztb}
  A.~B.~Abdikamalov, A.~A.~Abdujabbarov, D.~Ayzenberg, D.~Malafarina, C.~Bambi and B.~Ahmedov,
  arXiv:1904.06207 [gr-qc].

  \bibitem{A.deVries} 
 A. de Vries, Class.\ Quantum Grav.\ {\bf 17}, 123 (2000).

\bibitem{Abdujabbarov:2012bn} 
  A.~Abdujabbarov, F.~Atamurotov, Y.~Kucukakca, B.~Ahmedov and U.~Camci,
  Astrophys.\ Space Sci.\  {\bf 344}, 429 (2013)
  doi:10.1007/s10509-012-1337-6
  [arXiv:1212.4949 [physics.gen-ph]].

\bibitem{Wei:2013kza} 
  S.~W.~Wei and Y.~X.~Liu,
  JCAP {\bf 1311}, 063 (2013)
  doi:10.1088/1475-7516/2013/11/063
  [arXiv:1311.4251 [gr-qc]].

\bibitem{Atamurotov:2013sca} 
  F.~Atamurotov, A.~Abdujabbarov and B.~Ahmedov,
  Phys.\ Rev.\ D {\bf 88}, no. 6, 064004 (2013).
  doi:10.1103/PhysRevD.88.064004

\bibitem{Takahashi:2005hy} 
  R.~Takahashi,
  Publ.\ Astron.\ Soc.\ Jap.\  {\bf 57}, 273 (2005)
  doi:10.1093/pasj/57.2.273
  [astro-ph/0505316].

\bibitem{Bambi:2008jg} 
  C.~Bambi and K.~Freese,
  Phys.\ Rev.\ D {\bf 79}, 043002 (2009)
  doi:10.1103/PhysRevD.79.043002
  [arXiv:0812.1328 [astro-ph]].

\bibitem{Amarilla:2010zq} 
  L.~Amarilla, E.~F.~Eiroa and G.~Giribet,
  Phys.\ Rev.\ D {\bf 81}, 124045 (2010)
  doi:10.1103/PhysRevD.81.124045
  [arXiv:1005.0607 [gr-qc]].

\bibitem{Amarilla:2013sj} 
  L.~Amarilla and E.~F.~Eiroa,
  Phys.\ Rev.\ D {\bf 87}, no. 4, 044057 (2013)
  doi:10.1103/PhysRevD.87.044057
  [arXiv:1301.0532 [gr-qc]].


\bibitem{Shaikh:2018lcc} 
  R.~Shaikh, P.~Kocherlakota, R.~Narayan and P.~S.~Joshi,
  Mon.\ Not.\ Roy.\ Astron.\ Soc.\  {\bf 482}, no. 1, 52 (2019)
  doi:10.1093/mnras/sty2624
  [arXiv:1802.08060 [astro-ph.HE]].

\bibitem{Johannsen:2015mdd} 
  T.~Johannsen,
  Class.\ Quant.\ Grav.\  {\bf 33}, no. 11, 113001 (2016)
  doi:10.1088/0264-9381/33/11/113001
  [arXiv:1512.03818 [astro-ph.GA]].

\bibitem{Johannsen:2016uoh} 
  T.~Johannsen,
  Class.\ Quant.\ Grav.\  {\bf 33}, no. 12, 124001 (2016)
  doi:10.1088/0264-9381/33/12/124001
  [arXiv:1602.07694 [astro-ph.HE]].


\bibitem{Kubiznak:2007kh} 
  D.~Kubiznak and P.~Krtous,
  Phys.\ Rev.\ D {\bf 76}, 084036 (2007)
  doi:10.1103/PhysRevD.76.084036
  [arXiv:0707.0409 [gr-qc]].

\bibitem{Johannsen:2011dh} 
  T.~Johannsen and D.~Psaltis,
  Phys.\ Rev.\ D {\bf 83}, 124015 (2011)
  doi:10.1103/PhysRevD.83.124015
  [arXiv:1105.3191 [gr-qc]].

\bibitem{Hartle:1967he} 
  J.~B.~Hartle,
  Astrophys.\ J.\  {\bf 150}, 1005 (1967).
  doi:10.1086/149400

\bibitem{Hartle:1968si} 
  J.~B.~Hartle and K.~S.~Thorne,
  Astrophys.\ J.\  {\bf 153}, 807 (1968).
  doi:10.1086/149707

  \bibitem{Lukes} G. Lukes-Gerakopoulos, 
Phys.\ Rev. \ D {\bf 86}. (2012)  [arXiv:1206.0660].

\bibitem{Dolan:2019gsr} 
  S.~R.~Dolan,
  arXiv:1901.01202 [gr-qc].


\bibitem{Virbhadra:1999nm} K.~S.~Virbhadra and G.~F.~R.~Ellis, 
Phys.\ Rev.\ D {\bf 62}, 084003 (2000).
doi:10.1103/PhysRevD.62.084003
[astro-ph/9904193].

\bibitem{Cederbaum}
  C.~Cederbaum and G.~J.~Galloway,
  Class.\ Quant.\ Grav.\  {\bf 33}, 075006 (2016)
  doi:10.1088/0264-9381/33/7/075006
  [arXiv:1508.00355 [math.DG]].


\bibitem{Yazadjiev:2015hda} 
  S.~S.~Yazadjiev,
  Phys.\ Rev.\ D {\bf 91}, no. 12, 123013 (2015)
  doi:10.1103/PhysRevD.91.123013
  [arXiv:1501.06837 [gr-qc]].
  
\bibitem{Yazadjiev:2015jza}
  S.~Yazadjiev and B.~Lazov,
  Class.\ Quant.\ Grav.\  {\bf 32}, 165021 (2015)
  doi:10.1088/0264-9381/32/16/165021
  [arXiv:1503.06828 [gr-qc]].
  
\bibitem{Yazadjiev:2015mta} 
  S.~Yazadjiev and B.~Lazov,
  Phys.\ Rev.\ D {\bf 93}, no. 8, 083002 (2016)
  doi:10.1103/PhysRevD.93.083002
  [arXiv:1510.04022 [gr-qc]].
 
\bibitem{Rogatko}
  M.~Rogatko,
  Phys.\ Rev.\ D {\bf 93}, no. 6, 064003 (2016)
  doi:10.1103/PhysRevD.93.064003
  [arXiv:1602.03270 [hep-th]].

\bibitem{Cederbaumo}
  C.~Cederbaum,
  arXiv:1406.5475 [math.DG].
  
\bibitem{Yoshino:2016kgi} 
  H.~Yoshino,
  Phys.\ Rev.\ D {\bf 95}, no. 4, 044047 (2017)
  doi:10.1103/PhysRevD.95.044047
  [arXiv:1607.07133 [gr-qc]].


\bibitem{Claudel:2000yi} C.~M.~Claudel, K.~S.~Virbhadra and G.~F.~R.~Ellis,
J.\ Math.\ Phys.\ {\bf 42}, 818 (2001)
doi:10.1063/1.1308507
[gr-qc/0005050].

\bibitem{Gibbons}
  G.~W.~Gibbons and C.~M.~Warnick,
  Phys.\ Lett.\ B {\bf 763}, 169 (2016)
  doi:10.1016/j.physletb.2016.10.033
  [arXiv:1609.01673 [gr-qc]]
 
\bibitem{Paganini:2016pct} 
  C.~F.~Paganini, B.~Ruba and M.~A.~Oancea,
  arXiv:1611.06927 [gr-qc].

\bibitem{Teo} 
  E.~Teo,
  Gen.\ Rel.\ Grav {\bf 35}, 1909 (2003). 


\bibitem{Grenzebach}
  A.~Grenzebach, V.~Perlick and C.~Lammerzahl,
  Phys.\ Rev.\ D {\bf 89}, no. 12, 124004 (2014)
  doi:10.1103/PhysRevD.89.124004
  [arXiv:1403.5234 [gr-qc]].
 
\bibitem{Grenzebach:2015oea} 
  A.~Grenzebach, V.~Perlick and C.~L?mmerzahl,
  Int.\ J.\ Mod.\ Phys.\ D {\bf 24}, no. 09, 1542024 (2015)
  doi:10.1142/S0218271815420249
  [arXiv:1503.03036 [gr-qc]].

\bibitem{Charbulak:2018wzb} 
  D.~Charbul\'ak and Z.~Stuchlik,
  Eur.\ Phys.\ J.\ C {\bf 78}, no. 11, 879 (2018)
  doi:10.1140/epjc/s10052-018-6336-5
  [arXiv:1811.02648 [gr-qc]].


\bibitem{Pappas:2018opz} 
  G.~Pappas and K.~Glampedakis,
  arXiv:1806.04091 [gr-qc].
  
\bibitem{Glampedakis:2018blj} 
  K.~Glampedakis and G.~Pappas,
  arXiv:1806.09333 [gr-qc]. 

\bibitem{Cunha:2017eoe} 
  P.~V.~P.~Cunha, C.~A.~R.~Herdeiro and E.~Radu,
  Phys.\ Rev.\ D {\bf 96}, no. 2, 024039 (2017)
  doi:10.1103/PhysRevD.96.024039
  [arXiv:1705.05461 [gr-qc]].

\bibitem{Yoshino1}
  H.~Yoshino, K.~Izumi, T.~Shiromizu and Y.~Tomikawa,
  PTEP {\bf 2017}, no. 6, 063E01 (2017)
  doi:10.1093/ptep/ptx072
  [arXiv:1704.04637 [gr-qc]].

\bibitem{Galtsov:2019bty} 
  D.~V.~Gal'tsov and K.~V.~Kobialko,
  Phys.\ Rev.\ D {\bf 99}, no. 8, 084043 (2019)
  doi:10.1103/PhysRevD.99.084043
  [arXiv:1901.02785 [gr-qc]].

\bibitem{Zipoy} 
D.~M.~Zipoy, J. Math. Phys. {\bf 7}, 1137 (1966)

\bibitem{Voorhees:1971wh} 
  B.~H.~Voorhees,
  Phys.\ Rev.\ D {\bf 2}, 2119 (1970).
  doi:10.1103/PhysRevD.2.2119

\bibitem{Griffiths}  J.~B.~Griffiths and J.~Podolsky, ``Exact Space-Times in Einstein's General Relativity.''\ Cambridge University Press, \ 2009.

\bibitem{Kodama:2003ch} H.~Kodama and W.~Hikida,
Class.\ Quant.\ Grav.\ {\bf 20}, 5121 (2003)
doi:10.1088/0264-9381/20/23/011
[gr-qc/0304064].

  
\bibitem{Malafarina:2004yw} 
  D.~Malafarina,
  Conf.\ Proc.\ C {\bf 0405132}, 273 (2004).

\bibitem{Herrera:1998rj} 
  L.~Herrera, F.~M.~Paiva and N.~O.~Santos,
  Int.\ J.\ Mod.\ Phys.\ D {\bf 9}, 649 (2000)
  doi:10.1142/S021827180000061X
  [gr-qc/9812023].

\bibitem{Boshkayev:2015jaa} 
  K.~Boshkayev, E.~Gasperin, A.~C.~Gutierrez-Pineres, H.~Quevedo and S.~Toktarbay,
  Phys.\ Rev.\ D {\bf 93}, no. 2, 024024 (2016)
  doi:10.1103/PhysRevD.93.024024
  [arXiv:1509.03827 [gr-qc]].
  
 
\bibitem{Chazy}   
J.~Chazy,  1924. Bull. Soc. Math. Fr. {\bf52},  17 (1924)
 
\bibitem{Curzon} 
H.~E.~J.~Curzon,  Proc. Lond. Math. Soc. {\bf2}, 477 (1925)


\bibitem{Montero-Camacho:2014bza} 
  P.~Montero-Camacho, F.~Frutos-Alfaro and C.~Gutierrez-Chaves,
  Revista de Matem\'atica: Teor\'ia y Aplicaciones, 22 (2): 265-274,
  2015
  doi:10.15517/rmta.v22i2.20833
  [arXiv:1405.2899 [gr-qc]].


\bibitem{Okumura}
M.~Okumura,
 Kodai Math.\ Sem. Rep.\ {\bf19},  35 (1967).

\bibitem{Senovilla:2011np} 
  J.~M.~M.~Senovilla,
  arXiv:1111.6910 [math.DG].

\bibitem{Herrera:2013hm} 
  L.~Herrera, A.~Di Prisco, J.~Ibáñez and J.~Ospino,
  Phys.\ Rev.\ D {\bf 87}, no. 2, 024014 (2013)
  doi:10.1103/PhysRevD.87.024014
  [arXiv:1301.2424 [gr-qc]].

\bibitem{Synge}
J.~L.~Synge,
Mon.\ Not.\ R.\ astro. Soc.\ {\bf131}, 463 (1966).

	 \bibitem{Semerak} 
 O. Semerak, Helvetica\ Physica \ Acta,\ {\bf 69}, no. 1, 69 (1996).
	doi:10.5169/seals-116907

\bibitem{Bozza:2002zj} 
  V.~Bozza,
  Phys.\ Rev.\ D {\bf 66}, 103001 (2002)
  doi:10.1103/PhysRevD.66.103001
  [gr-qc/0208075].

\bibitem{Virbhadra:2008ws} 
  K.~S.~Virbhadra,
  Phys.\ Rev.\ D {\bf 79}, 083004 (2009)
  doi:10.1103/PhysRevD.79.083004

 \bibitem{Virbhadra:2002ju}
  K.~S.~Virbhadra and G.~F.~R.~Ellis,
  Phys.\ Rev.\ D {\bf 65}, 103004 (2002).
  doi:10.1103/PhysRevD.65.103004

\bibitem{Toshmatov:2019qih}
  B.~Toshmatov, D.~Malafarina and N.~Dadhich,
  arXiv:1905.01088 [gr-qc].




\end{thebibliography}
\end{document}